\documentclass[11pt,a4paper]{article}

\usepackage{jcappub}
\usepackage{graphicx}
\usepackage{amssymb}
\usepackage{amsmath}
\usepackage{epsfig,multicol,bbm}
\usepackage{amsmath,amsfonts,amssymb,longtable,mathtools}
\usepackage{array}
\usepackage{graphicx}
\usepackage{subfig}
\usepackage{enumerate}
\usepackage{verbatim}
\usepackage{bm}
\usepackage{latexsym}
\usepackage{epsfig}
\usepackage{psfrag}
\usepackage{ulem}
\usepackage{textcomp}
\usepackage{color}
\usepackage{colortbl} 
\usepackage{placeins}
\usepackage{longtable,booktabs,caption}
\usepackage{changepage}
\usepackage{float}
\usepackage[utf8]{inputenc}

\makeatletter
\gdef\@fpheader{}
\makeatother

\def\nn{\nonumber}

\def\d{{\rm d}}

\def\be{\begin{equation}}
\def\ee{\end{equation}} 
\def\bea{\begin{eqnarray}}
\def\eea{\end{eqnarray}} 
\def\bseq{\begin{subequations}}
\def\eseq{\end{subequations}}
\def\ba{\begin{align}}
\def\ea{\end{align}}

\def\Mp{M_{_{\rm Pl}}}

\def\a1{\alpha_{_1}}
\def\b1{\beta_{_1}}
\def\de1{\delta_{_1}}
\def\g1{\gamma_{_1}}

\def\Nt{\tilde{N}}

\newcommand{\ie}{\textit{i.e.~}}

\begin{document}
\hskip12.5cm  ET-0065A-22
\title{\huge The rise of the primordial tensor spectrum from an early  scalar-tensor epoch}
\author
{Debika Chowdhury,}
\author
{Gianmassimo Tasinato,}
\author
{Ivonne Zavala}
\affiliation
{Department of Physics, Swansea University, Swansea, SA2 8PP, U.K.}
\emailAdd{debika.chowdhury@swansea.ac.uk} 
\emailAdd{g.tasinato@swansea.ac.uk}
\emailAdd{e.i.zavalacarrasco@swansea.ac.uk}
\abstract{Primordial gravitational waves (PGW)  produced during inflation span a large range of frequencies, carrying information on the dynamics of the primordial universe. 
During an early scalar-tensor dominated epoch, the amplitude of the PGW spectrum can be enhanced  over a wide range of frequencies. To study this phenomenon,  we focus on a class of scalar-tensor theories, well motivated by high energy theories of dark energy and dark matter,  where the scalar is conformally and disformally coupled  to matter during the early cosmological evolution. For a conformally dominated epoch, the PGW  spectrum has a flat step-like shape. More interestingly, a  disformally dominated epoch is characterised by a peaked  spectrum with a broken power-law profile, with  slopes depending  on  the scalar-tensor theory considered.
We introduce a  graphical tool, called broken power-law sensitivity curve,  as a convenient visual indicator for understanding whether a given broken power-law profile can   be detected    by   GW experiments. We then analyse   the GW spectra for a variety of representative conformal and disformal models,   discussing their detectability prospects with the Einstein Telescope (ET), Laser Interferometer Space Antenna (LISA), DECi-hertz Interferometer Gravitational wave Observatory (DECIGO), and Big Bang Observer (BBO).
  }

\maketitle

\section{Introduction}
Scalar-tensor theories are ubiquitous in extensions of the Standard Model (SM) of particle physics and cosmology.   For example, string theory approaches to  particle physics and inflationary model building generically predict the presence of several new ingredients, in particular, new particles such as scalar fields with clear geometrical interpretations.  These scalars couple conformally and disformally to  matter living on  branes, extended objects where matter  is localised. In string D-brane constructions, longitudinal string fluctuations are identified with the matter fields such as the SM and/or dark matter (DM) particles, while transverse fluctuations correspond to scalar fields. Constraints on scalar fields and its couplings to matter today are extremely tight, for example from solar system tests \cite{Will:2014kxa,Sakstein:2014isa,Ip:2015qsa,Wang:2016lxa}, and recently from  strict bounds on the speed of gravitational waves \cite{LIGOScientific:2017zic}.

On the other hand, the thermal history of the  early Universe remains uncertain. Its evolution involves a sequence of epochs, each characterised by a certain expansion rate $H$, which 
 is key to understanding the physics of processes occurring during  these  eras. According to current observational evidence, the universe was radiation dominated at the time of Big Bang Nucleosynthesis (BBN) and, most likely, there was an early era of vacuum domination known as inflation.
 However, there might have been a non-standard cosmological history between the end of inflation and the onset of BBN. 

An epoch of a scalar-tensor modification to general relativity (GR) domination before the onset of BBN, can change the cosmological expansion rate, without violating present constraints on scalar fields and its couplings to matter. For example, a modified expansion rate felt by matter can change the standard predictions for the dark matter  relic abundance as studied in \cite{Catena:2004ba, Catena:2009tm, Gelmini:2013awa, Rehagen:2014vna, Wang:2015gua, Lahanas:2006hf, Pallis:2009ed, Salati:2002md, Arbey:2008kv, Iminniyaz:2013cla, Meehan:2014zsa, Meehan:2014bya, Meehan:2015cna,Dutta:2016htz} for the conformal case, and in \cite{Dutta:2017fcn} for the disformal case. 

In this work, we investigate the imprints of a scalar-tensor dominated epoch  with  conformal and disformal couplings, on the primordial gravitational wave  spectrum produced during inflation. 
This spectrum spans a large range of frequencies with an almost scale-invariant profile,  whose amplitude is
 too small to be detected by future  gravitational wave experiments. 
We show how this signal is enhanced during an era of scalar-tensor domination with distinct signatures in   the conformal and disformal cases. 

The conformal case, which characterises the Brans-Dicke class of scalar-tensor theories, was recently discussed in \cite{Bernal:2020ywq} for a particular choice of conformal coupling. In this case, the PGW flat spectrum is enhanced to a flat spectrum with a larger amplitude during the scalar-tensor epoch. We demonstrate that this step-like enhancement depends on the conformal factor (and initial conditions),  offering the possibility of probing its existence using correlations between different experiments. 

Remarkably,  during a  disformally dominated epoch, the PGW spectrum has a characteristic peak with a distinctive frequency profile, 
which offers a smoking-gun signature of the early scalar-tensor  epoch. In order to easily  understand in a visual manner whether any
given broken power-law GW spectrum can be detected by a GW experiment, we introduce the notion of broken power-law sensitivity curve (BPLS). 


The paper is organised as follows. In section~\ref{Sec:2}, we review the calculation of the primordial gravitational wave spectrum in standard cosmology, following closely \cite{Watanabe:2006qe,Saikawa:2018rcs,Bernal:2019lpc,Bernal:2020ywq}. In section~\ref{Sec:3},
 we  introduce the general modified cosmological setup due to scalar-tensor theories motivated by D-branes, as well as a  more phenomenological set-up, popular in the literature. 
 We focus on the calculation of the modified expansion rate for different cases of interest:  a purely conformal case, a purely disformal case, and a conformal-disformal case in D-brane-like scenarios.
 In section~\ref{Sec:4}, we  discuss the rise of the primordial gravitational wave spectrum due to a scalar-tensor epoch in the three cases discussed in section~\ref{Sec:3}.
  Specifically, subsection~\ref{Sec:phenoGW} focuses on the conformal rise of the PGW spectrum. Subsection~\ref{section_disfrise} discusses the interesting disformal case, which gives rise to a characteristic  peaked spectrum. In passing, we explain how  our results differ  with a similar peaked spectrum arising from a  short kination era due to a spinning axion. 
 In subsection~\ref{sec_BPLS}, we introduce and   discuss the
 concept of the broken power-law sensitivity curves.
 After our conclusions in section~\ref{sec:end}, appendix~\ref{sec:ICsDisform} contains more details on the calculations in section~\ref{Sec:3}.

\section{Primordial Gravitational Waves in Standard Cosmology}
\label{Sec:2}
In this section, we briefly review the PGW spectrum evolution in standard cosmologies \cite{Watanabe:2006qe,Saikawa:2018rcs,Bernal:2019lpc,Bernal:2020ywq}.  Given a model of inflation, a stochastic background of GWs arises inevitably from the primordial tensor fluctuations, whose equation of motion is given by 
\be\label{eq:hij}
\ddot h_{ij} +3 H \dot h_{ij} - \frac{\nabla^2}{a^2}h_{ij} =0\,,
\ee 
where $H$ is the Hubble expansion rate in GR and we consider that the evolution of the primordial tensor fluctuations is source-free for the frequencies we are interested in, $f \gtrsim 10^{-10}$ Hz, corresponding to temperatures $T\gtrsim 4$ MeV.  To solve the  tensor perturbation equation, one can write it in Fourier space as 
\be\label{eq:hijFourier}
h_{ij} (t,\vec{x}) = \sum_\lambda \int{\frac{d^3k}{(2\pi)^3} h^\lambda(t,\vec{k}) \epsilon_{ij}^\lambda(\vec{k}) e^{i\vec{k}\cdot \vec{x}}     }\,,
\ee
with $\lambda =+, \times$ corresponding to the two independent polarisations, and $\epsilon^\lambda$ being the spin-2 polarisation tensor satisfying the normalisation condition $ \epsilon_{ij}^\lambda \epsilon_{ij}^{\lambda'\,*} = 2\delta^{\lambda\lambda'}$. 
Using \eqref{eq:hijFourier}, the solution to \eqref{eq:hij} can be written as
\be\label{eq:hk}
h^{\lambda}(t,\vec{k}) = h^{\lambda}_{\rm inf}(\vec{k}) {\mathcal T}(t,\vec{k})\,,
\ee
where ${\mathcal T}$ is a transfer function  and $h^{\lambda}_{\rm inf}(\vec{k}) $ is the amplitude of the tensor perturbation.
The energy density of the relic GW today is given by 
\be
\rho_{GW}(t) = \frac{1}{16\pi G} \sum_\lambda\int{\frac{d^3k}{(2\pi)^3} |\dot h^{\lambda}(t,\vec{k}) |^2}\,.
\ee
The relic density of the PGWs from the tensor perturbation is calculated from 
\be
\Omega_{GW} (t, k )= \frac{1}{\rho_c(t)} \frac{d \rho_{GW}(t,k)}{d \ln k}\,, 
\ee
where $\rho_c$ is the critical density of the Universe. This can be further rewritten using \eqref{eq:hk} as \cite{Watanabe:2006qe,Saikawa:2018rcs,Bernal:2019lpc,Bernal:2020ywq}
\be\label{eq:Ogw}
\Omega_{GW} (\tau, k )= \frac{{\cal P}_t(k) \left({\cal T}_{_{\tau}}(\tau,k)\right)^2}{12 a^2(\tau) H^2(\tau)} \simeq \frac{1}{24}{\cal P}_t(k)\left(\frac{a_{\rm hc}}{a(\tau)}\right)^4\left(\frac{H_{\rm hc}}{H(\tau)}\right)^2,
\ee
where the subscript `$\tau$' denotes derivatives with respect to  conformal time, \ie $d\tau= dt/a$, and the subindex `hc' denotes horizon crossing \cite{Bernal:2020ywq}, \ie $k=~a_{\rm hc}\,H_{\rm hc}$. The primordial tensor spectrum ${\cal P}_t(k)$ is determined by
\be
{\cal P}_t(k) = \frac{2 H^2}{\pi^2\Mp^2}\Bigg|_{k=aH}\,.
\ee
The fractional energy density in primordial gravitational waves, as observed today, can be expressed as
\begin{equation}\label{eq:0gw-simple}
 \Omega^0_{\rm GW}(k) \simeq \frac{1}{24}\,{\cal P}_{t}(k)\,\left[\frac{a_{\rm hc}}{a_0}\right]^4\,\left[\frac{H_{\rm hc}}{H_0}\right]^2,
\end{equation}
where $H_0$ denotes the Hubble parameter today.
Using entropy conservation, we can write
\be
\frac{a_{\rm hc}}{a_0} = \left(\frac{g_{*s,0}}{g_{*s,{\rm hc}}}\right)^{1/3}
\frac{T_0}{T_{\rm hc}}\,.\,\label{eq:sf-temp}
\ee
where $g_{*s}$ represents the entropy degrees of freedom (see subsection~\ref{Sec:dof} for more details). 
Combining the above two equations, we obtain
\begin{equation}
 \Omega^0_{\rm GW}(k)\,h^2 \simeq \frac{1}{24}\,{\cal P}_{t}(k)\, \left(\frac{g_{*s,0}}{g_{*s,{\rm hc}}}\right)^{4/3}\,\left(\frac{T_0}{T_{\rm hc}}\right)^4\,\left(\frac{H_{\rm hc}}{H_0/h}\right)^2.\label{eq:omegagwh2-gr-rhototal}
\end{equation}
The amplitude of the primordial tensor power spectrum is given by ${\cal P}_{t}=r\,A_{\rm S}$, where $A_{\rm S}=2.1\times 10^{-9}$ is the amplitude of the primordial scalar power spectrum~\cite{Planck:2018jri}, and we consider the latest upper bound for the tensor-to-scalar ratio reported by BICEP/Keck, $r=0.036$~\cite{BICEP:2021xfz}. 

Now, the frequency, $f_0$, of GWs observed today, at which the corresponding mode $k=2\pi f_0 a_0$ (where $a_0$ is the scale factor today) reenters the horizon, can be related to the temperature $T_{\rm hc}$ via the following relation \cite{Kamionkowski:1993fg,Maggiore:1999vm,Watanabe:2006qe,Saikawa:2018rcs}:
\begin{equation}
 f_0 = 2.41473\times 10^{23}\,\left(\frac{T_0}{T_{\rm hc}}\right) \left(\frac{g_{*s,0}}{g_{*s,{\rm hc}}}\right)^{1/3} \sqrt{\frac{8\,\pi\,G\,\rho_{\rm hc}}{3}}\,\,{\rm Hz},\label{eq:freq-temp}
\end{equation}
where $\rho_{\rm hc}$ is the total energy density of the universe at horizon crossing.
Using this equation, we can obtain the evolution of $\Omega_{\rm GW}$ as a function of $f_0$ in section~\ref{Sec:4}.
%

\section{Scalar-tensor theories}\label{Sec:3}
In this section we  introduce the  general scalar-tensor setup describing a D-brane like scalar-tensor theory with conformal and disformal couplings motivated by D-brane world scenarios, as well as a more phenomenological setup, where these couplings are  in principle  not related to each other. We follow closely \cite{Dutta:2016htz,Dutta:2017fcn}, extending and generalising the results found there. 
The starting action  is given by:
\begin{equation}
    S = S_{\rm EH} + S_{\phi} + S_m,
\end{equation}
where
\begin{subequations}
 \begin{align}
     S_{\rm EH} &= \frac{1}{2\kappa^2}\,\int\,\d^4 x\,\sqrt{-g}\,R, \\
     S_{\phi} &= -\int\,\d^4 x\,\sqrt{-g}\left[\frac{b}{2} (\partial\phi)^2+ M^4 C^2(\phi)\,\sqrt{1+\frac{D(\phi)}{C(\phi)}\,\left(\partial \phi\right)^2} + V(\phi)\right], \label{eq:Sphi}\\
     S_m &= - \int\,\d^4 x\,\sqrt{-\tilde{g}}\,{\mathcal L}_M (\tilde{g}_{\mu\nu}),
 \end{align}
\end{subequations}
where $\kappa^2=\Mp^{-2}=8\,\pi\,G$, $b=0,1,$ depending on whether we are in a phenomenological ($b=1$) or D-brane ($b=0$) set-up. We will see below, how each different set-up affects the results.  $M$ is a mass scale, which can be related to the tension of the D-brane \cite{Dutta:2016htz,Dutta:2017fcn}, and we take $M=0$ when we consider the phenomenological model. 

%
%
The disformally coupled metric, $\tilde g_{\mu\nu}$, is given by\footnote{This relation is consistent with the most general physically consistent relation between
two metrics in the presence of a scalar field first discussed by Bekenstein in  \cite{Bekenstein:1992pj}. }
\begin{equation}\label{disformalg}
 \tilde{g}_{\mu\nu}=C(\phi)\,g_{\mu\nu} + D(\phi)\,\partial_\mu\phi\,\partial_\nu\phi,
\end{equation}
where  $C(\phi)$ and $D(\phi)$ are the conformal and disformal couplings of the scalar to the metric, respectively\footnote{In a (D-)brane scenario, \eqref{disformalg} corresponds to the induced metric on the brane where the scalar field describes  the (D-)brane position in the internal space. }. These functions are in principle arbitrary up to  a causality constraint, which requires $C > 0$ and  $C + 2DX> 0$ \cite{Bekenstein:1992pj}. In a  D-brane motivated set-up, these functions are related via $M^4CD=1$ (see Appendix C of \cite{Dutta:2016htz} for details) and automatically satisfy the causality constraint. 
\subsection*{Cosmological equations}
Consider  an homogeneous and isotropic FLRW metric given by
\begin{equation}
    \d s^2=-\d t^2 + a(t)^2\,\d x_i\,\d x^i,
\end{equation}
with $a(t)$ being the scale factor. In this background, the equations of motion become
\bea
&& H^2 =\frac{\kappa^2}{3} \left[\rho_\phi +\rho\right]\,, \label{friedmann1A}\\
&& \dot H + H^2 = -\frac{\kappa^2}{6}\left[ \rho_\phi+ 3P_\phi +\rho +3 P \right]\,,\label{eq:einstein1}\\
&& \ddot \phi \left[1+ \frac{b}{M^4CD\gamma^3}\right]+3H\dot\phi \left[ \gamma^{-2} + \frac{b}{M^4CD\gamma^3} \right]  \nonumber \\
&&\hskip0.5cm + \frac{C}{2D}\left(\gamma^{-2}\left[\frac{5 C'}{C}  - \frac{D'}{D}\right] 
 + \frac{D'}{D}- \frac{C'}{C} -4\gamma^{-3} \frac{C'}{C}  \right)
 + \frac{1}{M^4CD\gamma^{3}}\, ({\mathcal V}'+Q_0) =0 \label{eq:eom1} \,,  \nonumber \\
\eea
where $H= \dot{a}/a$, the dots indicate derivatives with respect to $t$, and the primes indicate derivatives with respect to the field $\phi$. The Lorentz factor is given by 
\be\label{eq:gamma}
\gamma= (1-D \,\dot\phi^2/C)^{-1/2},
\ee
while  
\bea
Q_0 = \rho \left[ \frac{D}{C} \,\ddot \phi + \frac{D}{C} \,\dot \phi \left(\!3H + \frac{\dot \rho}{\rho} \right) \!+ \!\left(\!\frac{D'}{2C}-\frac{D}{C}\frac{C'}{C}\!\right) \dot\phi^2 +\frac{C'}{2\,C} (1-3\,\omega)
\right], \nonumber \\
\eea
and we have used the equation of state,  $P=\omega\rho$, with $P$ and $\rho$ being the pressure and energy density, respectively. For the scalar field, the energy density $\rho_\phi$ and the pressure $P_\phi$ are given by 
\be\label{rhoPA2}
\rho_\phi = \left[\frac{b}{2}+\frac{M^4C D\gamma^2}{\gamma+1}\right] \dot\phi^2 + {\mathcal V}  \,, \qquad 
P_\phi =  \left[\frac{b}{2}+\frac{M^4C D\,\gamma}{\gamma+1}\right]\dot\phi^2 - {\mathcal V}  \,,
\ee
where  $\mathcal{V} \equiv V + C^2 M^4$ (see \cite{Dutta:2016htz,Dutta:2017fcn} for details).  

The  energy-momentum conservation  equation gives $\nabla_\mu T^{\mu\nu}_{tot} = \nabla_\mu \left(T^{\mu\nu}_{\phi}+  T^{\mu\nu}\right)=0$. That is, the scalar field and matter are not separately conserved in the Einstein frame. The time component of this constraint yields the equations
\bea 
&&\dot\rho_\phi + 3H(\rho_{\phi}+P_{\phi}) = -Q_0\dot\phi\,, \label{contA}\\
&&\dot\rho + 3H(\rho+P) = Q_0\,\dot\phi\,.\label{cont1A}
\eea
Using the last equation \eqref{cont1A}, $Q_0$ can be rewritten as 
\be\label{Q0A}
Q_0 = \rho\left( \frac{\dot \gamma}{\dot \phi\, \gamma} + \frac{C'}{2C}  (1-3\,\omega \,\gamma^2) -3H\omega\,\frac{(\gamma^2 -1) }{\dot \phi}\right) \,.
\ee
Plugging this into  the  
(non)conservation equation for  matter \eqref{cont1A}, we obtain
\be\label{conservaDMA}
\dot \rho + 3H (\rho + P\,\gamma^{2}) = \rho \left[\frac{\dot \gamma}{\gamma} + \frac{C'}{2C} \,\dot\phi\, (1-3\,\omega \gamma^2)\right]\,.
\ee

We are interested  in the disformal, or Jordan, frame, which is defined as the frame where matter and entropy are conserved.  
That is, $\tilde{\nabla}_\mu\,\tilde{T}^{\mu\nu}=0$, where $\tilde{\nabla}_\mu$ is computed with respect to the disformal metric, and the energy-momentum tensor is defined  by
\begin{equation}
    \tilde{T}^{\mu\nu}=\frac{2}{\sqrt{-\tilde{g}}}\,\frac{\delta\,S_M}{\delta\,\tilde{g}_{\mu\nu}},
\end{equation}
where the tilde denotes the Jordan frame. 
This can be related to the energy-momentum tensor in the Einstein frame through the following relation:
\begin{equation}
    \tilde{T}^{\mu\nu}=C^{-3}\,\gamma\,T^{\mu\nu}.
\end{equation}
Then the energy densities, pressures, equations of state, and the scale factors in the Einstein and Jordan frames are  related through the following expressions:
\begin{subequations}\label{eq:einstjor}
 \begin{align}
     \tilde{\rho} &= C^{-2}\,\gamma^{-1}\,\rho, \\
     \tilde{p} &= C^{-2}\,\gamma\,P, \\
     \tilde{w} &= w\,\gamma^2, \\
     \tilde{a} &= C^{1/2}\,a.
 \end{align}
\end{subequations}
The Hubble parameter in the Jordan frame is given by
\be\label{eq:Hubble-gen-1}
\tilde H \equiv \frac{d \ln{\tilde a}}{d\tilde t} = \frac{\gamma}{C^{1/2}}\left[ H + \frac{C'}{2C}\dot \phi \right]\,,
\ee
with $d\tilde t = C^{1/2} \gamma^{-1} d t$, so that it  is computed from $H$ and $\phi$. As indicated above, in the Jordan frame, the continuity equation for  matter takes the standard form, that is \cite{Dutta:2016htz}:
\be\label{eq:tilderhocons}
\frac{d{\tilde \rho}}{d\tilde t}  +  3 \tilde H( \tilde \rho +\tilde P)  =0 \,.
\ee
Note that if we consider a purely disformal case, with $C=1$, then the scale factor in both frames coincide. Moreover, at the onset of BBN, whrn $C=\gamma=1$, the two frames coincide. 
 
\subsection{Modified expansion rate}\label{Sec:MER}
We are interested in computing the modified expansion rate in the Jordan frame to compare it with the standard GR evolution.  For this purpose, it is convenient to introduce the dimensionless scalar  $\varphi=\kappa\,\phi$, and   swap time derivatives with derivatives with respect to the number of e-folds, \ie $N=\ln(a/a_0)$, so that $dN=Hdt$. 
With these changes, we can rewrite  equations \eqref{friedmann1A}-\eqref{eq:eom1} as follows 
\bea
&&H^2 = \frac{\kappa^2}{3}\frac{C^2\,\gamma\,(1+\lambda)}{B}\, \tilde\rho \label{H2}   \\
&& H_N=- H\left[\frac{3\,B}{2\,(1+\lambda)} \left(1+ \tilde w \,\gamma^{-2}\right) + \frac{\varphi_{_N}^2}{2}\left(b +M^4CD\,\gamma \right)\right], 
\label{HN} \\
&& \varphi_{_{NN}}\left[1+ \frac{b}{M^4CD\gamma^3} + \frac{\gamma^{-1}}{M^4C^2}\frac{3\,B\,H^2}{\kappa^2\,(1+\lambda)}\right] 
 + 3\,\varphi_{_N}\left[\gamma^{-2}+\frac{b}{M^4CD\gamma^3} 
 - \frac{\tilde w}{M^4C^2\gamma^3}\,\frac{3\,B\,H^2}{\kappa^2\,(1+\lambda)}\right]  \nn\\
&&\hskip0.5cm + \frac{H_{_N}}{H}\,\varphi_{_N} \left[1+  \frac{b}{M^4CD\gamma^3}+   \frac{\gamma^{-1}}{M^4C^2}\frac{3\,B\,H^2}{\kappa^2\,(1+\lambda)}\right]     
 + \frac{\gamma^{-3}}{M^4CD}\frac{3\,B}{(1+\lambda)}\,\alpha(\varphi)\left(1-3\,\tilde w\right)
 \nn\\
&& \hskip0.1cm   
+ \frac{\gamma^{-1}}{M^4C^2} \frac{3\,B\,H^2}{\kappa^2\,(1+\lambda)}\, \varphi_{_N}^2
\left[\delta(\varphi)-\alpha(\varphi)\right]\!
+ \!\frac{\kappa^2C}{H^2D}\left[ \gamma^{-2}(5\alpha(\varphi)-\delta(\varphi))+\delta(\varphi)-\alpha(\varphi)(1+4\gamma^{-3})\right]
\nn\\
&&\hskip9cm 
+\frac{3\,B\,\lambda}{M^4CD\,\gamma^3(1+\lambda)}\,\frac{{\mathcal V}_{,\varphi}}{\mathcal V}  =0\,, \label{fiNN}
\eea
where the subscript $N$ denotes a derivative with respect to the number of e-folds, we have used \eqref{eq:einstjor} to replace the energy density and equation of state in the Jordan frame,   and we have  defined   
\bea
&&B= 1-\varphi_N^2\left(\frac{b}{6}+\frac{M^4CD\,\gamma^2}{3\,(\gamma+1)}\right)\label{B}  \,,\\
&& \alpha(\varphi) \equiv \frac{d\ln C^{1/2}}{d\varphi}\,,\label{eq:alphadef}\\
&&\delta(\varphi) \equiv \frac{\d\,{\rm ln}\,D^{1/2}}{\d \varphi} \,,\label{eq:deltadef} 
\eea
and $\lambda = {\cal V}/\rho\, (=\tilde {\cal V}/\tilde \rho)$. Also, in terms of $\varphi$ and  $N$-derivatives, the Lorentz factor is  given by
\be\label{eq:gamma2}%
\gamma^{-2} = 1-\frac{H^2}{\kappa^2}\frac{D}{C}\,\varphi_N^2\,.
\ee

\smallskip

Now, the expansion rate  in General Relativity (GR), which we express is given by  ($\kappa_{GR}$ will be specified below)
\be\label{HGR}
H_{GR}^2 = \frac{\kappa_{GR}^2}{3} \tilde \rho\,,
\ee
using  \eqref{eq:einstjor} we can  write $H_{GR}$ entirely as a function of $H, \varphi, \varphi_N$ as follows (see \cite{Catena:2004ba,Dutta:2016htz,Dutta:2017fcn}):
\be\label{eq:H-HGR}
H_{GR}^2 = \frac{\kappa_{GR}^2}{\kappa^2} \frac{C^{-2}B\,\gamma^{-1}H^2}{(1+\lambda)}\,.
\ee
Therefore, once we find a solution for $H$ and  $\varphi$, we can compare the expansion rates $\tilde H$ with $H_{GR}$, by introducing the following parameter $\xi$, which measures  the departure from the standard expansion:
\be\label{xi}
\xi\equiv \frac{\tilde H}{H_{GR}} = \frac{\kappa}{\kappa_{GR}} \frac{\gamma^{3/2} C^{1/2}\left[1+\alpha(\varphi) \,\varphi_N \right] (1+\lambda)^{1/2}}{B^{1/2}}\,,
\ee
where 
\begin{equation}\label{kappas}
    \kappa^2_{\rm GR}=\kappa^2\,C(\varphi_{_0})\left[1+\alpha^2(\varphi_{_0})\right],
\end{equation}
with $\kappa_{\rm GR}^2$ being the gravitational constant and $\kappa^2$ being the value of the gravitational constant measured by local experiments for conformally coupled theories, and $\varphi_{_0}$ is the value of the scalar field at the present time.
Notice that $\xi$ can be larger or smaller than one, indicating an enhancement or reduction of  $\tilde H$ with respect to $H_{GR}$. This means that $\tilde H$ can grow during the cosmological evolution\footnote{This does not imply a violation of the the null energy condition (NEC) because  the Einstein frame expansion rate $H$ is dictated by the energy density $\rho$ and pressure $p$, which obey the NEC and therefore $\dot H<0$ during the whole evolution, as it should be.}.

Below we  solve the system of equations \eqref{H2}, \eqref{HN}, and \eqref{fiNN} numerically. For this, we first change the  number of e-folds to the Jordan frame,  $\tilde N$, which is given by  \cite{Dutta:2016htz}
 \begin{align}
    N \equiv {\rm ln}\,\frac{a}{a_{_0}} &= {\rm ln}\biggl[\frac{\tilde{T_{_0}}}{\tilde{T}}\biggl(\frac{g_{*s}(\tilde{T_{_0}})}{g_{*s}(\tilde{T})}\biggr)^{1/3}\biggr] + {\rm ln}\biggl[\frac{C_{_0}}{C(\varphi)}\biggr]^{1/2} \nn\\
    &= \tilde{N} + {\rm ln}\biggl[\frac{C_{_0}}{C(\varphi)}\biggr]^{1/2}.
 \end{align}
Thus, in terms of $\Nt$, the derivatives of $\varphi$ can be expressed as
\begin{subequations}
 \begin{align}
     \varphi_{_N} &= \frac{1}{\left[1-\alpha(\varphi)\,\varphi_{_{\Nt}}\right]}\,\varphi_{_{\Nt}},\\
     \varphi_{_{NN}} &= \frac{1}{\left[1-\alpha(\varphi)\,\varphi_{_{\Nt}}\right]^3}\,\left[\varphi_{_{\Nt\Nt}}+\frac{\d\,\alpha}{\d\,\varphi}\,\varphi_{_{\Nt}}^3\right].
 \end{align}\label{eq:deriv-transformations}
\end{subequations}

\subsubsection{Degrees of freedom and the kick function}\label{Sec:dof}
When the conformal factor is turned on, $\alpha(\varphi)\ne 1$ and thus the equation of motion for the scalar field \eqref{fiNN}, has a term, $\alpha(\varphi)(1-3\,\tilde\omega)$, which depends non-trivially on the equation of state $\tilde \omega$, and which acts as an effective potential, ${\rm V}_{\rm eff}= \ln C^{1/2}$,  when $\tilde \omega\ne 1/3$.
Thus, deep in the radiation dominated era, when $\tilde{w} \sim 1/3$, this term drops out.  As the universe cools down, when the temperature of the universe drops below the rest mass of a particle species, the particle becomes non-relativistic, and there arise small departures in the value of $\tilde{w}$ from $\tilde{w}=1/3$. This leads to non-zero contributions due to  this term, $(1-3\,\tilde{w})$  in  Eq.~(\ref{fiNN}), which generates a kick in the scalar field as we will see.  This factor is referred to as the `kick function'.

\begin{figure}[!h]
 \begin{center}
 \includegraphics[width=7.50cm]{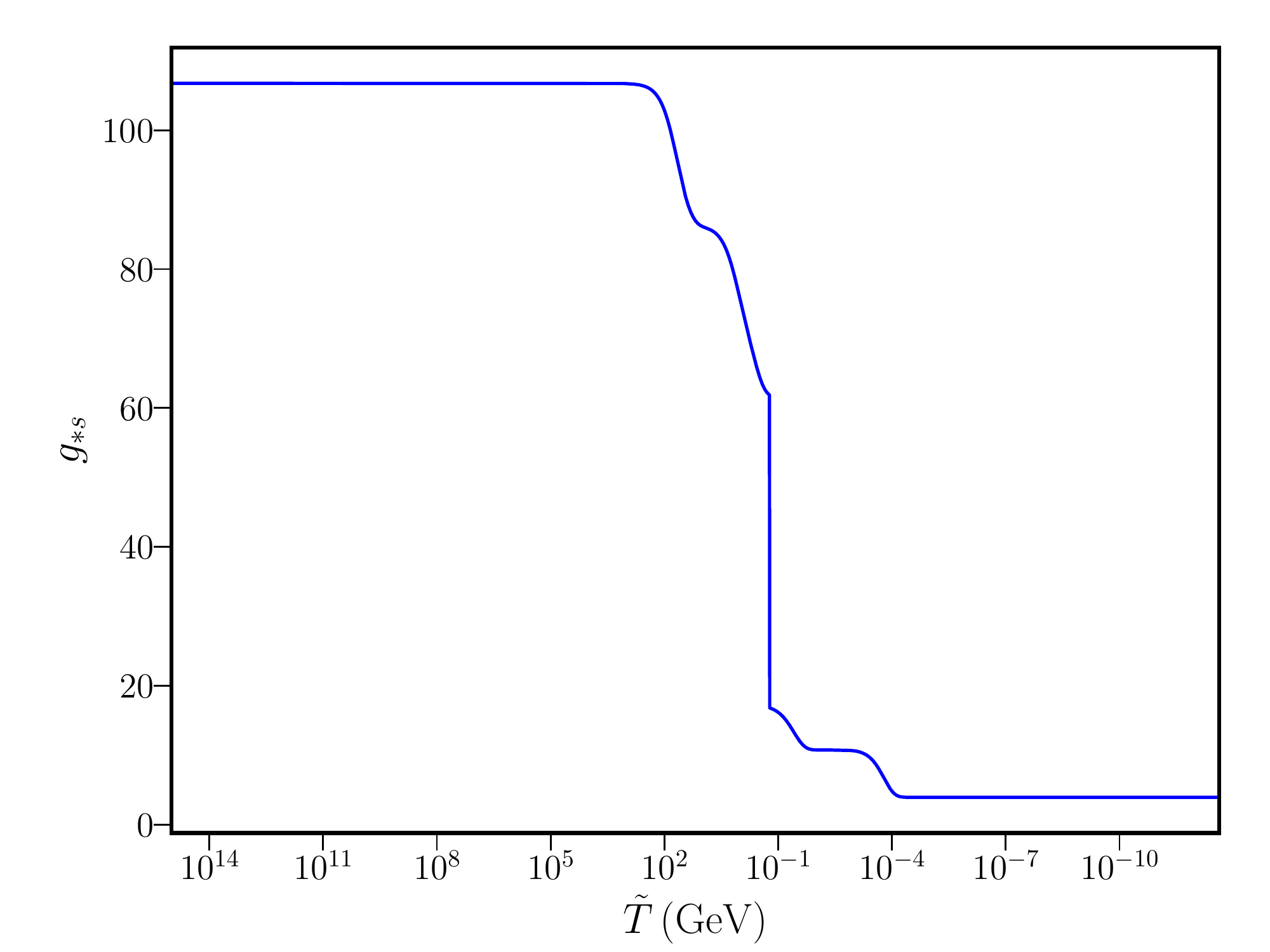}
 \end{center}
 \vskip -15pt
 \caption{The evolution of the total number of entropy degrees  of freedom $g_{*s}$ has been plotted with respect to temperature. This quantity starts from the value of $106.75$ and undergoes successive changes as the different particles become non-relativistic. The value finally settles to around $3.9$.}\label{fig:gtot}
 \end{figure}

We now compute the kick function referred to above, which we shall use to solve Eqs.~\eqref{HN} and \eqref{fiNN} numerically. 
For a particle species `A' that is in thermal equilibrium with the radiation bath during the radiation dominated era, its energy density and pressure are given by \cite{Watanabe:2006qe, Saikawa:2018rcs, Erickcek:2013dea, Coc:2006rt}
\begin{subequations}
\begin{align}
 \tilde{\rho}_A^R &= \frac{g_A \tilde{T}^4}{2 \pi^2}\,\int_{0}^{\infty} \frac{x^2 \sqrt{x^2+y_A^2}\, \d x}{{\rm e}^{\sqrt{x^2+y_A^2}} \pm 1}, \\
 \tilde{p}_A^R &= \frac{g_A \tilde{T}^4}{6 \pi^2}\,\int_{0}^{\infty} \frac{x^4\, \d x}{\sqrt{x^2+y_A^2}\left({\rm e}^{\sqrt{x^2+y_A^2}} \pm 1\right)},
\end{align}
\end{subequations}
where $x=\sqrt{(E/\tilde{T})^2-y_A^2}$, $y_A=m_A/\tilde{T}$, and $g_A$ is the number of degrees of freedom for each species `A'.
The total energy density and pressure is equal to the sum of energy densities and pressures of all these particles:
\begin{subequations}
\begin{align}
\tilde{\rho}^R=\sum_A\,\tilde{\rho}_A^R, \label{eq:rho-r} \\
\tilde{p}^R=\sum_A\,\tilde{p}_A^R \label{eq:p-r}.
\end{align}
\end{subequations}
The total number of entropy degrees of freedom is given by 
(see Fig.~\ref{fig:gtot})
\begin{equation}
 g_{*s}=\sum_A\,\frac{15}{4 \pi^4}\,g_A \int_{0}^{\infty} \frac{x^2\left(4x^2+3y_A^2\right)\, \d x}{\sqrt{x^2+y_A^2}\left({\rm e}^{\sqrt{x^2+y_A^2}} \pm 1\right)}.
\end{equation}
The evolution of this quantity is essential to  relate the scale factor and the temperature of the universe through entropy conservation~\eqref{eq:sf-temp}.

We evaluate the energy density, pressure, and the total number of entropy degrees of freedom numerically for all the particles in Table~\ref{table:particles}. We then add each of these quantities to the corresponding quantities of the relativistic particles. Before the QCD phase transition, the relativistic species of interest are the gluons ($g_A=16$), photons ($g_A=2$), light quarks ($g_A=36$), and neutrinos ($g_A=6$). After the QCD phase transition, \ie below $170$ MeV, the relativistic species that remain are the photons and neutrinos. Also, after neutrino decoupling, \ie below $1$ MeV, we take into account the fact that the neutrino temperature evolves differently compared to the photon temperature. This can be estimated using conservation of entropy.

\begin{table}
\begin{center}
\begin{tabular}[!t]{|c|c|c||c|c|c|}
 \hline
 \multicolumn{3}{|c||}{fermions} & \multicolumn{3}{|c|}{bosons} \\
 \hline
 particle & $g_A$ & $m_A$(GeV) & particle & $g_A$ & $m_A$(GeV) \\
 \hline
 \multicolumn{6}{|c|}{before QCD phase transition} \\
 \hline
 top & 12 & 173.2 & Higgs & 1 & 125.1 \\
 \hline
 bottom & 12 & 4.18 & Z & 3 & 91.19 \\
 \hline
 charm & 12 & 1.27 & W$^{\pm}$ & 6 & 80.39 \\
 \hline 
 tau & 4 & 1.78 & & & \\
 \hline
 \multicolumn{6}{|c|}{after QCD phase transition} \\
 \hline
 muon & 4 & 0.106 & $\pi^0$ & 1 & 0.140 \\
 \hline 
 electron & 4 & $5.11\times 10^{-4}$ & $\pi^{\pm}$ & 2 & 0.135 \\
 \hline
\end{tabular}
\caption{The number of degrees of freedom and masses of the particles that contribute to the kick function. For each of the fermion species, the contributions due to antiparticles are included in the number of degrees of freedom.}\label{table:particles}
\end{center}
\end{table}

The kick function, $\tilde{\Sigma}$, during the radiation,  matter dominated and $\Lambda$ dominated epochs is given by
\begin{eqnarray}
 \tilde{\Sigma} &=& \frac{\tilde{\rho}^R - 3 \tilde{p}^R + \tilde{\rho}^M - 3 \tilde{p}^M + \tilde{\rho}^\Lambda - 3 \tilde{p}^\Lambda}{\tilde{\rho}^R + \tilde{\rho}^M + \tilde{\rho}^\Lambda} \nn\\
 &=& \frac{\tilde{\rho}^R - 3 \tilde{p}^R + \tilde{\rho}^M + 4 \tilde{\rho}^\Lambda}{\tilde{\rho}^R + \tilde{\rho}^M + \tilde{\rho}^\Lambda},\label{eq:kick-function}
\end{eqnarray}
where we have considered $\tilde{p}^M=0$, $\tilde{p}^\Lambda=-\tilde{\rho}^\Lambda$, and the equation of state parameter is given by $\tilde{w}=(1-\tilde{\Sigma})/3$.
So the  Hubble parameter  in GR that we use below, is given by the total energy density as
\begin{equation}
 H_{\rm GR}^2 = \frac{\kappa^2_{\rm GR}}{3}\,\tilde{\rho}_{\rm total} = \frac{\kappa^2_{\rm GR}}{3} \left(\tilde{\rho}^R + \tilde{\rho}^M + \tilde{\rho}^\Lambda\right),\label{eq:hgr}
\end{equation}
where
\begin{subequations}
 \begin{align}
  \tilde{\rho}^M &= \Omega^M_0\,\rho_0\left(\frac{\tilde{a}_{_0}}{\tilde{a}}\right)^3,\\
  \tilde{\rho}^\Lambda &= \Omega^\Lambda_0\,\rho_0,
 \end{align}
\end{subequations}
and $\tilde{\rho}^R$ is given by Eq.~(\ref{eq:rho-r}), while $\rho_0$, $\Omega^M_0$, and $\Omega^\Lambda_0$ are the energy density, the matter density parameter, and the dark energy density parameter evaluated today, respectively.
 Note that there is no contribution from the scalar field energy density in the Jordan frame. In the Einstein frame on the other hand, $\rho_{\rm total} = \rho^R+\rho^M+\rho^{\Lambda}+\rho_\varphi$.

\begin{figure}[!h]
\begin{center}
\includegraphics[width=8.0cm]{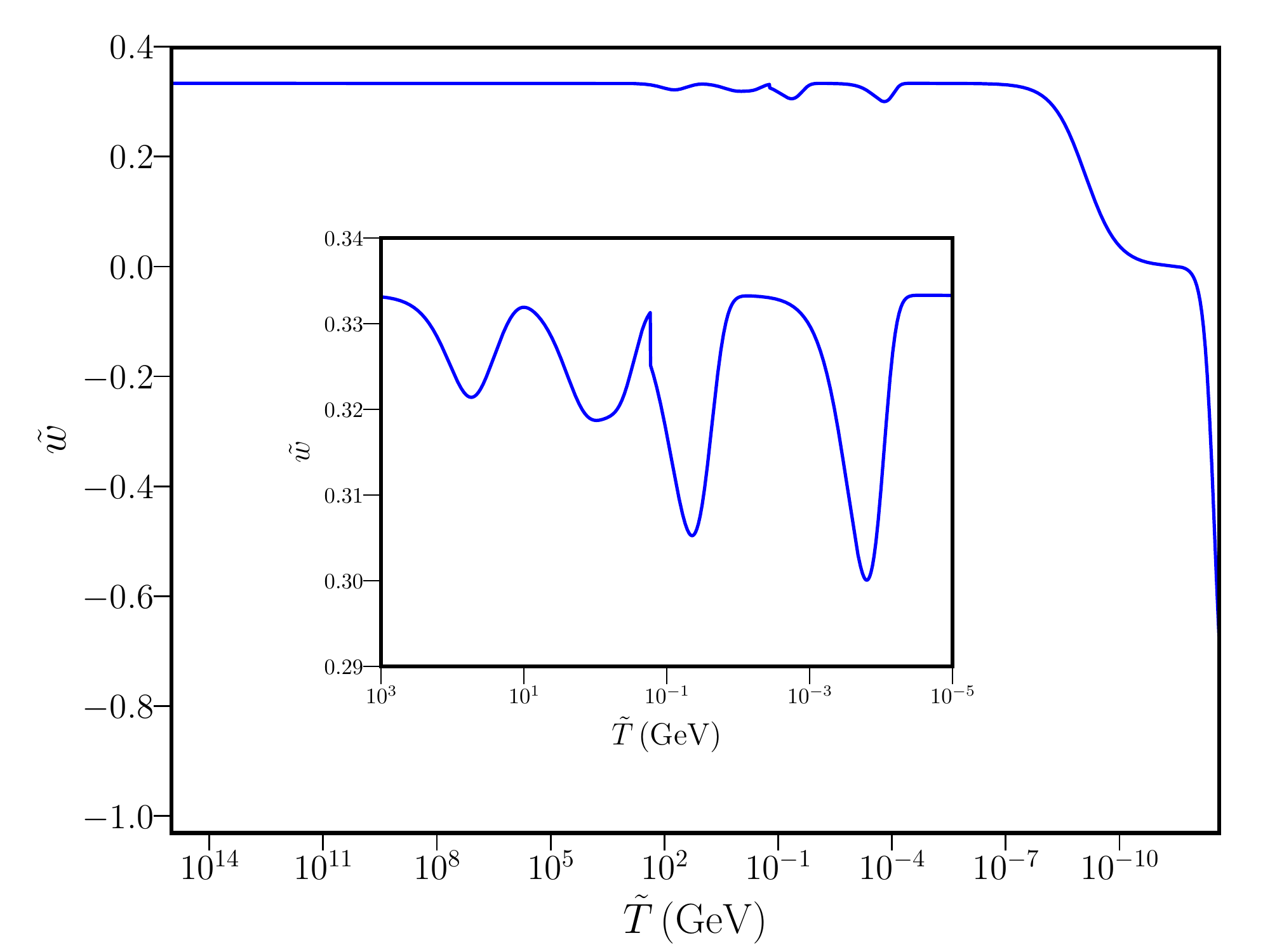}
\end{center}
\vskip -15pt
\caption{The evolution of the   equation of state parameter as function of the temperature. The inset highlights the non-trivial modifications to these function as different particles become non-relativistic during the radiation-dominated era.}\label{fig:sigmamod-wmod}
\end{figure}


We now focus on the two cases of interest separately, namely $(b=1,M=0)$ and $(b=0,M\ne 0)$, to study the expansion rate modification during an early scalar-tensor theory domination in the  universe's evolution. 

\subsection{Phenomenological case }\label{sec:Pheno1}
In this case, we take $M=0$ and $b=1$ in \eqref{eq:Sphi}. That is, the scalar field has a standard kinetic term and there is in principle no particular relation between $C$ and $D$, except the causality constraint $C>0$, $C+2D X>0$, $X\equiv\frac12(\partial\phi)^2$ \cite{Bekenstein:1992pj}.
%
%
Below  we consider the cosmological evolution for the following cases: a purely conformal case, that is $D=0$ ($\gamma=1$), and a purely disformal case, that is $C=1, D\ne 0$. 

\subsubsection{Purely conformal enhancement  }\label{section:purely-conformal}

In the purely conformal case, we have $D(\varphi)=0$, hence $\gamma=1$. 
We further take $\lambda\sim 0$, such that the dark energy today is fully dominated by a cosmological constant. In this case, the equations \eqref{HN}, \eqref{fiNN}
can be simplified into a single master equation, given by \cite{Damour:1993id,Coc:2006rt,Catena:2004ba}
\begin{equation}
    \frac{2}{3\,\left(1-\frac{\varphi_{_N}^2}{6}\right)}\,\varphi_{_{NN}} + \left(1-\tilde{w}\right)\varphi_{_N} + 2\left(1-3\,\tilde{w}\right)\alpha(\varphi)=0.\label{eq:conformal-diffeq-N}
\end{equation}
As we discussed above, deep in the radiation dominated era, $\tilde{w} \sim 1/3$, hence the last term in (\ref{eq:conformal-diffeq-N}) drops out 
and the equation can be solved analytically, to give $\varphi_{_N} \propto {\rm e}^{-N}$
 \ie the field velocity decreases      rapidly in the radiation dominated era. 
 Further integration shows that $\Delta \varphi=\varphi(N)-\varphi^i \simeq \varphi_N^i$ %
for $\sqrt{6}  \gg \varphi^{i}_{_N} $ and $N_i \to -\infty$ (the   Friedmann equation implies that $\varphi_{_N} \in (-\sqrt{6}, \sqrt{6})$) .
Thus  the field settles to a constant value after a few e-folds \cite{Damour:1993id,Coc:2006rt,Catena:2004ba}. This behaviour holds, even for larger values of  $\varphi^{i}_{_N}$, however for values close $\sqrt{6}$, the approximation $\Delta \varphi\simeq \varphi_N^i$ stops being valid, and the constant value to which the field settles differs largely from $\varphi_N^i$. 
We shall encounter this behaviour explicitly in the full numerical solution of (\ref{eq:conformal-diffeq-N}) (see Fig.~\ref{fig:phi-Cphi-conformal}).

As the universe cools down, when the temperature of the universe drops below the rest mass of a particle species, the particle becomes non-relativistic, and there arise small departures in the value of $\tilde{w}$ from $\tilde{w}=1/3$. This leads to non-zero contributions due to  $(1-3\,\tilde{w})$  in the last term in Eq.~(\ref{eq:conformal-diffeq-N}), which generate a kick in the scalar field with  $\alpha(\varphi)$ acting as an effective potential: ${\rm V}_{\rm eff}= \ln C^{1/2}$.

We now consider a set of  conformal functions motivated by the choice in \cite{Catena:2004ba}, which satisfy the requirement that the standard cosmological evolution is recovered at the onset of BBN. Moreover, this choice will allow us to demonstrate the dependence on this choice (and the initial conditions) on the enhancement of the expansion rate $\tilde H$, compared to the standard GR case, $H_{GR}$. In section \ref{Sec:4}, we will discuss how this impacts the PGW signal and the potential for its detection.  

\section*{Conformal expansion rate enhancement}

We consider a suitable modification of the conformal factor used in \cite{Catena:2004ba,Dutta:2016htz}, given by 
\begin{equation}\label{eq:conformal-factor}
C(\varphi)=(1+b\,{\rm e}^{-\beta\varphi})^{2n},
\end{equation}
with $b=0.1$ and $\beta=8$. In \cite{Catena:2004ba,Dutta:2016htz}, $n=1$ was chosen  such that the evolution of the Hubble parameter matches that of the standard GR evolution after BBN. 
Here we choose $n=1,2,4$ in order to demonstrate the dependence of the signal on the conformal factor. 

The effective potential is a runaway of the form $V_{\rm eff} = n\ln (1+b\,{\rm e}^{-\beta\varphi})$. As discussed above, deep in the radiation era, any initial velocity $\varphi_{N}$, goes rapidly to zero and  $\varphi\to $ constant value. As soon as $\tilde\omega$ differs slightly from 1/3, the effective potential kicks in, and the field rolls along it until $\tilde\omega \sim 1/3$ again \cite{Damour:1993id,Coc:2006rt,Catena:2004ba,Dutta:2016htz}. During the following  matter  and dark energy dominated eras, $\tilde \omega\ne 1/3$ and the field keeps  rolling down its effective potential, weighted by $2(1-3\,\tilde\omega)$ (see \eqref{eq:conformal-diffeq-N}).  

The enhancement of the expansion rate with respect to  GR depends on the initial conditions and the conformal factor, and thus can in principle be probed depending on its effects, e.g.~on the dark matter relic abundance \cite{Catena:2004ba,Dutta:2016htz}, or on the stochastic gravitational wave background, as we shall explore in the next section. 
In general, both the initial position and velocity of the scalar field can take any value, positive or negative. We choose initial conditions for the scalar field and its velocity to be less than the Planck scale, that is, in the range $(\varphi^i,\varphi^i_N)\in (\pm1,\pm1)$.  In the runaway effective potential dictated  by $\ln C^{1/2}$ for the conformal factor \eqref{eq:conformal-factor}, there are the following possibilities:
 \begin{enumerate}[{\bf (a)}]
 \item The scalar field starts somewhere up in the runaway effective potential with zero or positive initial velocity, thus staying at or reaching a constant value until the 
first particles become non-relativistic, giving a kick to the scalar field, so that it rolls down its potential until $\tilde \omega\sim 1/3$ again, and thereafter stays constant until the kick function releases it again during the radiation era. It  then evolves  rapidly until  $C$  reaches $1$, well before  BBN.
It is thus clear that the largest enhancement will occur when the field starts as high as possible in the effective potential, where the conformal factor will be the largest. 
 
\item The second  possibility  arises when the initial velocity is  negative. In this case, deep in the radiation era, the  field quickly reaches a constant negative value until the first particles become non-relativistic, turning on the effective potential $V_{\rm eff}$. At this point, the field  starts rolling down its potential with  kicks dictated  by $\tilde \Sigma$ \eqref{eq:kick-function}. The field subsequently rolls down its  effective potential during the matter and dark energy dominated eras. The enhancement is thus dictated by the smallest constant negative value reached by the scalar deep in the radiation era. The subsequent evolution proceeds as in the previous case. 
\end{enumerate}

\begin{figure}[H]
\begin{center}
\includegraphics[width=7.0cm]{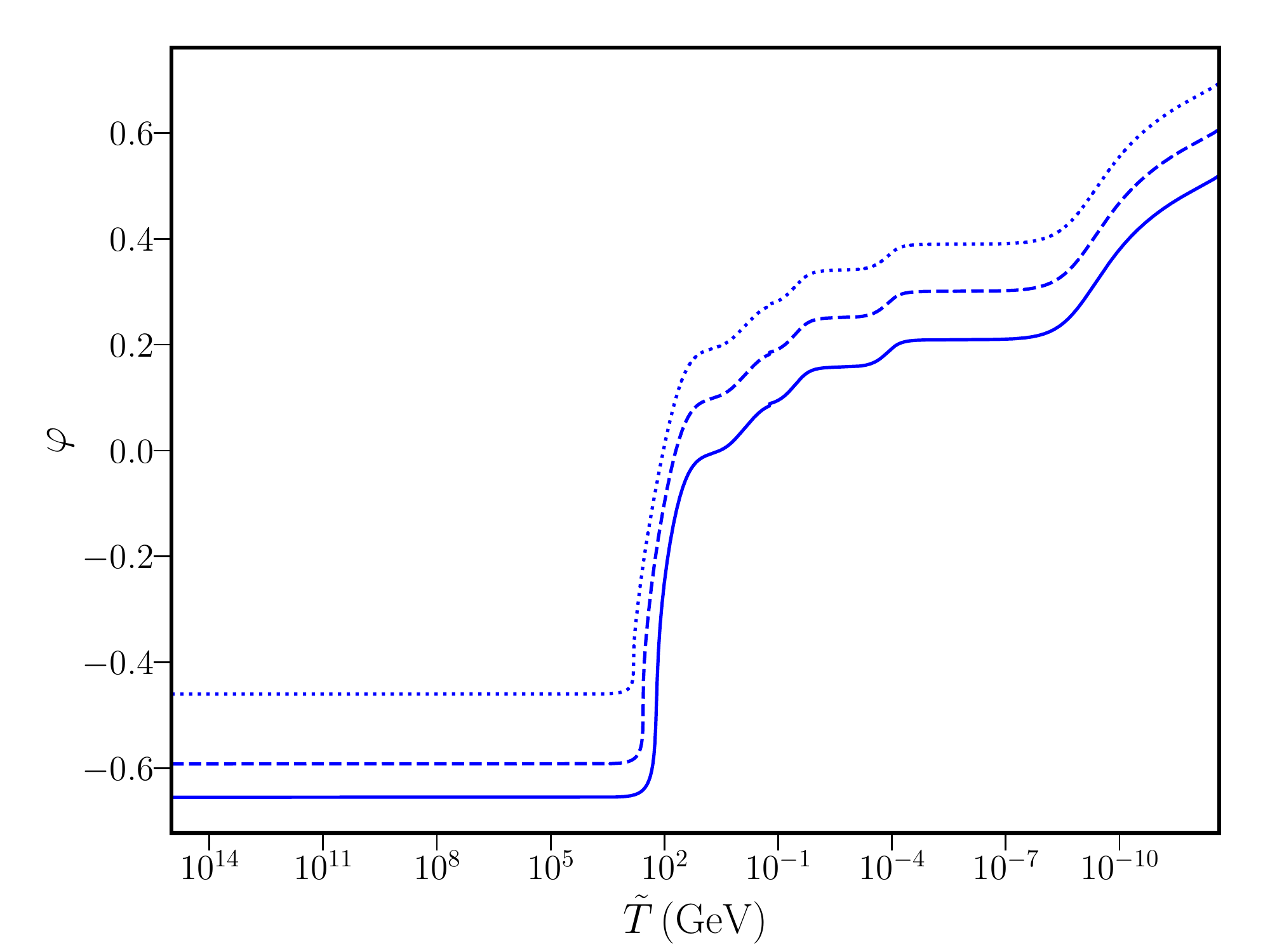}
\includegraphics[width=7.0cm]{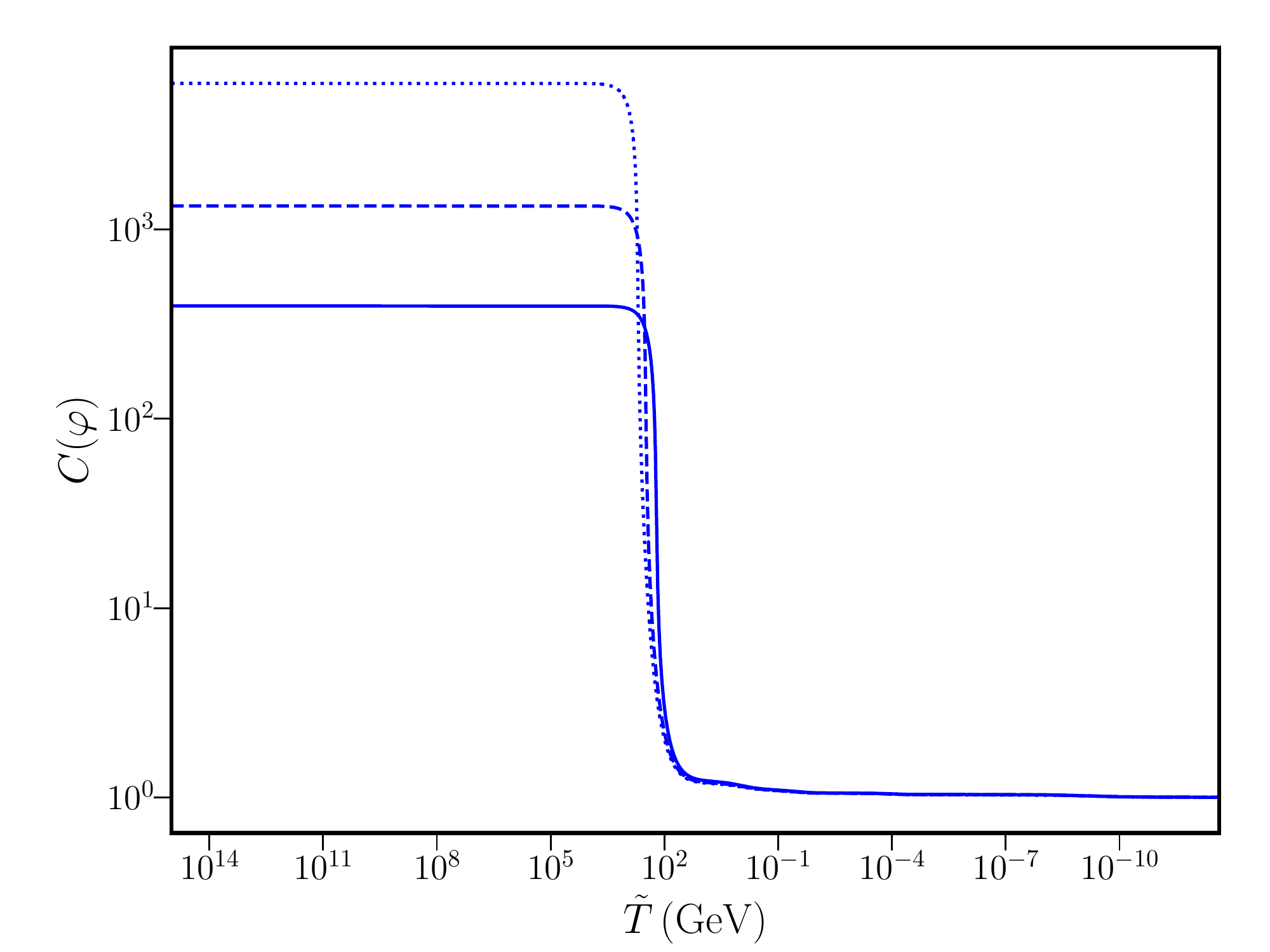}
\end{center}
\vskip -15pt
\caption{The evolution of the scalar field $\varphi$ (left panel) and the conformal factors (right panel) have been plotted as  functions of temperature.  The initial conditions have been chosen as mentioned in Table~\ref{Table3}. The solid, dashed, and dotted lines correspond to $n=1$, $n=2$, and $n=4$ respectively.}
\label{fig:phi-Cphi-conformal-zerovel}
\end{figure}

\begin{figure}[H]
\begin{center}
\includegraphics[width=10.00cm]{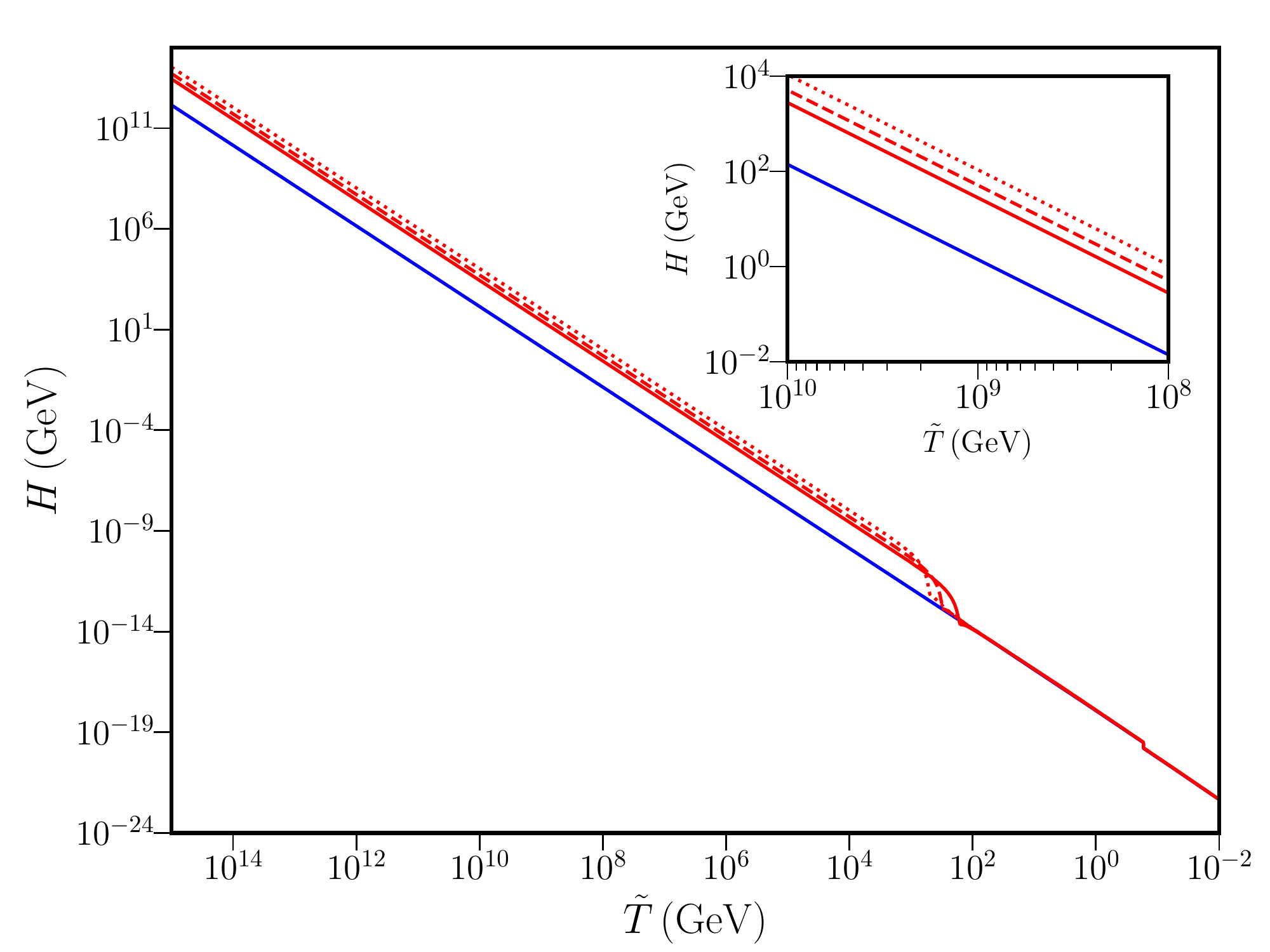}
\end{center}
\vskip -15pt
\caption{The evolution of the Hubble parameters in the Jordan  frame (red lines) and GR (blue solid line), have been plotted as functions of temperature for the conformal factors and initial conditions mentioned in Table~\ref{Table3}. The solid, dashed, and dotted lines correspond to $n=1$, $n=2$, and $n=4$ respectively.}\label{fig:H-conformal-zerovel}
\end{figure}

We numerically  calculate  the evolution and enhancement using both types of  initial conditions, which are summarised in Tables~\ref{Table3} and~\ref{Table4} below. 
In the first case, we choose the initial conditions with  $\varphi_{_{\Nt}}^i=0$  at the initial temperature of $10^{15}\,{\rm GeV}$, with  $\varphi^i$ as in Table~\ref{Table3}. We start the evolution of the scalar field deep in the radiation era.   The  evolution of $\varphi$ and the corresponding behaviour of the conformal factor~(\ref{eq:conformal-factor}) are shown in  Fig.~\ref{fig:phi-Cphi-conformal-zerovel}. As discussed before,  deep in the radiation era, the scalar stays at a constant value, after which, when the first particles become non-relativistic, the effective potential turns on as $\tilde \omega\ne 1/3$, and the field starts to roll down the runaway. The conformal factor starts at a large  value set by the initial value of $\varphi$, eventually dropping back  to unity well before the onset of BBN. We see that  larger powers of $n$, give a larger value of $C$ and thus larger enhancement. This is also reflected in the Hubble parameter, Fig.~\ref{fig:H-conformal-zerovel}.

\begin{table}[H] 
\begin{center}
\centering
\begin{tabular}{| l | c | c | c|}
\hline
\cellcolor[gray]{0.9}$(1+be^{-\beta\varphi})^{2n}$ & \cellcolor[gray]{0.9} $\varphi^i$   \\
\hline \hline
$n=1$  & $ -0.655$    \\
\hline
$n=2$  & $-0.490$  \\
\hline
$n=4$  & $-0.372$    \\
\hline
\end{tabular}
\end{center} 
\caption{Type {\bf (a)} initial conditions with $\varphi^i_{_{\Nt}}=0$, for the purely conformal case with conformal function  \eqref{eq:conformal-factor}. The initial temperature in all the cases is $\tilde T_i=10^{15}\,{\rm GeV}$.}
\label{Table3}
\end{table}   

%

\begin{figure}[!h]
\begin{center}
\includegraphics[width=7.50cm]{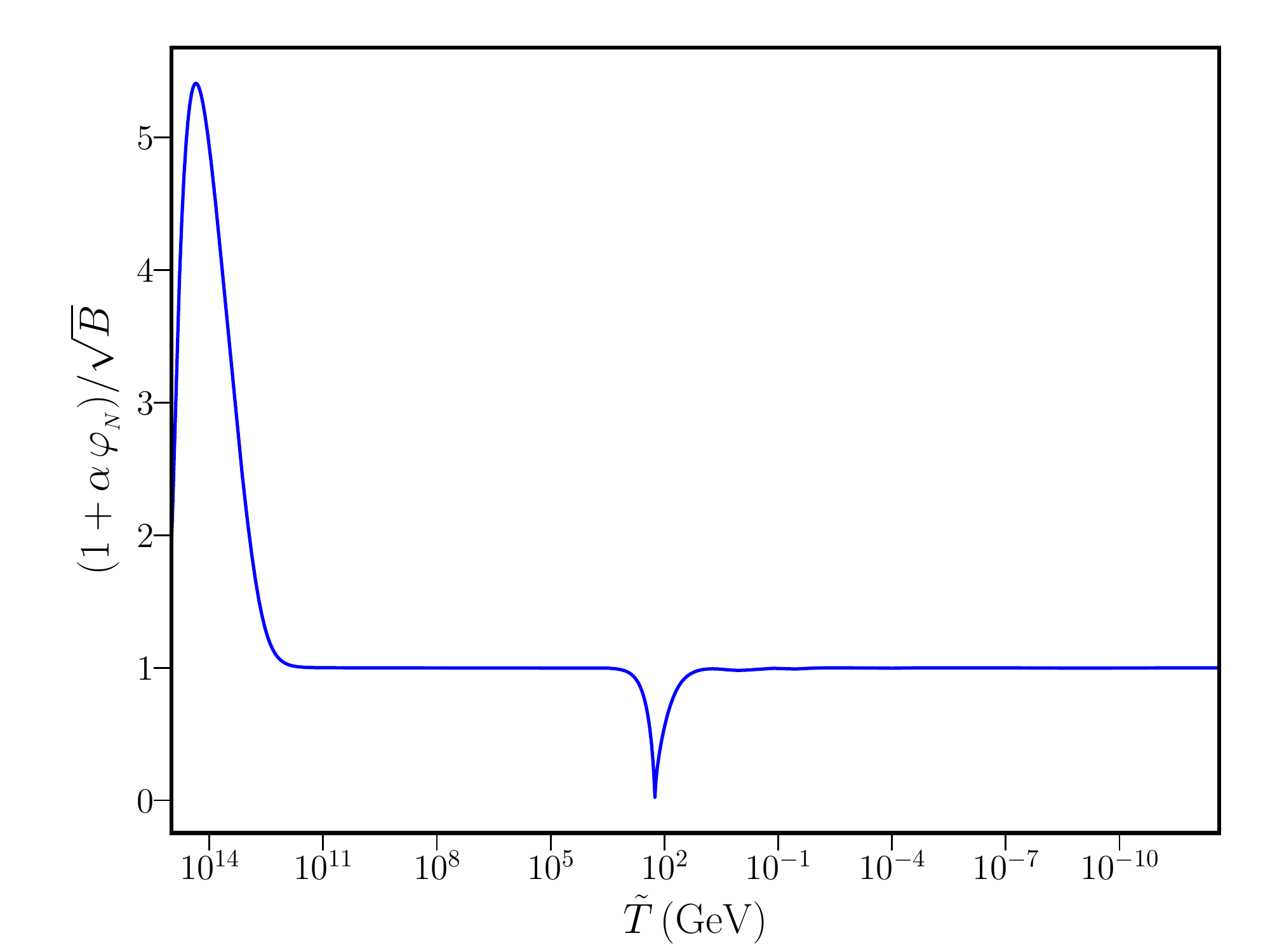}
\end{center}
\vskip -15pt
\caption{Behaviour of the quantity $\left(1+\alpha(\varphi)\,\varphi_{_N} \right)/\sqrt{B}$ in the purely conformal phenomenological scenario for the case $n=1$.
}\label{fig:factor-conformal}
\end{figure}

Using the  solutions for $\varphi$ and $C$, we  compute the modified expansion rate in the Jordan frame  and compare it to $H_{GR}$ (see \eqref{xi}) to get:
\begin{equation}
 \xi =\frac{C^{1/2}(\varphi)}{C^{1/2}(\varphi_{_0})}\,\frac{1}{[1-\alpha(\varphi)\varphi_{_{\Nt}}]\,\sqrt{B}}\,\frac{1}{\sqrt{1+\alpha^2(\varphi_{_0})}}
 = \frac{C^{1/2}(\varphi)}{C^{1/2}(\varphi_{_0})}\,\frac{[1+\alpha(\varphi)\varphi_{_N}]}{\sqrt{B}}\,\frac{1}{\sqrt{1+\alpha^2(\varphi_{_0})}}\,,\label{eq:H-conformal}
\end{equation}
where $\varphi_{_0}$ is the value of the field today, and $B$ is given by \eqref{B} with $M=0$.
The behaviour of the Hubble parameters is shown in  Fig.~\ref{fig:H-conformal-zerovel}. The notch observed in this plot is due to the fact that, in the Jordan frame, the Hubble parameter can become smaller than $H_{\rm GR}$  for a brief period of time when $(1+\alpha(\varphi)\,\varphi_{_N})/\sqrt{B}<1$, as observed in \cite{Dutta:2016htz} (see Fig.~\ref{fig:factor-conformal}). As we can see from Fig.~\ref{fig:H-conformal-zerovel}, the relative enhancement due to the different powers of $n$ in \eqref{eq:conformal-factor} is not very prominent.


\begin{figure}[H]
\begin{center}
\includegraphics[width=7.50cm]{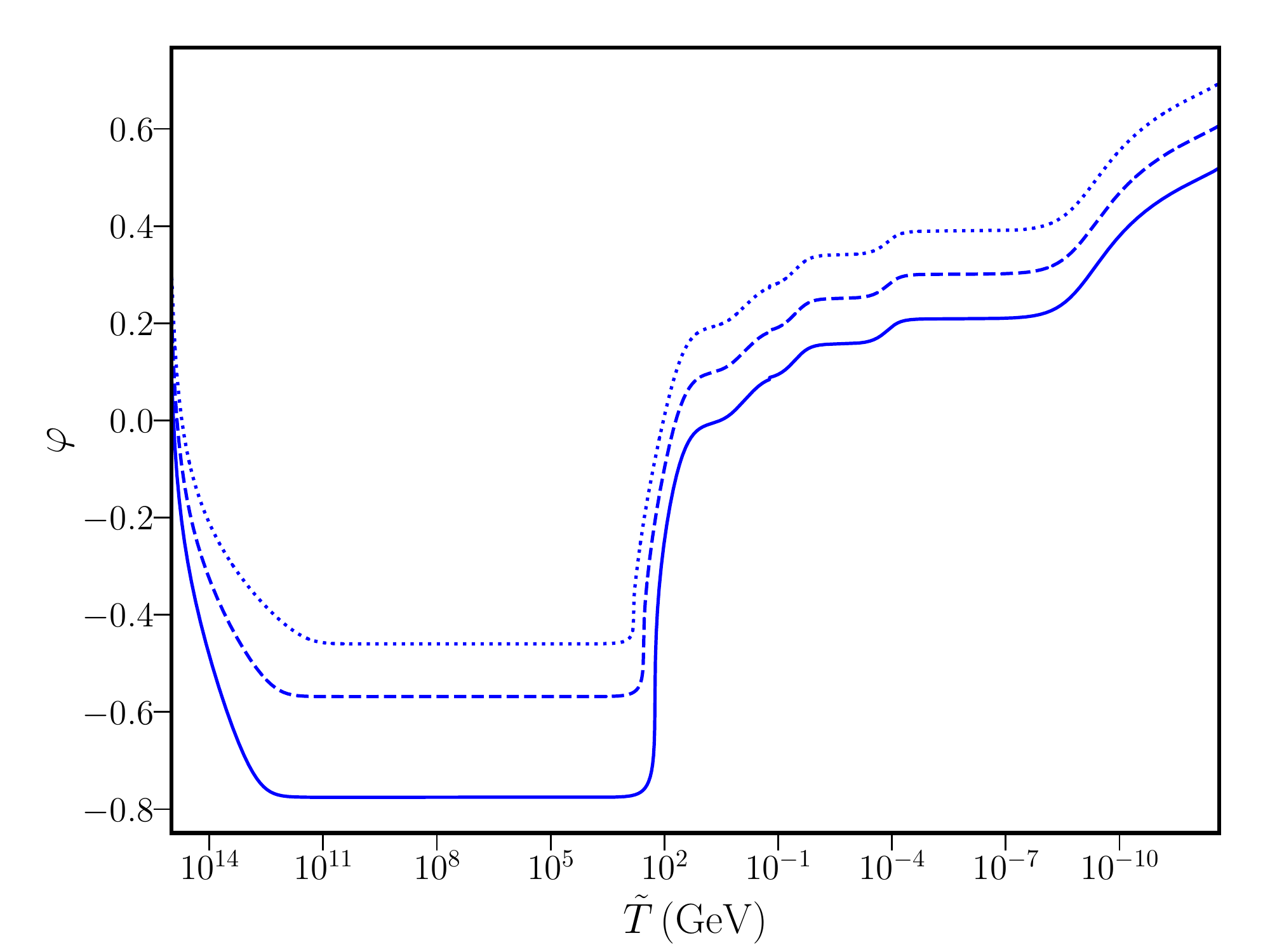}
\includegraphics[width=7.50cm]{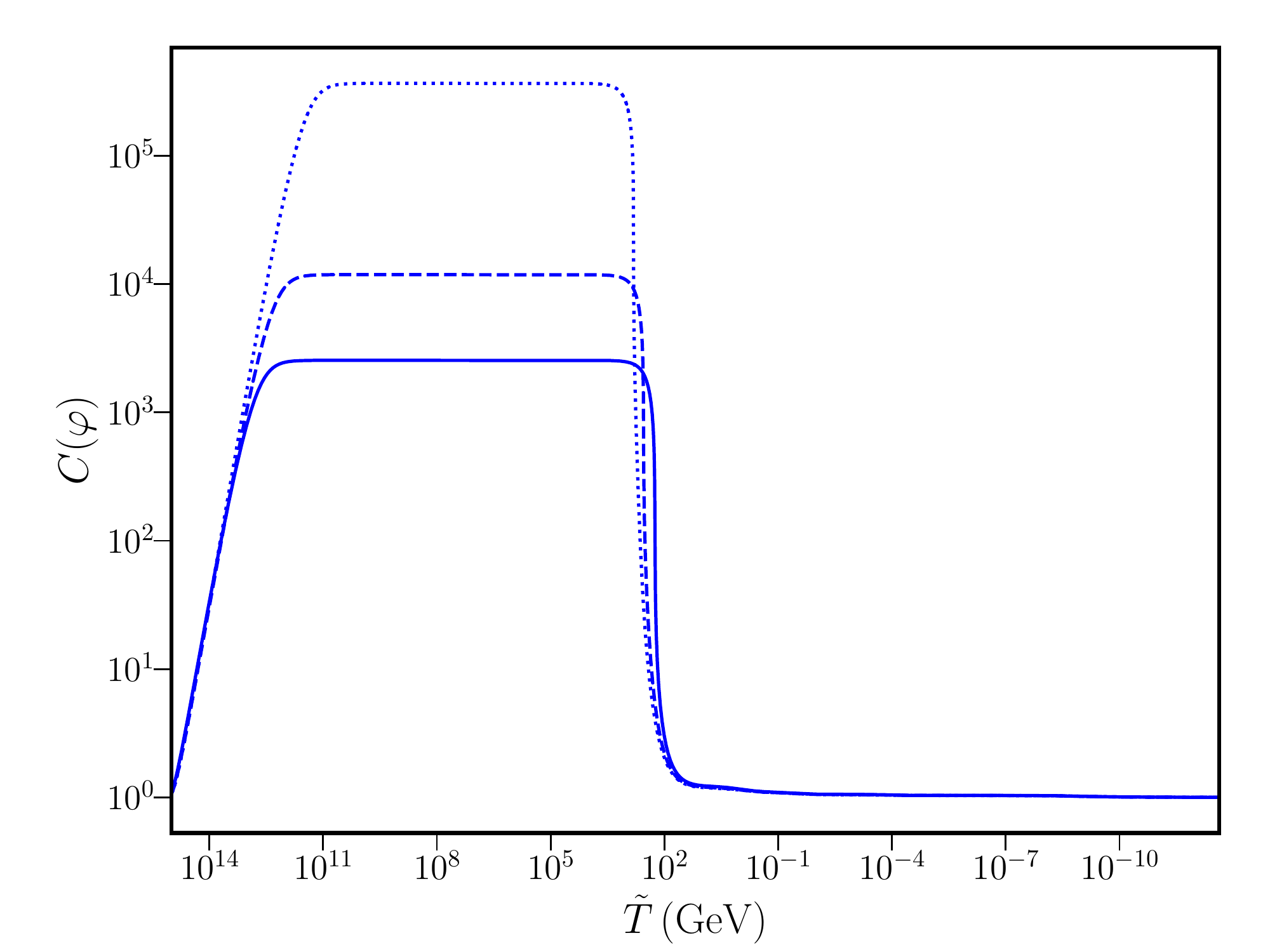}
\end{center}
\vskip -15pt
\caption{The evolution of the scalar field $\varphi$ (left panel) and the conformal factors as  functions of temperature for the set of  initial conditions mentioned in Table~\ref{Table4}. The solid, dashed, and dotted lines correspond to $n=1$, $n=2$, and $n=4$ respectively.}
\label{fig:phi-Cphi-conformal}
\end{figure}

Next, we consider  the initial conditions with non-zero initial velocity   $\varphi_{_{\Nt}}^i\ne 0$, and  initial temperature of $10^{15}\,{\rm GeV}$ as shown in Table~\ref{Table4}.  The resultant evolution of $\varphi$ and the corresponding behaviour of the conformal factor~(\ref{eq:conformal-factor}) are shown in Fig.~\ref{fig:phi-Cphi-conformal}. After an initial negative velocity, the scalar field settles down to a constant value\footnote{Note that the approximation $\Delta\varphi\sim \varphi^i_N$ is not valid as the initial velocity is of order $\sqrt{6}$.}, after which, when the first particles become non-relativistic, the effective potential turns on as $\tilde \omega\ne 1/3$, and the field starts to roll down the effective runaway potential. The conformal factor reaches a larger maximum value than in the case with zero initial velocity, during the radiation era, eventually dropping back  to unity well before the onset of BBN.

\begin{table}[H] 
\begin{center}
\centering
\begin{tabular}{| l | c | c | }
\hline
\cellcolor[gray]{0.9}$(1+be^{-\beta\varphi})^{2n}$ & \cellcolor[gray]{0.9} $\varphi^i$  & \cellcolor[gray]{0.9} $\varphi^i_{\tilde N}$  \\
\hline \hline
$n=1$  & $ 0.101$  &  $-0.994$  \\
\hline
$n=2$  & $0.220$   &   $-1.010$   \\
\hline
$n=4$  & $0.313$   &   $-1.010$   \\
\hline
\end{tabular}
\end{center} 
\caption{Type (b) initial conditions for the purely conformal case with conformal function \eqref{eq:conformal-factor}. The initial temperature in all the cases is  $\tilde T_i=10^{15}\,{\rm GeV}$. }
\label{Table4}
\end{table}   

\begin{figure}[H]
\begin{center}
\includegraphics[width=10.00cm]{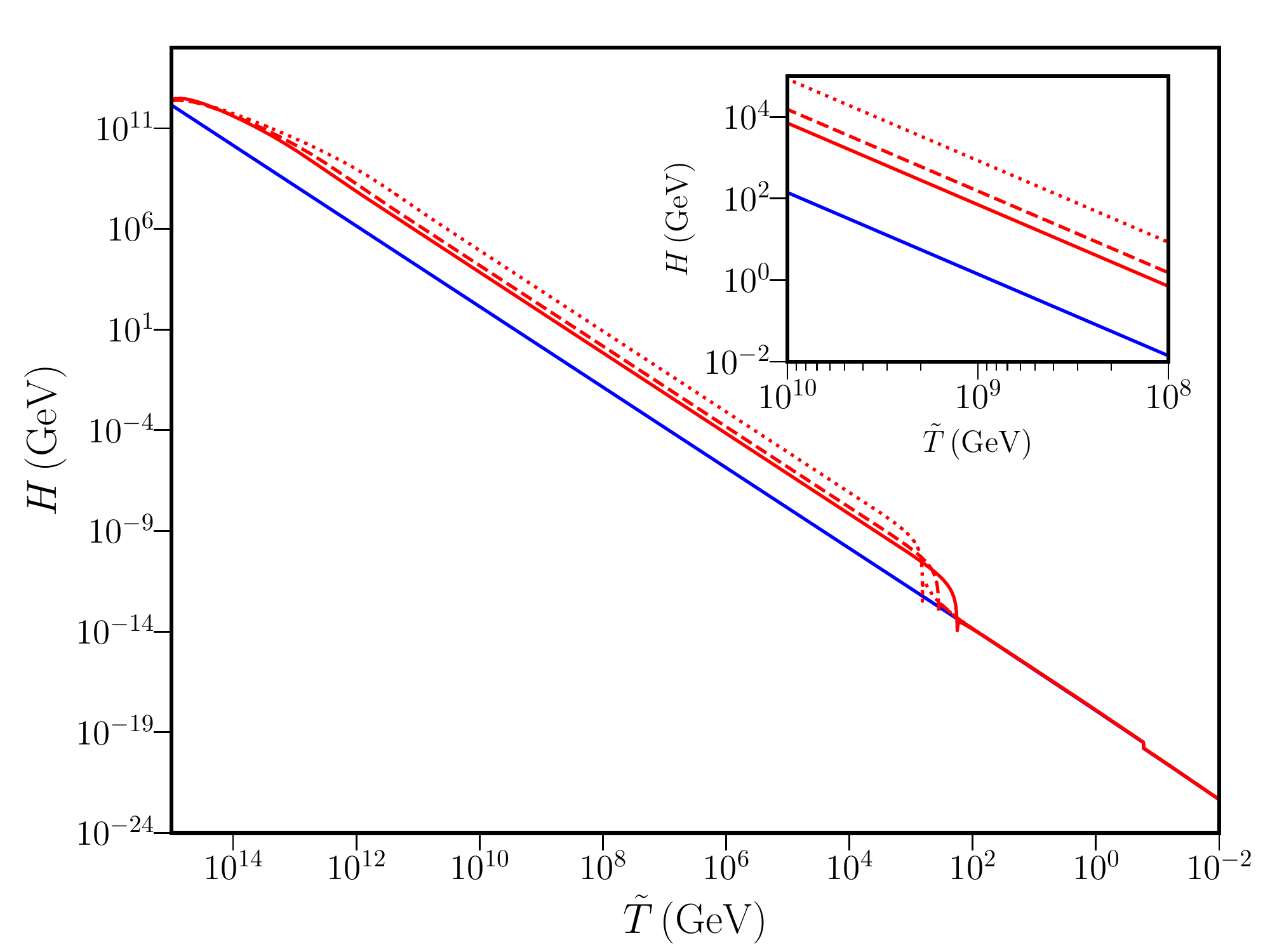}
\end{center}
\vskip -15pt
\caption{The evolution of the Hubble parameters in the Jordan  frame (red lines) and GR (blue solid line), have been plotted as functions of temperature for the conformal factors and initial conditions mentioned in Table~\ref{Table4}. The solid, dashed, and dotted lines correspond to $n=1$, $n=2$, and $n=4$ respectively.}\label{fig:H-conformal}
\end{figure}

The evolution of the Hubble parameter for the three cases ($n=1, 2, 4$) is shown  in Fig.~\ref{fig:H-conformal}.  Here again we observe the notch  due to the factor $(1+\alpha(\varphi)\,\varphi_N)/\sqrt{B}$ becoming less than one for a brief period of time, thus making the Hubble parameter in the Jordan frame smaller than $H_{\rm GR}$  \cite{Dutta:2016htz}.
From Eq.~(\ref{eq:H-conformal}), it is evident that the amplitude of the Hubble parameter at any temperature is directly proportional to the conformal factor. This effect  directly impacts the enhancement of the amplitude of the gravitational waves as we discuss this in the next section. 
From Fig.~\ref{fig:H-conformal}, one can easily see that the enhancement is largely amplified. Further, we can see that the larger the conformal factor, the larger  the enhancement of the expansion rate.

\subsubsection{Purely disformal case }\label{sec:pureDpheno}
In the purely disformal case, we have $C(\varphi)=1$. We  study two different scenarios -- one wherein the disformal factor is a constant, and one wherein the disformal factor is a function of the scalar field $\varphi$. Since in the phenomenological case, the only relation between the conformal and disformal factors is given by the causality constraint, we have to ensure that we choose the relevant parameters such that the following condition is always satisfied:
\begin{equation}\label{eq:causality}
 C(\phi) - D(\phi)\,\dot{\phi}^2 > 0.
\end{equation}
In terms of  $\phi=\kappa \varphi$  and $\tilde N$ derivatives, this  simplifies to
\begin{equation}
 \left[D(\varphi)\,H^2\,\varphi_{_{\Nt}}^2/\kappa^2\right] -1 < 0,\label{eq:causality-condition}
\end{equation}
which will need to be satisfied by our choice of $D(\varphi)$. 

The equations of motion (see Eqs.~\eqref{HN}, \eqref{fiNN}) in this case simplify to (note that in the pure disformal case, $N=\tilde N$):
\begin{subequations}
 \begin{align}
     & H_{_{\Nt}} = -H\left[\frac{3\,B}{2}\left(1+\tilde{w}\gamma^{-2}\right)+\frac{\varphi_{_{\Nt}}^2}{2}\right], \\ 
     &\label{eq:disf-feno-phi} \varphi_{_{\Nt\Nt}}\left[1+\frac{3 H^2 \gamma^2 B D}{\kappa^2}\right] + \varphi_{_{\Nt}}\,\frac{H_{_{\Nt}}}{H}\left[1+\frac{3 H^2 
     \gamma^2 B D}{\kappa^2}\right] + 3\,\varphi_{_{\Nt}}\left[1-\frac{3 B D H^2 \tilde{w} }{\kappa^2}\right] \nn\\
     &\hskip 9cm 
    + \frac{3 H^2 \gamma^2 B D}{\kappa^2}\,\delta(\varphi)\,\varphi_{_{\Nt}}^2= 0,\nn\\
 \end{align} 
\end{subequations}
where $B=1-(\varphi_{_{\Nt}}^2/6)$. 
%

\section*{Constant disformal factor: $D=D_0$ }

We start by considering the simplest possibility of a constant disformal function, $D=D_0$. In this case, $\delta=0$ and the last term in \eqref{eq:disf-feno-phi} vanishes. 
The initial condition for the Hubble parameter is found by finding a real positive solution to the cubic equation for $H$ in \eqref{eq:H-HGR}  for which $\gamma_i\sim 1$ (see  \cite{Dutta:2016htz} and Appendix \ref{sec:ICsDisform} for details). This imposes the following condition on $D_0$:
\begin{equation}
    D_0 \leq \left(\frac{2}{\varphi_{_{\Nt}}^2}-\frac{1}{3}\right)\frac{30}{\sqrt{3} \pi^2 g_* \tilde{T}^4}, 
\end{equation}
which is further complemented by the causality condition~(\ref{eq:causality-condition}) on $D_0$.
Using these two constraints, we choose the parameters and initial conditions as given in Table~\ref{Table5}.

\begin{figure}[H]
\begin{center}
\includegraphics[width=7.50cm]{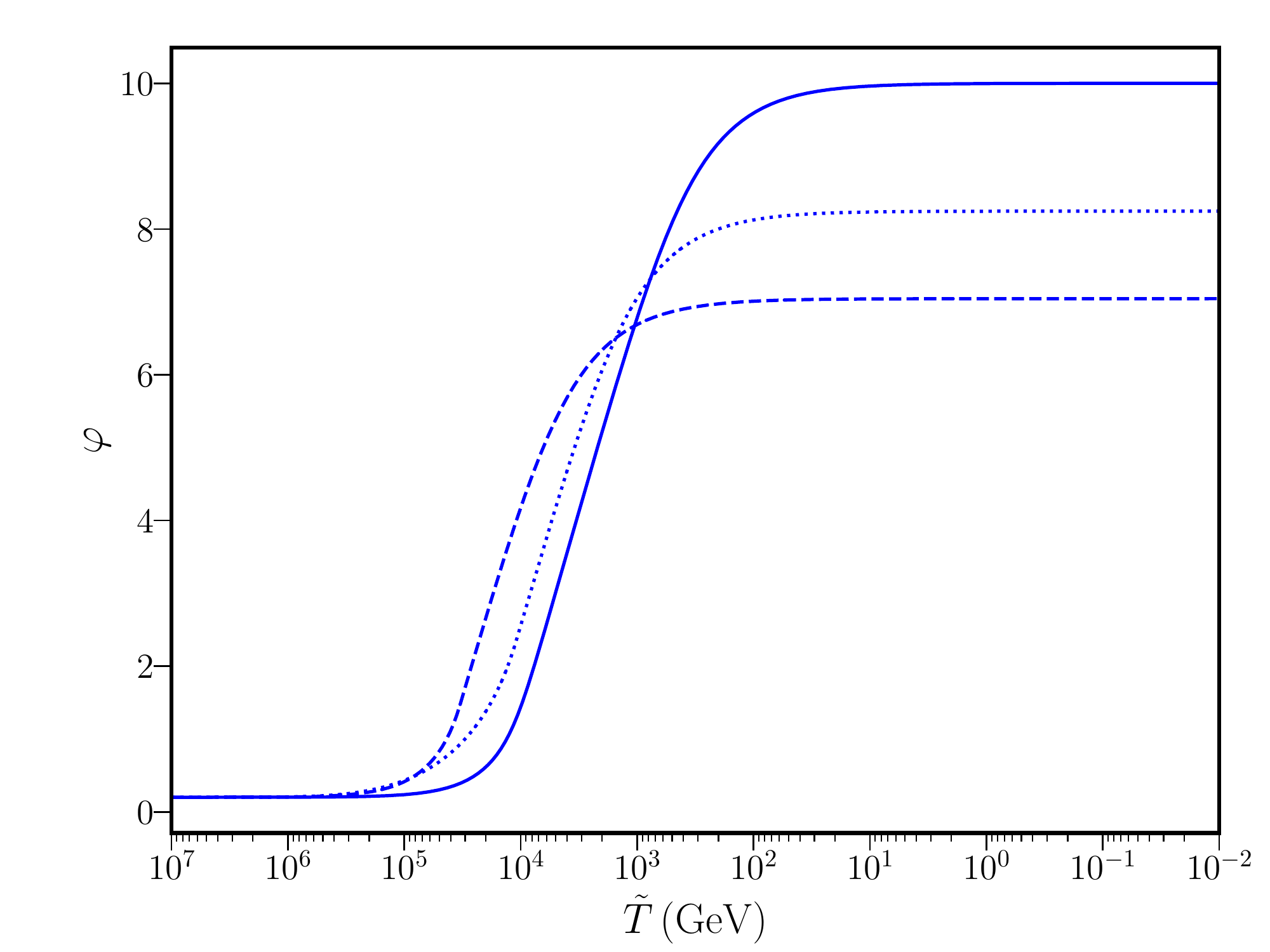}
\includegraphics[width=7.50cm]{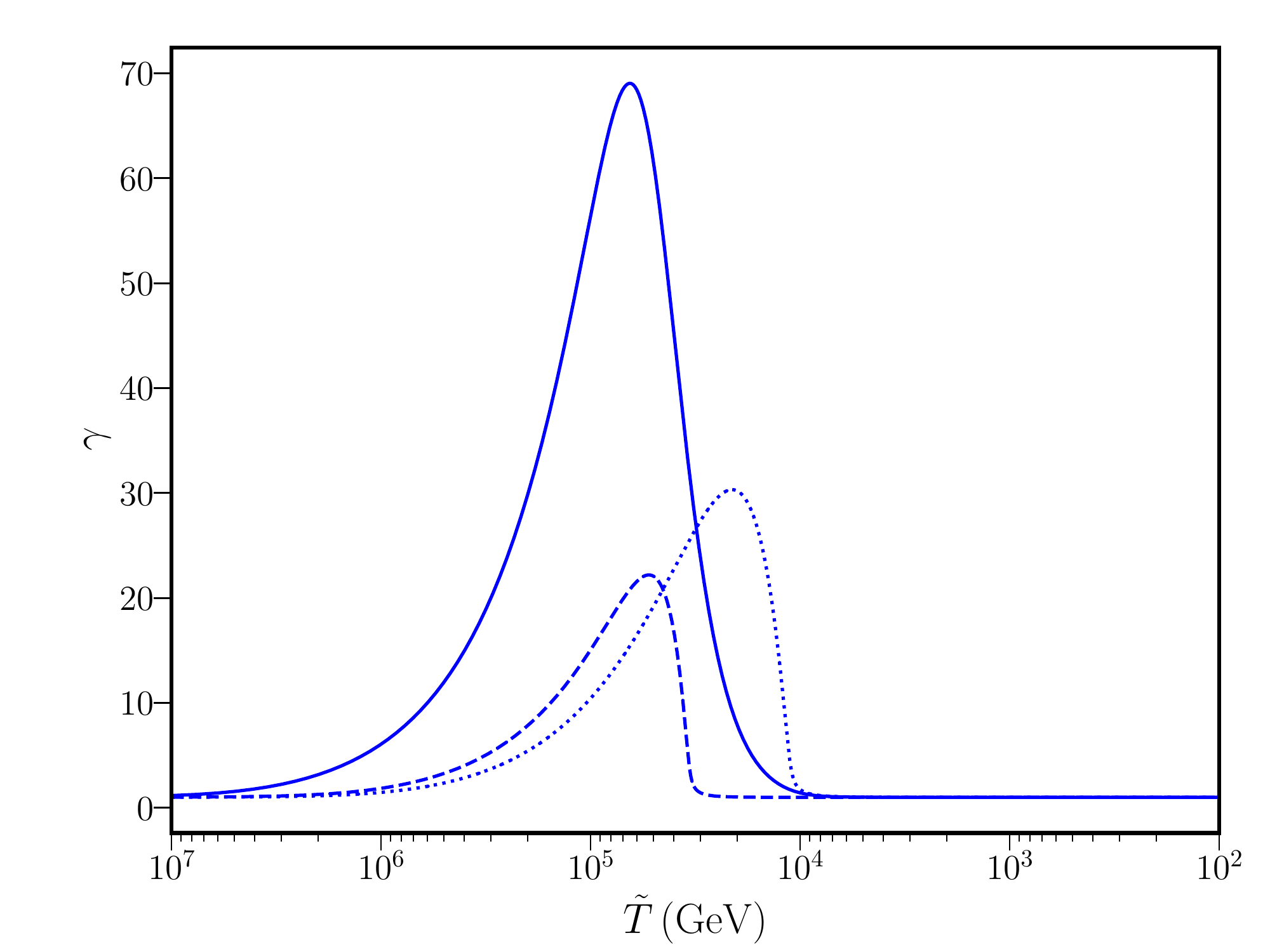}
\end{center}
\vskip -15pt
\caption{The evolution of the scalar field $\varphi$ (left panel) and  Lorentz factor $\gamma$ (right panel) for the  disformal coupling in the phenomenological and D-brane cases. 
The dashed, dotted, and solid lines correspond to the following choices of the disformal factor: $D=D_0$, $D=D_0\varphi^2$, and the D-brane case with $D=1/M^4$, respectively. The values of $D_0$ and the initial conditions are given in Table~\ref{Table5}. }\label{fig:phi-gammafunc-pheno1}
\end{figure}

The evolution of the scalar field $\varphi$ and the  Lorentz factor $\gamma$ are shown  in Fig.~\ref{fig:phi-gammafunc-pheno1} (dashed lines). As we see there, the scalar field stays constant for a few e-folds, but quickly evolves towards larger values, causing $\gamma$ to increase, before going back to one well before the end of BBN.  This  causes an enhancement of the  Hubble parameter, according to $\xi= \gamma^{3/2}/B^{1/2}$, as  shown in Fig.~\ref{fig:Hubbles-pheno-disformal}. It is important to point out that the choice of the initial temperature impacts the moment at which the enhancement occurs. Thus, for a larger initial temperature, the enhancement will occur earlier \cite{Dutta:2017fcn}. Interestingly, as we shall discuss in the next section, we can probe this with gravitational waves.

\section*{Field dependent disformal factor: $D=D_0\,\varphi^2$ }
\begin{figure}[H]
\begin{center}
\includegraphics[width=10.00cm]{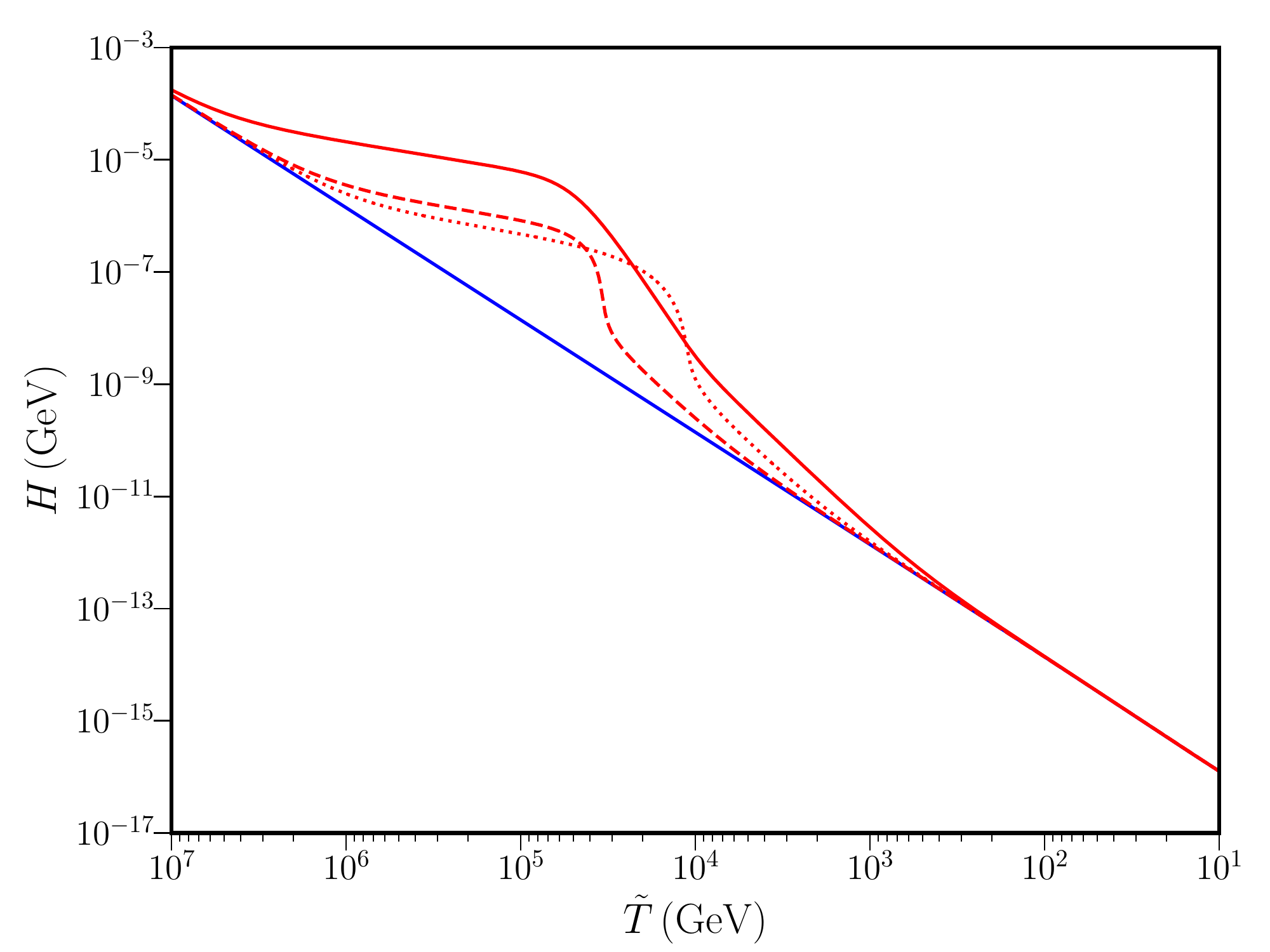}
\end{center}
\vskip -15pt
\caption{The evolution of the Hubble parameters  in the Jordan frame (red lines), and GR (blue solid line) have been plotted as functions of temperature, for the  disformal coupling in the phenomenological and D-brane cases. 
The dashed, dotted, and solid lines correspond to the following choices of the disformal factor: $D=D_0$, $D=D_0\varphi^2$, and the D-brane case with $D=1/M^4$, respectively. The values of $D_0$ and the initial conditions are given in Table~\ref{Table5}.}\label{fig:Hubbles-pheno-disformal}
\end{figure}

In this case, $\delta(\varphi) \neq 0$, and we cannot neglect the last term in \eqref{eq:disf-feno-phi}. We again proceed to set the initial condition for the Hubble parameter by finding the real positive solution of  \eqref{eq:H-HGR}  for which $\gamma_i\sim 1$ (see  \cite{Dutta:2016htz} and Appendix \ref{sec:ICsDisform} for details). This imposes the following condition 
\begin{equation}
    D_0 \leq \left(\frac{2}{\varphi_{_{\Nt}}^2}-\frac{1}{3}\right)\frac{30}{\sqrt{3} \pi^2 \varphi^2 g_* \tilde{T}^4}.
\end{equation}
Further, the causality condition~(\ref{eq:causality-condition}) implies that we must have 
\begin{equation}
    \left(D_0\,\varphi^2\,H^2\,\varphi_{_{\Nt}}^2/\kappa^2\right) -1 < 0.\label{eq:causality-pheno2}
\end{equation}
Taking into account these constraints, we choose the parameters and initial conditions as given in Table~\ref{Table5}. As in the previous case, by starting the evolution at a higher temperature, the enhancement of the Hubble parameter will occur earlier, which will impact when the GW background will be enhanced. We shall see this explicitly in the next section. 
The evolution of the field $\varphi$ and the Lorentz factor $\gamma$ are shown in Fig.~\ref{fig:phi-gammafunc-pheno1}  (dotted line). The Hubble parameters for the constant and field dependent cases are shown in Fig.~\ref{fig:Hubbles-pheno-disformal}, (dashed and dotted   lines respectively). Compared to the constant case, the field dependent example results in a slight increase in the maximum of $\gamma$, which thus is reflected in a larger enhancement in the Hubble parameter with a slightly different profile, which will be reflected also in the PGW effect.  

\begin{table}[H] 
\begin{center}
\centering
\begin{tabular}{| l | c | c | }
\hline
\cellcolor[gray]{0.9}      & \cellcolor[gray]{0.9} $D_0 ({\rm GeV}^{-4})$  &  \cellcolor[gray]{0.9} $H_i$(GeV)  \\
\hline \hline
Pheno case: $D=D_0$  & $5.000\times 10^{-22}$  &   $1.413\times 10^{-4}$   \\
\hline
Pheno case: $D=D_0\varphi^2$  & $6.000\times 10^{-21}$   &   $1.408\times10^{-4}$  \\
\hline
D-brane case: $D=1/M^4$ & $4.822\times 10^{-21}$   &   $1.516\times10^{-4}$  \\
\hline
\end{tabular}
\end{center} 
\caption{Initial conditions for the disformal coupling models in the phenomenological and D-brane scalar-tensor theories. The other initial conditions in all cases are $\varphi^i=0.200$, $\varphi^i_{\tilde N}=2.000\times 10^{-5}, $ $\tilde  T_i = 10^7$(GeV).    }
\label{Table5}
\end{table}   

\subsection{D-brane scalar tensor theories}\label{sec:Dbrane1}
This case corresponds to  $b=0$ in \eqref{eq:Sphi}, and it arises in D-brane cosmology scenarios. The scalar field has a non-standard kinetic term dictated by the Dirac-Born-Infeld (DBI) action, \eqref{eq:Sphi} and the modifications to the expansion rate and its effect on the dark matter relic abundance was discussed in \cite{Dutta:2017fcn}.  This type of scenario  can arise from a post-string inflationary scenario. At this scale, the universe is already four-dimensional and moduli associated to the compactification have been  stabilised\footnote{In some scenarios, compactification moduli  might be displaced from their minima, giving rise to a matter dominated regime before the onset of BBN, with interesting consequences (see e.g. Ref.~\cite{Allahverdi:2013noa}).}. However, in what follows, our study  is purely phenomenological and can be used as a first step to understand the effects for gravitational waves in D-brane scalar tensor theories in the early universe. As shown in Appendix C of \cite{Dutta:2016htz}, the canonical normalisation of $\phi$, obtained by expanding the DBI action, implies a relation between the conformal and disformal factors through $M^4 CD =1$. %
Thus, in this section, we study the solutions for the D-brane conformally and disformally coupled matter with the choice above, which implies $\delta(\phi) = -\alpha(\phi)$ (see Eqs.~\eqref{eq:alphadef} and \eqref{eq:deltadef}).

\subsubsection*{Evolution equations}

In this case, the evolution equations \eqref{HN}, \eqref{fiNN} simplify to 
\begin{subequations}
 \begin{align}
     &\!\!\!H_{_N} = -H\left[\frac{3}{2}\left(1+\tilde{w}\,\gamma^{-2}\right)B + \frac{\varphi_{_N}^2}{2}\,\gamma\right],\\
 &\!\!\!\varphi_{_{NN}}\left[1+\frac{3\,H^2\,\gamma^{-1}\,B}{M^4\,C^2\,\kappa^2}\right] + 3\,\varphi_{_N}\,\gamma^{-2}\left[1-\frac{3\,H^2\,\gamma^{-1}\,B}{M^4\,C^2\,\kappa^2}\,\tilde{w}\right] + \frac{H_{_N}}{H}\,\varphi_{_N}\left[1+\frac{3\,H^2\,\gamma^{-1}\,B}{M^4\,C^2\,\kappa^2}\right]\nn\\
 &\!\!\!-\frac{6\,H^2\,\gamma^{-1}\,B}{M^4\,C^2\,\kappa^2}\,\alpha(\varphi)\,\varphi_{_N}^2 + 3B\,\gamma^{-3}\alpha(\varphi)(1-3\,\tilde{w}) - \frac{2\,M^4\,C^2\,\kappa^2}{H^2}\!\left[2\,\gamma^{-3}-3\,\gamma^{-2}+1\right]\!\alpha(\varphi)=0,
 \end{align}\label{eq:conformal-disformal-diffeqs}
\end{subequations}
where $\gamma$ is given by \eqref{eq:gamma2}, and here $B$ is given by 
\be\label{eq:B-dbrane}
B= 1-\frac{\gamma^2}{3(\gamma+1)}\varphi_N^2 \,.
\ee

In what follows we consider the expansion rate modification in the purely disformal case, wherein $C=1$ and therefore $D=1/M^4$, and in the conformal plus disformal case wherein $C\ne 1$ and $D=1/CM^4$ with $C$ given by \eqref{eq:conformal-factor}. Remember that, in this case, the conformal and disformal couplings are related and thus we cannot have a purely conformal case as in the phenomenological example above. Moreover, the causality constraint is always satisfied. 

\subsubsection{Purely disformal case }\label{section:purely-disformal}
In the purely disformal case,  $C=1$ and thus $D=1/M^4(=D_0)$. 
As in the phenomenological case, the initial condition for the Hubble rate $H$ can be obtained by looking for the positive real solution of the cubic equation for $H$ (see Appendix \ref{sec:ICsDisform} and \cite{Dutta:2016htz,Dutta:2017fcn} for details), given the initial values for $\left(\varphi_i, \varphi_{_{\Nt}}^i, M\right)$.
These initial conditions can be used together to obtain the initial value of the Lorentz factor, $\gamma_i$, which needs to be of order of one, in order to satisfy the constraints on the present value of the Hubble parameter. 

With this in mind, we have chosen $M=1.200\times10^{5}\,{\rm GeV}$, as well as the other parameter values and initial conditions as given in Table~\ref{Table5}. The  evolution of $\varphi$ and  $\gamma$ are shown in Fig.~\ref{fig:phi-gammafunc-pheno1} (solid line), while the Hubble parameter is shown in Fig.~\ref{fig:Hubbles-pheno-disformal} (solid line). We  see that, compared to the phenomenological examples above, the enhancement in the pure D-brane disformal scenario is larger, for the same initial conditions.  Moreover, the $\gamma$ profile also differs slightly. This behaviour has interesting implications for the PGW spectrum discussed in the next section. 
%

\subsubsection{Conformal-disformal case }\label{section:conformaldisformal}
We now turn on a non-trivial conformal factor, $C(\varphi)$. We choose the same conformal function as in the phenomenological case, namely, \eqref{eq:conformal-factor}. In the present case, this immediately fixes $D$ through  the condition $M^4\,C\,D=1$. That is, in this case, both conformal and disformal factors are turned on. 
Therefore, we shall have to solve the full  set of coupled equations~(\ref{eq:conformal-disformal-diffeqs}). 
\subsubsection*{Initial conditions and evolution}\label{subsec:ic-choice2}

As in the phenomenological case, we can either choose a suitable initial field value  and a zero initial velocity (see discussion in section \ref{section:purely-conformal}) or a non-zero initial velocity. In the present case, if the initial velocity is zero, the initial condition for the Hubble rate $H$ can be obtained from the positive real solution of the quadratic equation for $H$ (see Appendix \ref{subsec:ic-choiceD+C}). Interestingly, in this case, the initial conditions do not depend on the choice of $M$. On the other hand, if the initial field velocity is non-zero, the equation for $H$ becomes a cubic equation and the initial conditions depend on the choice of $M$ (see  \ref{subsec:ic-choiceD+C} and \cite{Dutta:2017fcn}).

\begin{figure}[H]
\begin{center}
\includegraphics[width=7.50cm]{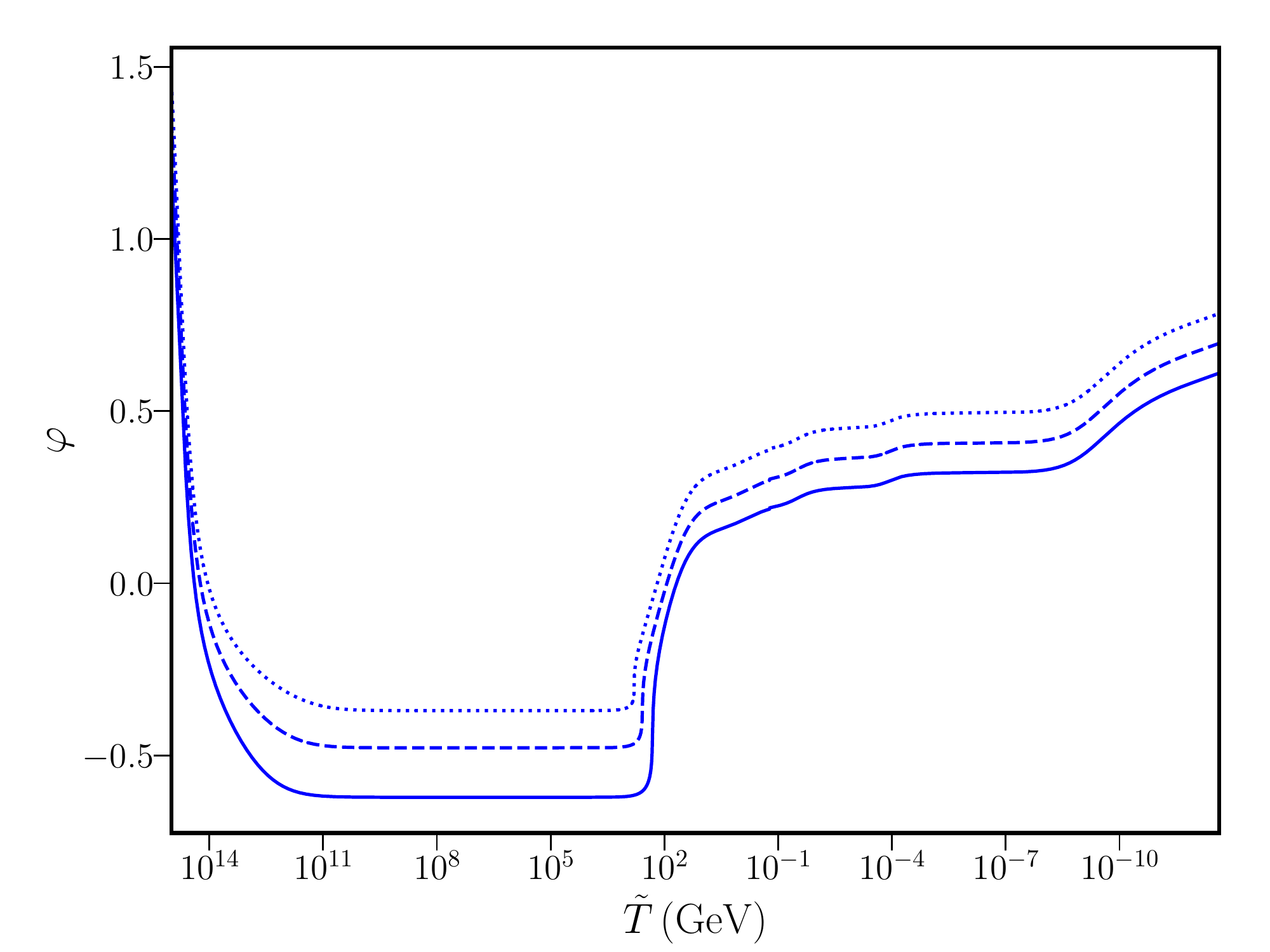}
\includegraphics[width=7.50cm]{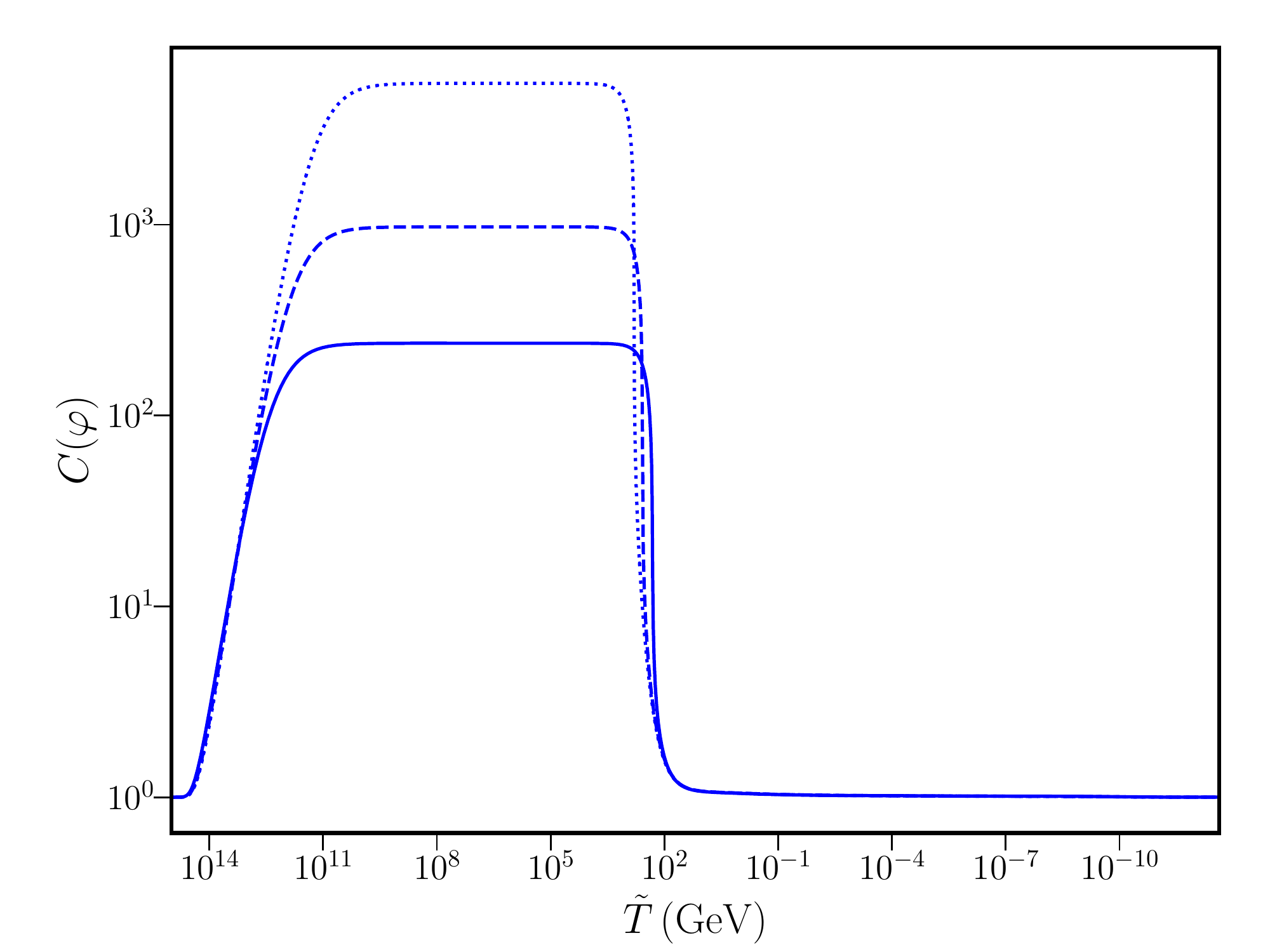}
\end{center}
\vskip -15pt
\caption{The evolution of the scalar field $\varphi$ (left panel) and the conformal factors for the conformal-disformal case have been plotted as functions of temperature for the set of  initial conditions mentioned in Table~\ref{Table6}. The solid, dashed, and dotted lines correspond to $n=1$, $n=2$, and $n=4$ respectively.}\label{fig:phi-Cphi-confdisf}
\end{figure}

\begin{figure}[H]
\begin{center}
\includegraphics[width=10.00cm]{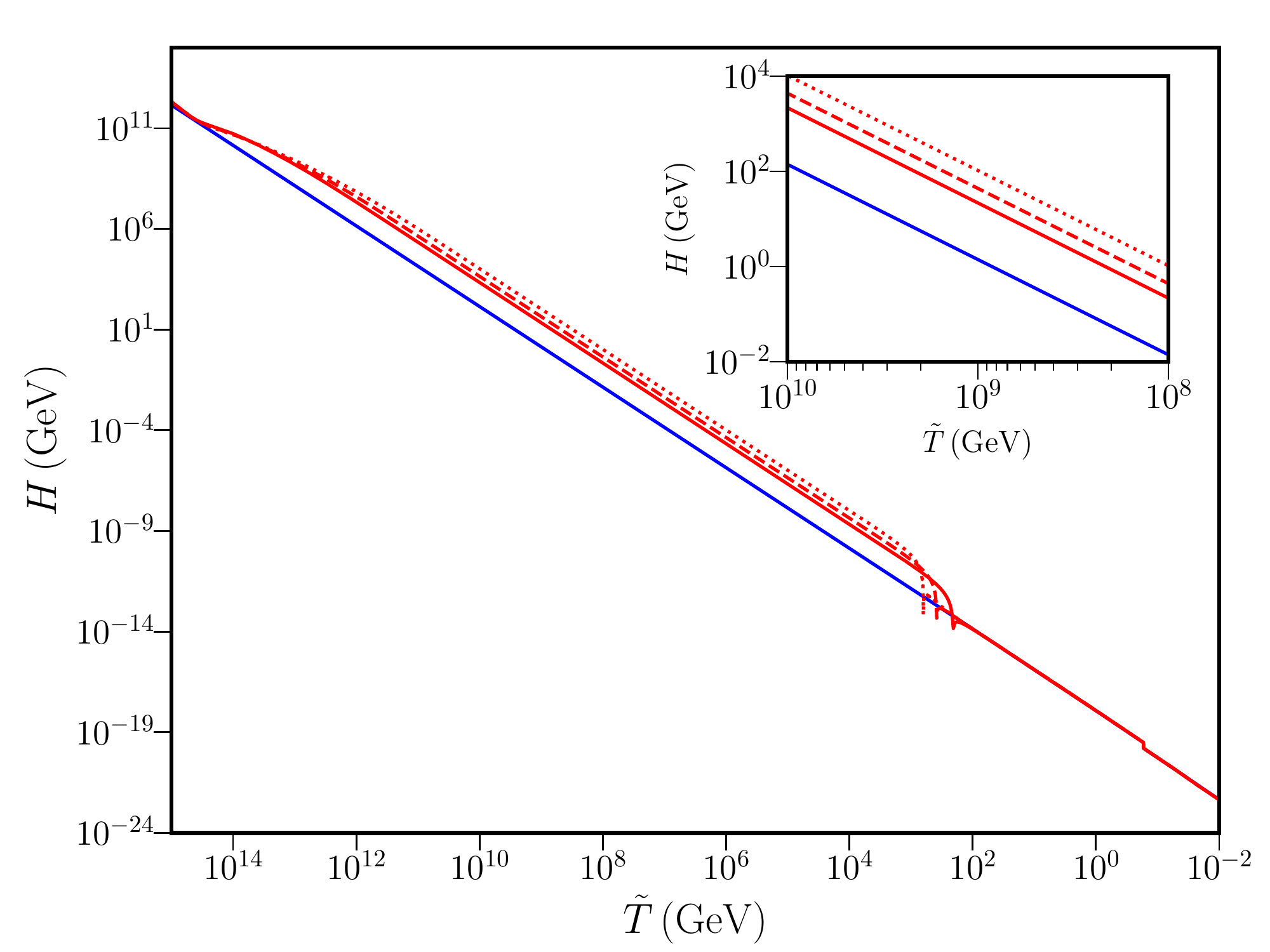}
\end{center}
\vskip -15pt
\caption{The evolution of the Hubble parameters in the Jordan  frame (red lines) and GR (blue solid line), have been plotted as functions of temperature in the conformal-disformal case for the conformal factors and initial conditions mentioned in Table~\ref{Table6}. The solid, dashed, and dotted lines correspond to $n=1$, $n=2$, and $n=4$ respectively.}\label{fig:H-confdisf}
\end{figure}

With this in mind, we again consider two sets of initial conditions: 

\begin{enumerate}[{\bf (a)}]

\item Zero initial velocity:  Analogous as  to the phenomenological case,  we can start with a zero velocity and a suitable value for the scalar field. We choose the same initial values as in the phenomenological case (see Table~\ref{Table3}) with $M=2.600\times 10^{15}\,{\rm GeV}$. 
The resulting evolution for the scalar field and the conformal factors have been plotted in Fig.~\ref{fig:phi-Cphi-conf-disf-zerovel}, while the Hubble parameters for the D-brane case are shown in Fig.~\ref{fig:H-conf-disf-zerovel}. Comparing these plots with those in Figs.~\ref{fig:phi-Cphi-conformal-zerovel} and \ref{fig:H-conformal-zerovel}, we can see that although the evolution of the scalar field and the conformal factors look very similar, the Hubble parameters in the Jordan frame have slightly  different behaviour. Remember that, in the present case, the disformal coupling is turned on as well through the relation $M^4 C D=1$. 
%

\begin{figure}[H]
\begin{center}
\includegraphics[width=7.0cm]{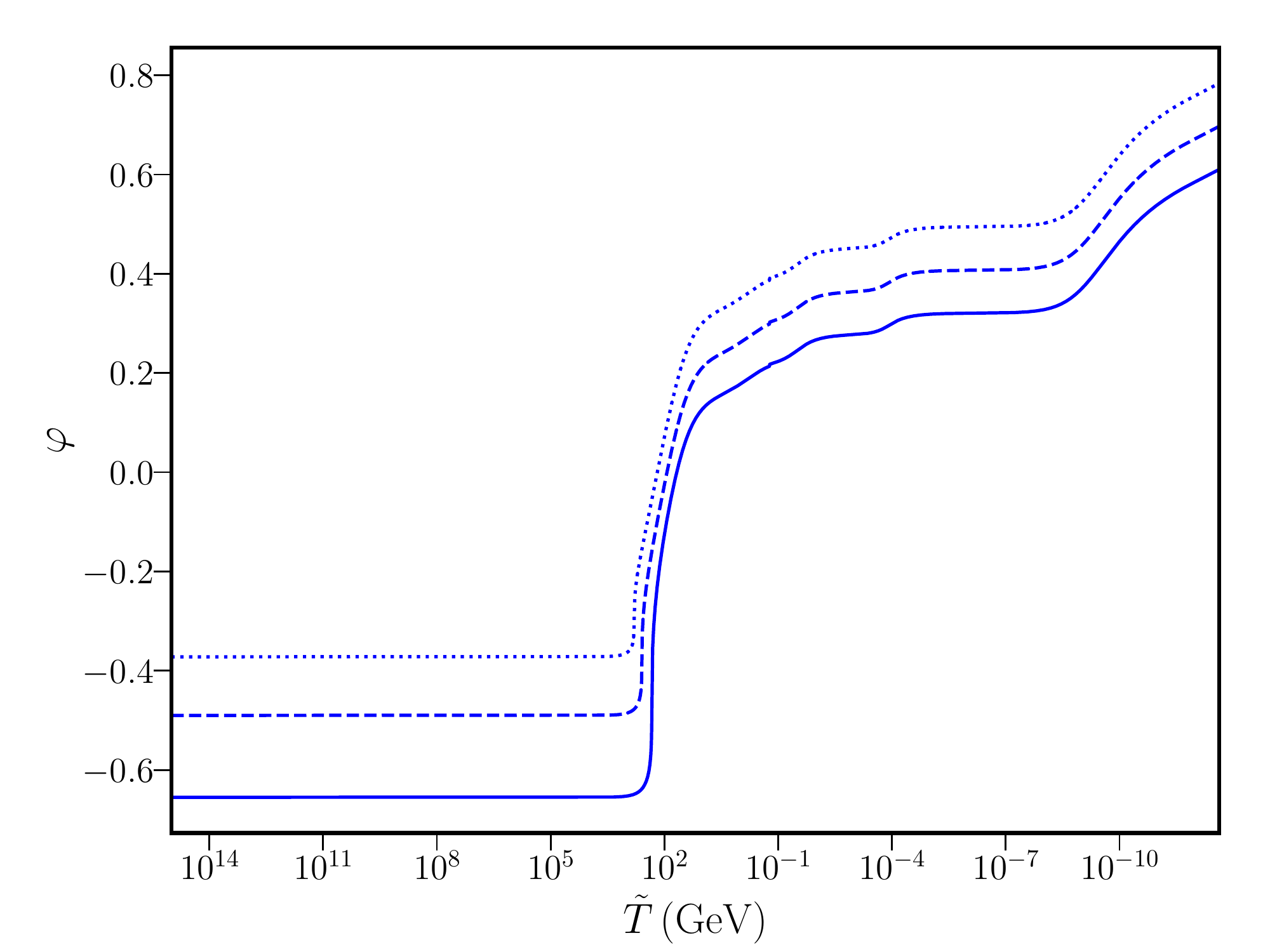}
\includegraphics[width=7.0cm]{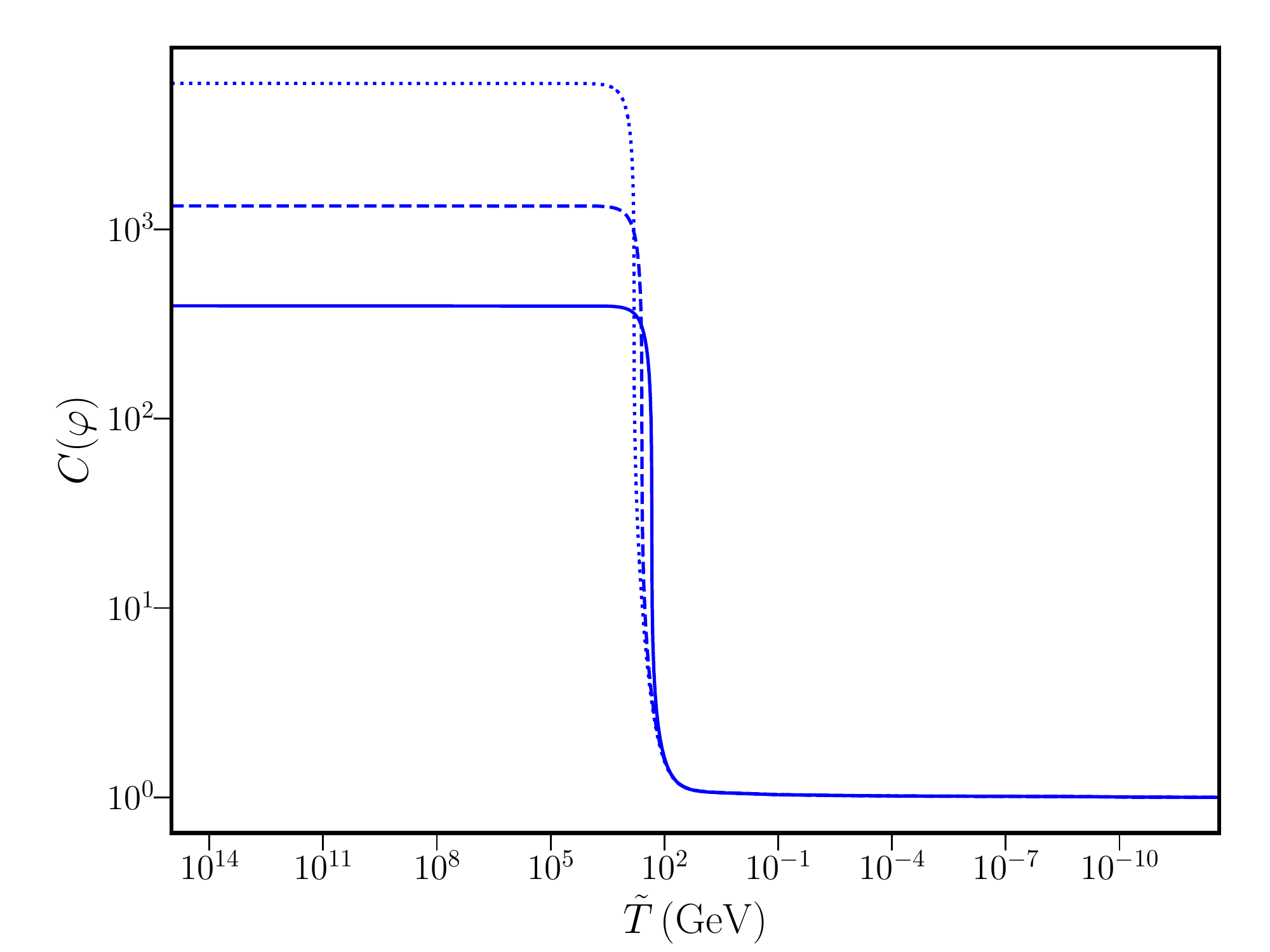}
\end{center}
\vskip -15pt
\caption{The evolution of the scalar field $\varphi$ (left panel) and the conformal factors (right panel) have been plotted as functions of temperature in the conformal-disformal case.  The initial conditions are the same as in the phenomenological case in Table~\ref{Table3} with 
$M=2.600\times 10^{15}\,{\rm GeV}$. The solid, dashed, and dotted lines correspond to $n=1$, $n=2$, and $n=4$ respectively.}
\label{fig:phi-Cphi-conf-disf-zerovel}
\end{figure}
\begin{figure}[!h]
\begin{center}
\includegraphics[width=7.50cm]{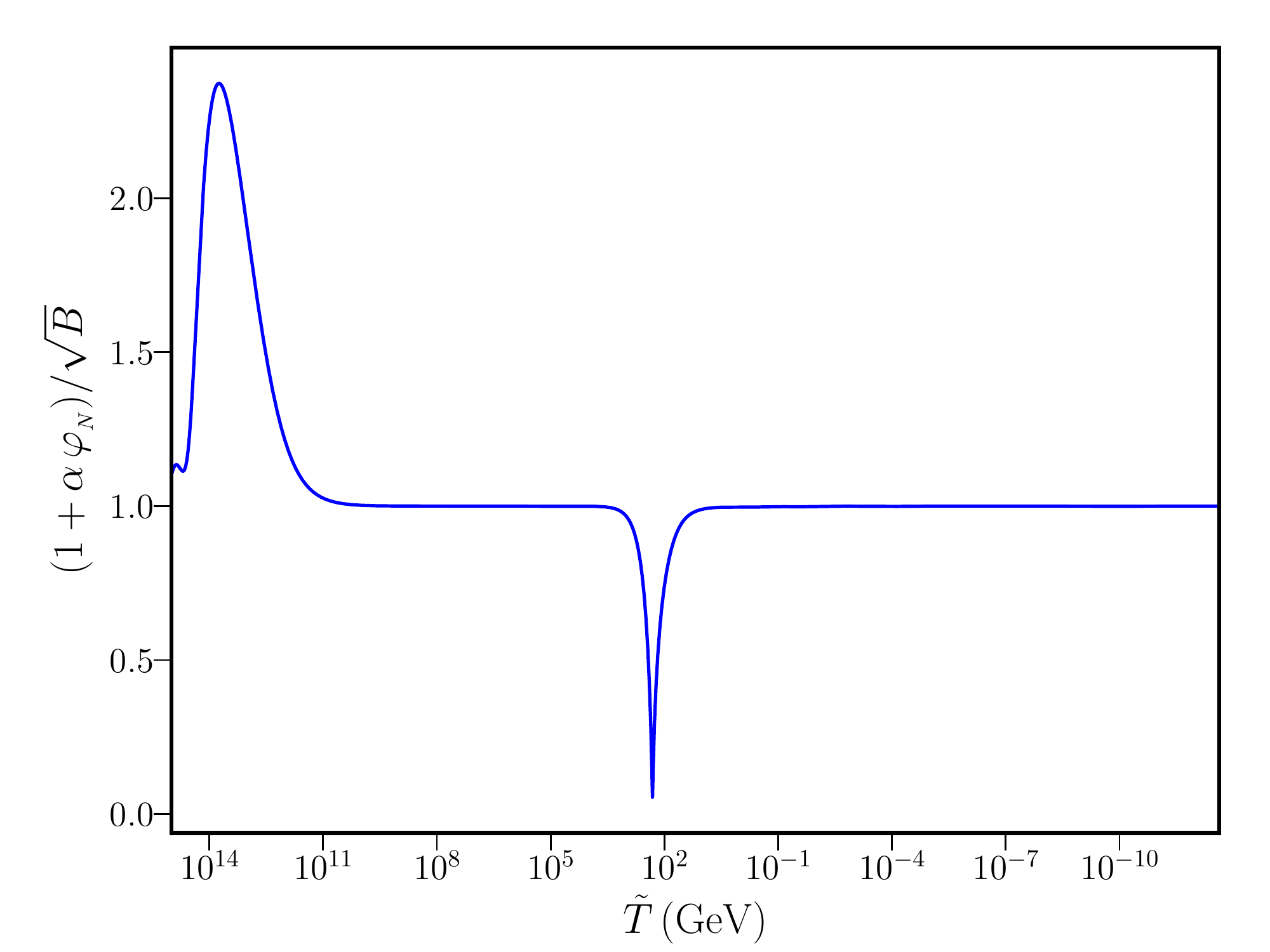}
\end{center}
\vskip -15pt
\caption{Behaviour of the quantity $\left(1+\alpha(\varphi)\,\varphi_{_N} \right)/\sqrt{B}$ in the conformal-disformal D-brane scenario for the case $n=1$.
}\label{fig:factor-conf-disf}
\end{figure}
\begin{figure}[H]
\begin{center}
\includegraphics[width=10.00cm]{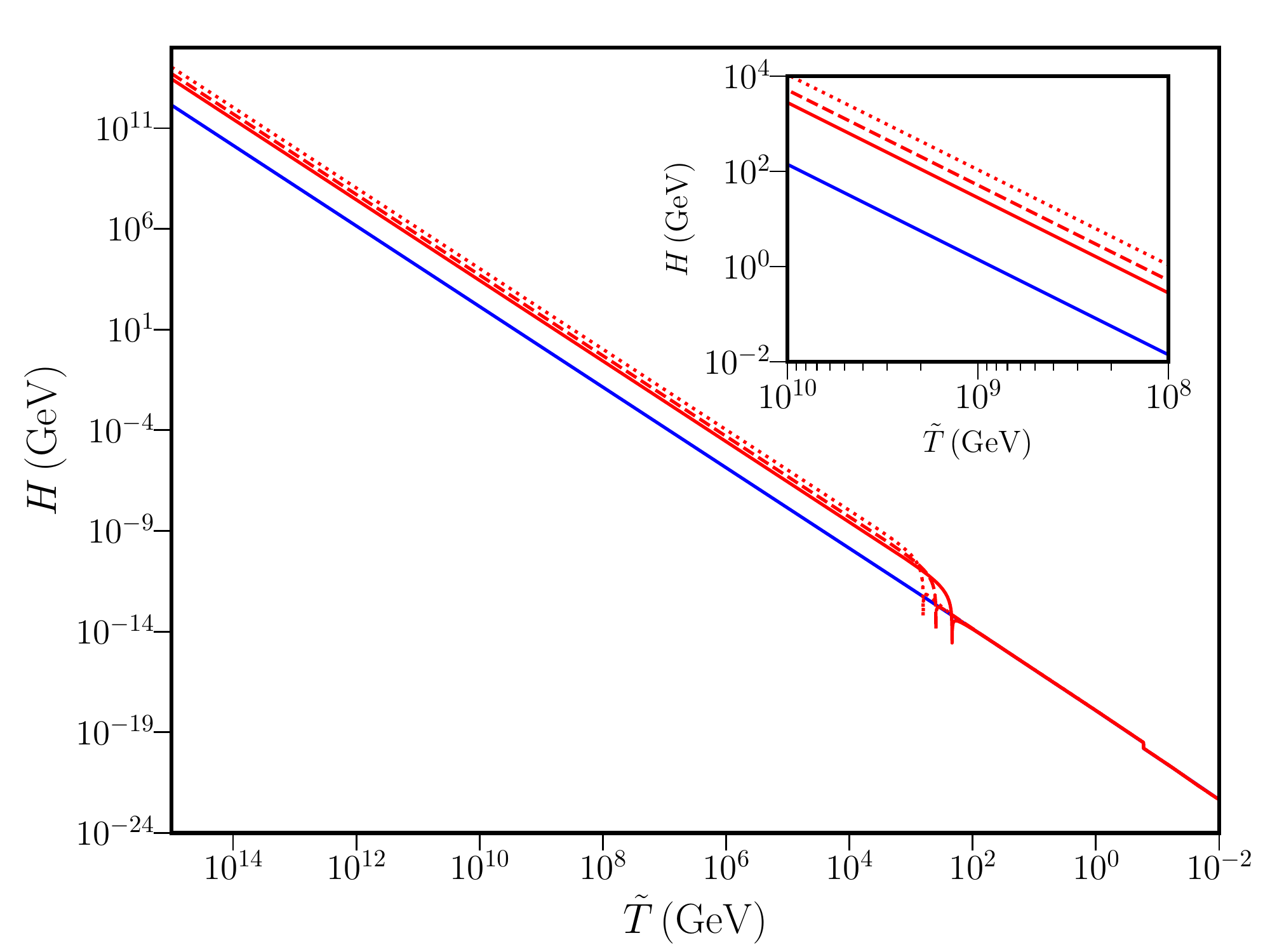}
\end{center}
\vskip -15pt
\caption{The evolution of the Hubble parameters in the Jordan  frame (red lines) and GR (blue solid line), have been plotted as functions of temperature in the conformal-disformal case for the conformal factors and initial conditions mentioned in Table~\ref{Table3}. The solid, dashed, and dotted lines correspond to $n=1$, $n=2$, and $n=4$ respectively.}\label{fig:H-conf-disf-zerovel}
\end{figure}
\item Non-zero initial velocity: In this case, for a particular set of values of $\left(\varphi^i, \varphi_{_N}^i, M\right)$, and using the normalization condition $M^4 C D=1$, we  obtain the real positive solutions for $H$ (see Eq.~(\ref{eq:H-cubiceqn})). Note that, in order to solve the equations~(\ref{eq:conformal-disformal-diffeqs}) in terms of $\Nt$, we have to convert the initial conditions obtained by the above procedure into initial conditions on the set of values of $\left(\varphi^i, \varphi_{_{\Nt}}^i, M\right)$. 
With this in mind, we consider the initial conditions as mentioned in Table~\ref{Table6}.   The initial conditions on $\varphi$ and $\varphi_{_N}$ are different from what we had chosen for the same conformal factor in the phenomenological case. This is because the equations~(\ref{eq:conformal-disformal-diffeqs}) are different from the phenomenological case, and the same choice of initial conditions is not numerically viable in both cases. Since starting with a large negative initial velocity leads to a greater enhancement in the conformal factor, we choose the initial velocity accordingly and then choose the initial condition on the field such that the equations~(\ref{eq:conformal-disformal-diffeqs}) can be numerically solved.
The evolution of the field and  the conformal factor is shown in Fig.~\ref{fig:phi-Cphi-confdisf}, while the Hubble expansion rates are shown in Fig.~\ref{fig:H-confdisf}.  
As we can see, compared to the previous choice of starting with zero initial velocity, in this case, $C\sim 1$ at the start of the evolution. It subsequently increases to its maximum value and then drops back quickly to one well before the onset of BBN. The behaviour is thus very similar to the phenomenological case, and it is dominated by the conformal evolution. In fact, during the whole evolution, $\gamma\sim 1$. 

\end{enumerate}

\begin{table}[H] 
\begin{center}
\centering
\begin{tabular}{| l | c | }
\hline
\cellcolor[gray]{0.9}$(1+be^{-\beta\varphi})^{2n}$ & \cellcolor[gray]{0.9} $\varphi^i$   \\
\hline \hline
$n=1$  & $ 1.200$    \\
\hline
$n=2$  & $1.344$   \\
\hline
$n=4$  & $1.452$   \\
\hline
\end{tabular}
\end{center} 
\caption{Initial conditions for $\varphi^i$ for D-brane scalar-tensor model with conformal and disformal couplings turned on (see section~\ref{section:conformaldisformal}). The other initial conditions are $\varphi^i_N = -0.900$, $\tilde T_i = 10^{15}\,{\rm GeV}$, $H_i=1.692\times 10^{12}\,{\rm GeV}$, $M=2.600\times 10^{15}\,{\rm GeV}$. The evolution of $\varphi$ and $C$ is shown in Fig.~\ref{fig:phi-Cphi-confdisf}.}
\label{Table6}
\end{table}

\begin{figure}[H]
\begin{center}
\includegraphics[width=7.0cm]{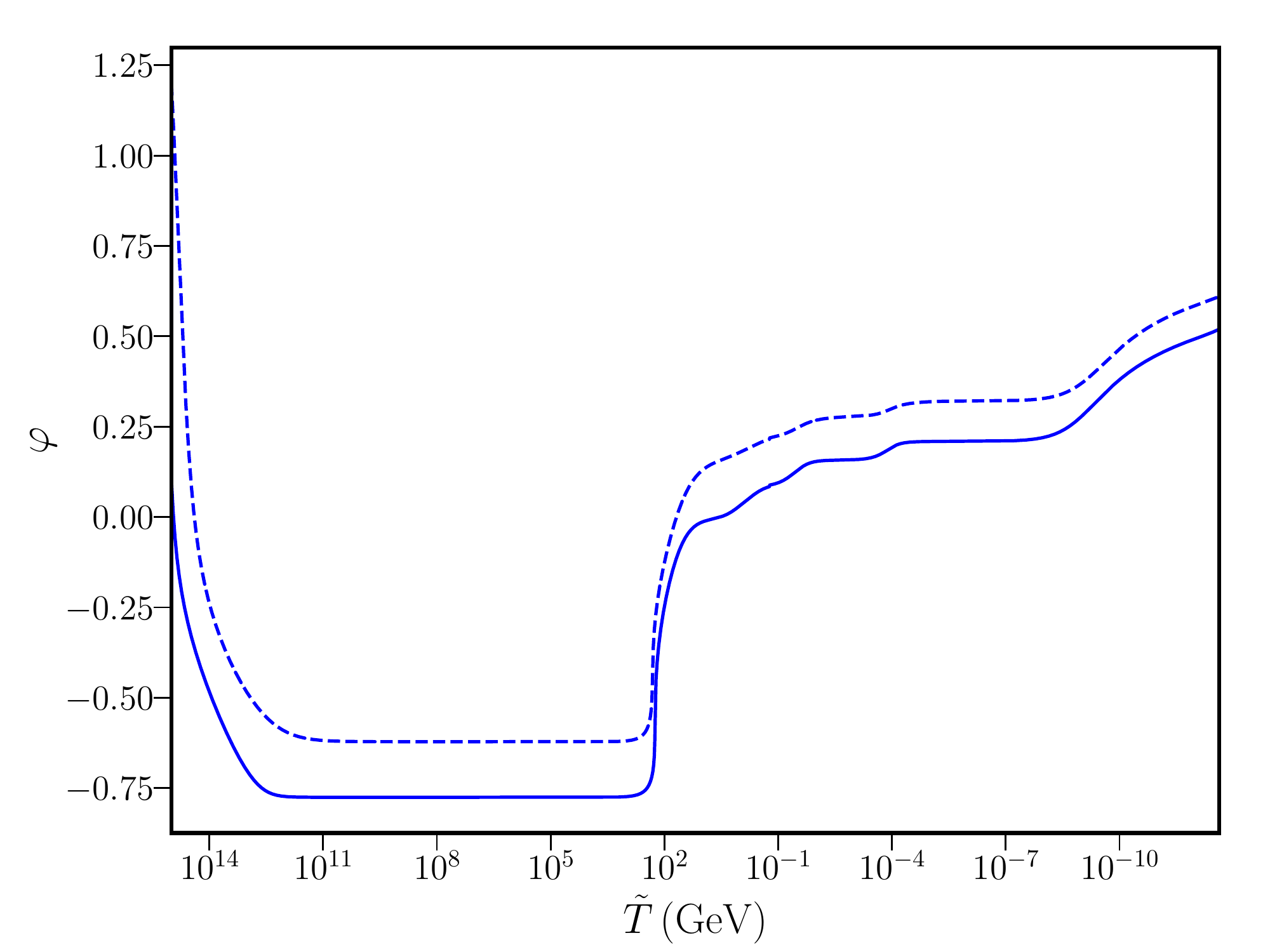}
\includegraphics[width=7.0cm]{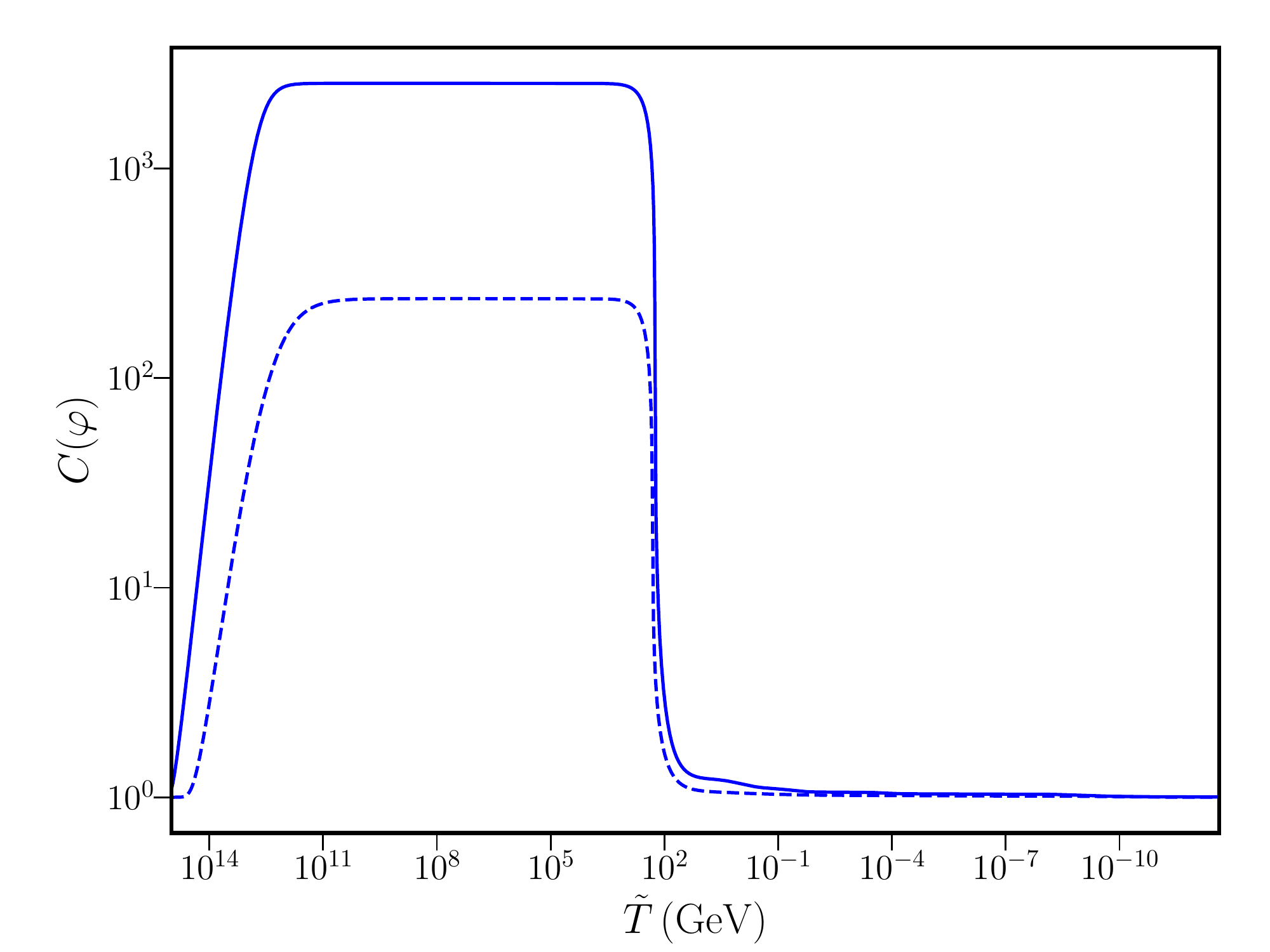}
\end{center}
\vskip -15pt
\caption{The evolution of the scalar field $\varphi$ (left panel) and the conformal factors for the purely conformal phenomenological case (solid lines), and the conformal-disformal D-brane case (dashed lines), have been plotted as functions of temperature for $n=1$, and the corresponding sets of initial conditions as mentioned in Tables~\ref{Table4} and~\ref{Table6} respectively.}\label{fig:conf-conf-dis}
\end{figure}

%
Comparing Figs.~\ref{fig:phi-Cphi-conf-disf-zerovel} and \ref{fig:phi-Cphi-confdisf} with Figs.~\ref{fig:phi-Cphi-conformal-zerovel} and \ref{fig:phi-Cphi-conformal}, we find that the evolution of the field, the conformal factors, and the expansion rates are similar overall in the purely conformal phenomenological scenario and the conformal-disformal D-brane scenario. This implies that, even in the presence of a disformal factor, the field evolution is dominated by the conformal factor. In order to better illustrate the different behaviour in the two scenarios, in Fig.~\ref{fig:conf-conf-dis}, we show  the evolution of the scalar field and the conformal factor for the purely conformal phenomenological case (solid line), and the conformal-disformal D-brane case (dashed line), for $n=1$ and initial conditions as in Tables~\ref{Table4} and~\ref{Table6} respectively. As we see, the phenomenological case seems to be more effective in enhancing the Hubble rate and thus the PGW as we will see. 
 
\section{The rise of the primordial tensor spectrum}\label{Sec:4}
In section~\ref{Sec:2}, we briefly described  the estimation of the fractional energy density of gravitational waves in a standard cosmological scenario, given by  Eq.~\eqref{eq:0gw-simple}. In  the previous section \ref{Sec:3}, we saw  that during an epoch of scalar-tensor domination,  the expansion rate is modified, thus inducing a non-trivial modification in   the PGW  energy density as follows:
\begin{equation}
 \tilde{\Omega}^0_{\rm GW}(k)\,h^2 \simeq \frac{1}{24}\,{\cal P}_{\rm T}(k)\,\left(\frac{\tilde{a}_{\rm hc}}{\tilde{a}_{_0}}\right)^4\,\left(\frac{\tilde{H}_{\rm hc}}{H_0/h}\right)^2,\label{eq:0gw-nongr}
\end{equation}
where $\tilde a$ and $\tilde H$ are the modified  scale factor and Hubble parameter in the Jordan frame, which is the frame in which energy density and entropy are conserved. As we have seen in the previous section, the Jordan  frame is  indistinguishable from $H_{GR}$ at the onset of BBN, when the conformal and/or disformal factors become one. 
Using \eqref{xi} and \eqref{kappas}, \eqref{eq:0gw-nongr} becomes 
\be\label{PGWmodified}
 \tilde{\Omega}_{\rm GW}^0\,h^2 \simeq  \left(\frac{{\cal P}_{t}}{24}\right) \left(\frac{\tilde{a}}{\tilde{a}_0}\right)^4\,\frac{\gamma^3 \,C(\varphi) \left(1+\alpha(\varphi)\,\varphi_{ N} \right)^2\,H_{\rm GR}^2}{B\,C(\varphi_0)\,\left[1+\alpha^2(\varphi_0)\right] \left(H_0/h\right)^2}\,,
\ee
where $\gamma$ is given by \eqref{eq:gamma2} and $B$  by \eqref{B}. 

We now   discuss the rising of the PGW spectrum due to the modified expansion rate  during a period of    scalar-tensor domination epoch, in the post-inflationary evolution.
In subsection~\ref{Sec:phenoGW}, we consider the characteristic step-like enhancement of the PGW due to  a conformal coupling in  the  phenomenological and D-brane scalar-tensor theories discussed in subsections.~\ref{section:purely-conformal} and \ref{section:conformaldisformal}. Then, in  subsection~\ref{Sec:DbraneGW}, we  discuss the non-trivial rise of the PGW spectrum due to a disformal coupling in the phenomenological and D-brane scalar-tensor theories discussed in subsections~\ref{sec:pureDpheno} and \ref{section:purely-disformal}. 

\subsection{Conformal enhancement of the PGW spectrum}\label{Sec:phenoGW}
In this section, we  consider the effect of the conformal coupling on the expansion rate and thus in the PGWs. As discussed previously, in the phenomenological model, the disformal and conformal couplings are not related, except via the causality condition, \eqref{eq:causality} and  this case corresponds to setting $D=0$. On the other hand, for the D-brane case, these couplings are  related through $M^4CD =1$. We have also seen in section \ref{section:conformaldisformal} that, when both couplings are turned on, the disformal effect is negligible and the behaviour of the expansion rate is dominated by the conformal term.  

As also discussed in the previous section, in the presence of a conformal coupling, an effective potential is turned on when  the temperature of the universe drops below the rest mass of a particle species so that it becomes non-relativistic, thereby causing a small departure from $\tilde{w}=1/3$, which generates a `kick' in the scalar field. 
The temperature when the kick function starts to trigger a change in the evolution of the field is approximately given by $173.2\,\,{\rm GeV}$. This temperature corresponds to a frequency value given by
\begin{equation}
\tilde f_0 \simeq 4.6\times\,10^{-6}\,\,{\rm Hz}.
\end{equation}
Hence, for frequencies above this, there is a rise and eventual enhanced  flat  spectrum for $\tilde{\Omega}^0_{\rm GW}(k)\,h^2$ for both sets of initial conditions  discussed in sections \ref{sec:Pheno1} and \ref{section:conformaldisformal},
 namely, the scalar field starting with  zero and non-zero initial velocities.  As we have already illustrated in these sections, the enhancement in the Hubble parameter can be increased (or decreased) depending on the choice of the conformal factor. 
 
\subsubsection{Phenomenological case }\label{subsec:pheno-conf-GW}

In the phenomenological  conformal scenario,  $D=0$ and thus $\gamma=1$. Therefore, Eq.~\eqref{PGWmodified} simplifies to 
\be
 \tilde{\Omega}_{\rm GW}^0\,h^2 \simeq  \left(\frac{{\cal P}_{t}}{24}\right) \left(\frac{\tilde{a}}{\tilde{a}_0}\right)^4\,\frac{C(\varphi)\,H_{\rm GR}^2 \left(1+\alpha(\varphi)\,\varphi_{ N} \right)^2}{B\,C(\varphi_0)\,\left(1+\alpha^2(\varphi_0)\right)\left(H_0/h\right)^2},
\ee
where $B$ is given by \eqref{B} with $b=1, M=0$. From this equation, we  see that  an enhancement is determined mostly by the amplitude of the conformal factor $C(\varphi)$. The quantity $\left(1+\alpha(\varphi)\,\varphi_{ N} \right)^2/B$ reaches a large value $\sim 29$ at very early temperatures but then drops rapidly to ${\cal O}(1)$. As shown in Fig.~\ref{fig:factor-conformal}, this quantity exhibits a dip in its value when the equation of state starts changing, which also results in a dip in the PGW spectrum before it gets amplified due to the conformal factor.

In Fig.~\ref{fig:sensitivity-curves-conformal} (left panel), we
 show the resulting step-like rise of the gravitational wave amplitude for the three choices of $n$ and the corresponding initial conditions in Table~\ref{Table4}.
   
\subsubsection{D-brane conformal-disformal enhancement }\label{subsec:Dbrane-conf-disf-GW}
When the conformal coupling is turned on in the D-brane scalar-tensor case, the disformal coupling is also turned on according to the relation $M^4 C D=1$. Thus, in this case, $\gamma\ne 1$, and \eqref{PGWmodified} can be rewritten as 
\be
\tilde{\Omega}_{\rm GW}^0\,h^2= \left(\frac{{\cal P}_{t}}{24}\right) \left(\frac{\tilde{a}}{\tilde{a}_0}\right)^4\,\frac{C(\varphi)\,\gamma^3\,H_{\rm GR}^2 \left(1+\alpha(\varphi)\,\varphi_N\right)^2}{B\,C(\varphi_0)\,\left(1+\alpha^2(\varphi_0)\right)\left(H_0/h\right)^2},
  \ee
where $B$ is given by \eqref{B} with $b=0$.
However, as we have seen in section \ref{section:conformaldisformal}, the modification of the expansion rate  is driven mostly by the conformal coupling. As we saw, throughout the entire evolution, $\gamma\sim 1$, and thus the rise in the PGW spectrum is driven by the conformal factor. The dip in the value of $(1+\alpha(\varphi)\varphi_N)^2/B$ (cf.~Fig.~\ref{fig:factor-conf-disf}) is reflected in a dip also in the corresponding PGW spectrum.

\begin{figure}[!h]
\begin{center}
\includegraphics[width=7.50cm]{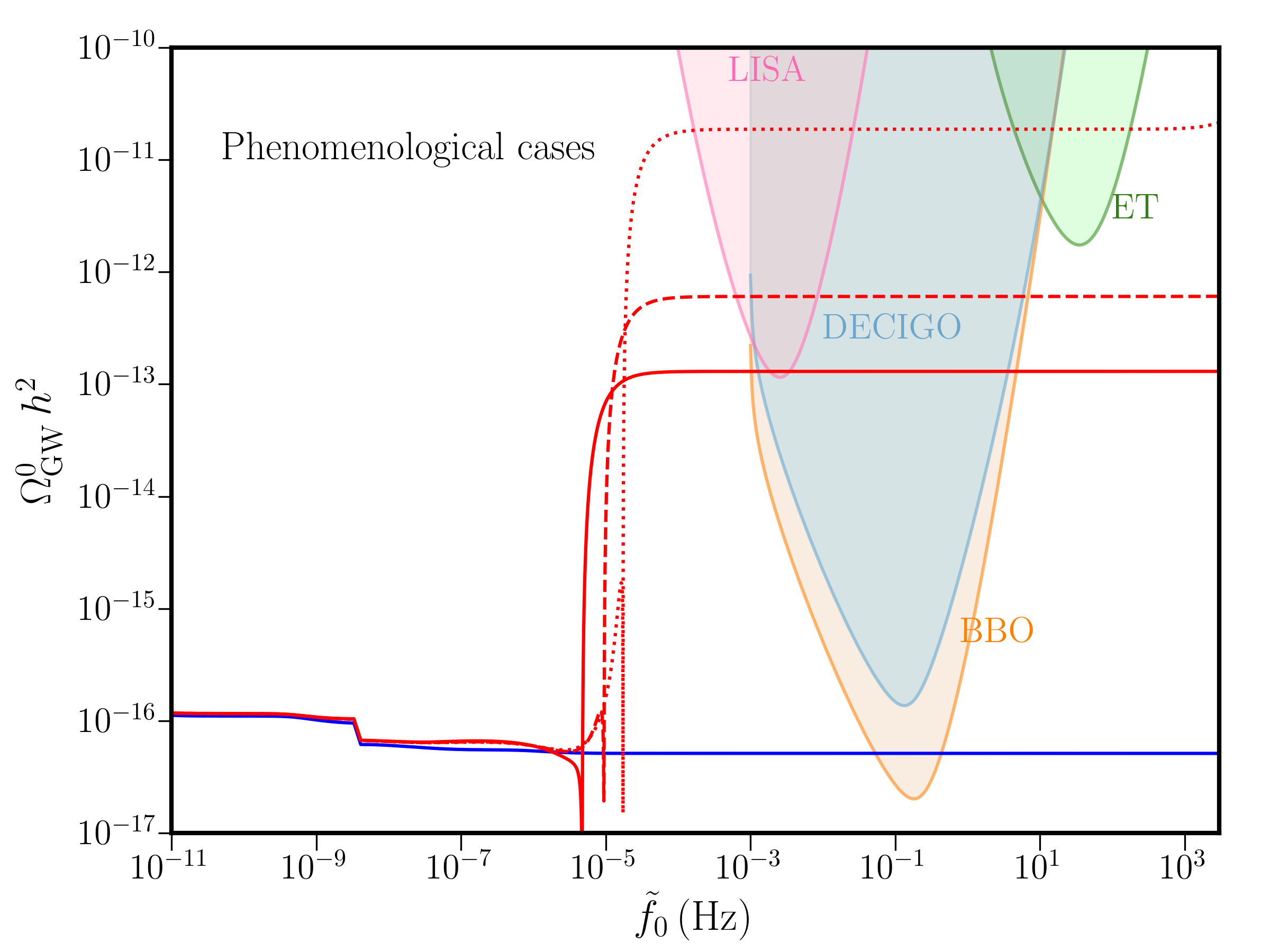}
\includegraphics[width=7.50cm]{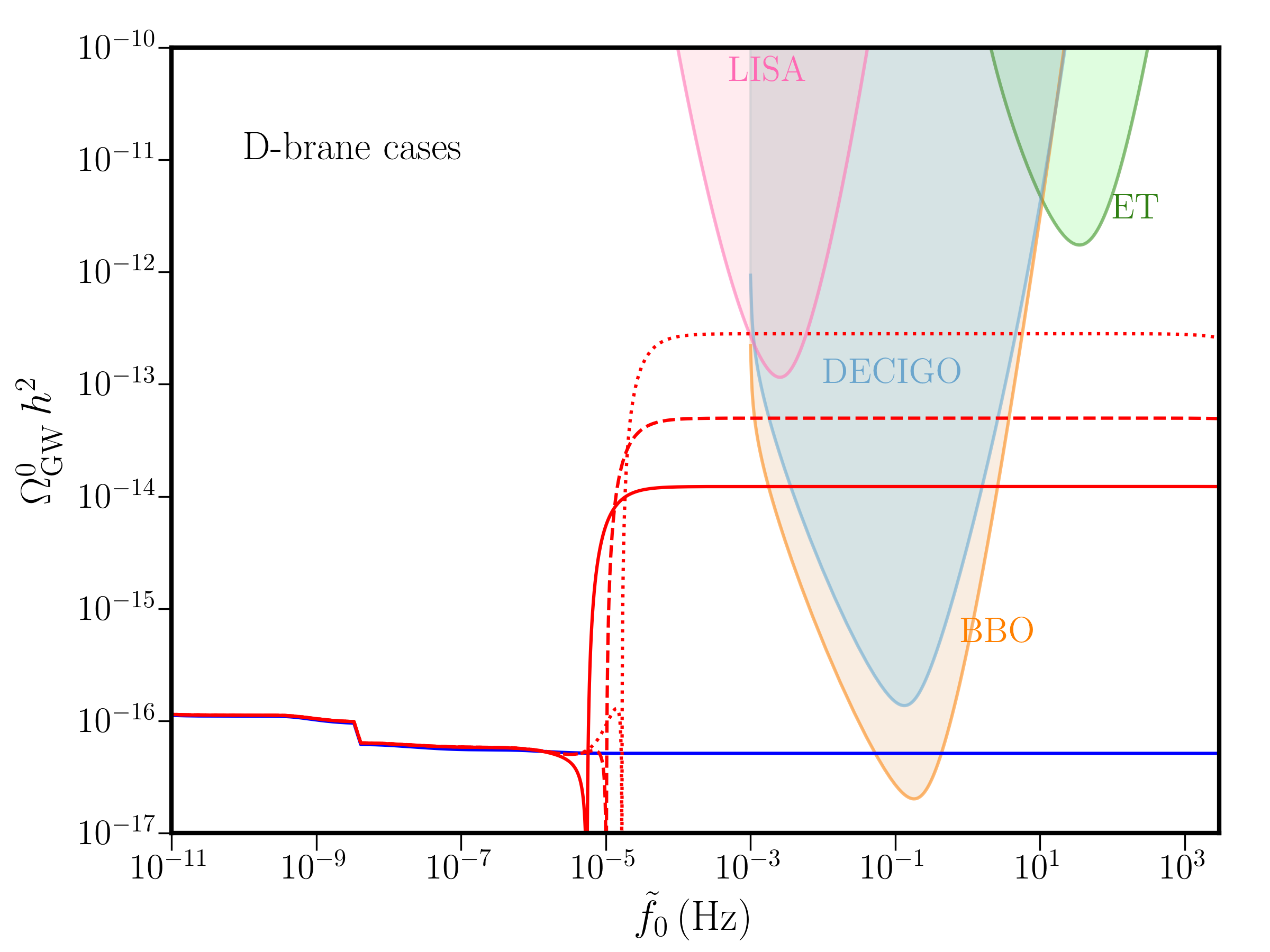}
\end{center}
\vskip -15pt
\caption{The conformally enhanced gravitational wave spectra have been plotted as functions of frequency. On the left panel, we show the PGW spectra for the phenomenological pure conformal case for the  set of initial conditions  in Table~\ref{Table4}. On the right panel, we show the PGW spectra for the D-brane conformal-disformal case for the set of initial conditions  in Table~\ref{Table6}. For both cases, the solid, dashed, and dotted lines correspond to $n=1$, $n=2$, and $n=4$ respectively. The  power-law sensitivity curves have been plotted  using the method described in~\cite{Thrane:2013oya} (more on this in section \ref{sec_BPLS}).
}\label{fig:sensitivity-curves-conformal}
\end{figure}

In Fig.~\ref{fig:sensitivity-curves-conformal} (right panel), we show the rise of the  PGW spectrum for the three choices of $n$ in \eqref{eq:conformal-factor} and the corresponding initial conditions  in Table~\ref{Table6}. 
As in the phenomenological case, the enhancement has a step-like flat spectrum. 
Thus the profile for both, phenomenological and D-brane like is the same. Note  that  in \cite{Bernal:2020ywq} the profile for the PGW had also a step-like flat behaviour, for a the  different conformal function, $C=e^{\beta \phi^2}$.  We thus expect a  step-like flat  behaviour for other choices of the conformal function.

From Fig.~\ref{fig:sensitivity-curves-conformal}, we can see that, in the phenomenological pure conformal example, where $C$  and $D$ are unrelated (except via the causality condition), the amplification of the PGW spectrum is larger compared to the D-brane model, where $C$ and $D$ are related through $M^4CD=1$. Recall that the choice of the conformal function is exactly the same in both cases. This effect may be caused by the disformal contribution.

 \subsection{Disformal rise of the PGW spectrum }
 \label{section_disfrise}
When only the disformal coupling is turned on, that is, $C=1$, 
Eq.~\eqref{PGWmodified} simplifies to:
\be
 \tilde{\Omega}_{\rm GW}^0\,h^2 \simeq  \left(\frac{{\cal P}_{\rm T}}{24}\right) \left(\frac{\tilde{a}}{\tilde{a}_0}\right)^4\,\frac{\gamma^3 \,H_{\rm GR}^2}{B\left(H_0/h\right)^2}\,,
\ee
where $\gamma$ is given by \eqref{eq:gamma2} and $B$  by \eqref{B}. Thus, the amplitude of the gravitational waves depends only on the Lorentz factor $\gamma$ and $B$ as does the modified Hubble parameter.
As we discussed  in sections \ref{sec:pureDpheno} and \ref{section:purely-disformal},  the position of the maximum enhancement of the Hubble parameter -- and thus the gravitational wave amplitude -- depends on the initial temperature. The scalar field quickly evolves, producing a large peak in  $\gamma$,  eventually settling to a constant value well before the onset of BBN. Thus, the earlier the initial temperature is chosen, the earlier the enhancement of $H$ occurs, which is reflected in the characteristic peaked  enhancement at a higher frequency in the gravitational wave spectrum.  

As we shall discuss now, the disformal coupling in both these cases gives rise to a characteristic peak in the PGW spectrum, with a peculiar frequency dependence. Moreover, the phenomenological and disformal cases give rise to  different frequency profiles, thus making the two cases potentially distinguishable by future experiments.

\subsubsection{Phenomenological disformal enhancement}\label{subsec:pheno-disf-GW}

\begin{figure}[H]
\begin{center}
\includegraphics[width=10.00cm]{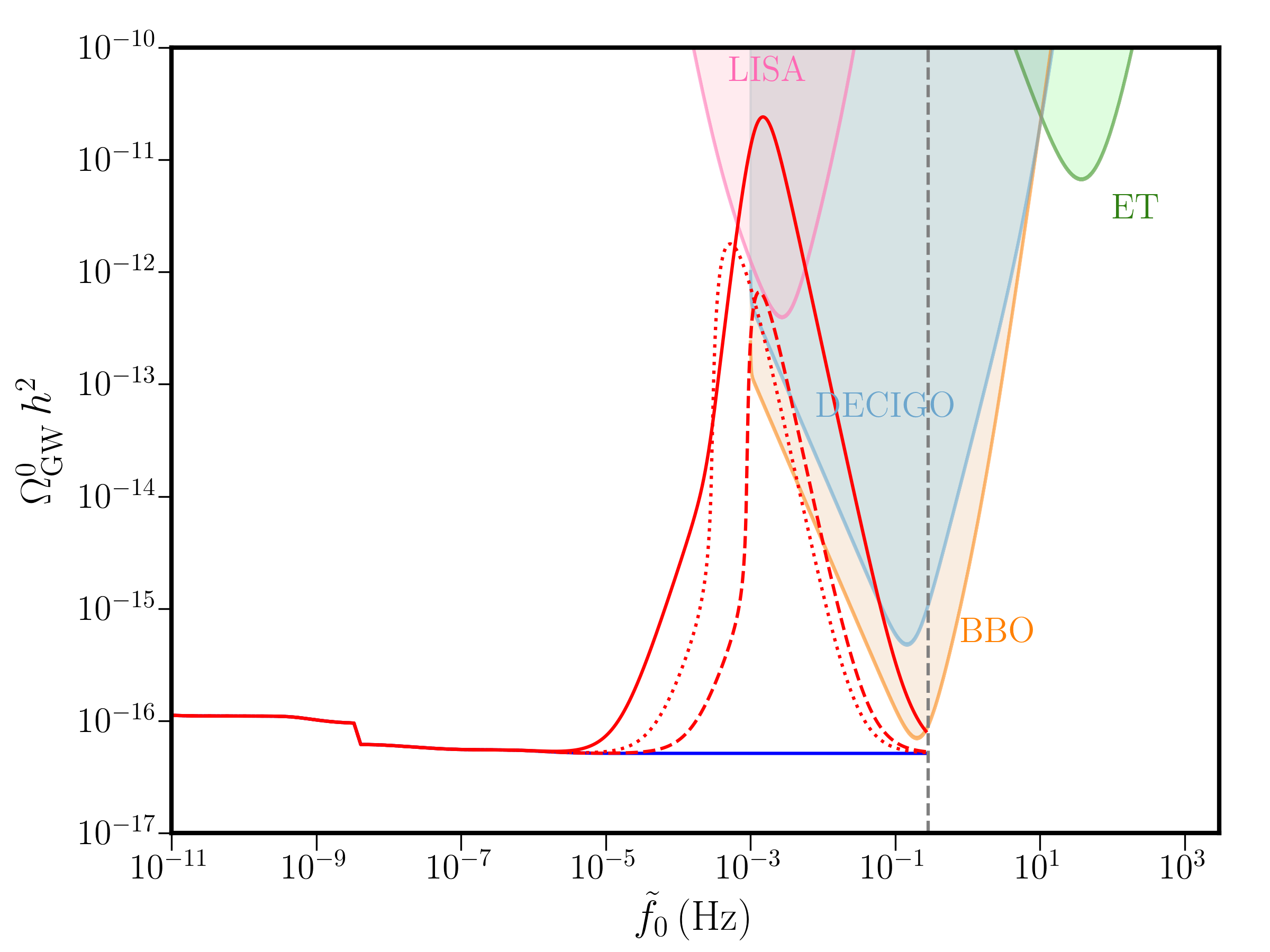}
\end{center}
\vskip -15pt
\caption{The gravitational wave spectra for the  purely disformal scenarios ($C=1$)  in the phenomenological and D-brane cases have been plotted as functions of frequency. 
The dashed and dotted lines correspond to the phenomenological cases with $D=D_0$ and $D=D_0\varphi^2$ respectively, while the solid line corresponds to the D-brane case with $D=1/M^4$.
The initial conditions have been chosen as mentioned in Table~\ref{Table5}.
The vertical line indicates the frequency corresponding to the initial temperature ($10^{7}\,{\rm GeV}$).
The experimental sensitivity curves have been plotted using the notion of broken power-law sensitivity curve (see section \ref{sec_BPLS} for details).}\label{fig:sensitivity-curves-pheno1}
\end{figure}

In section \ref{sec:pureDpheno}, we discussed two choices of the disformal function: a constant $D=D_0$ and a field dependent one, $D=D_0\,\varphi^2$ (see Table~\ref{Table5}).  As we have seen, the field dependent coupling gives rise to a larger  maximum value of the Lorentz factor $\gamma$  compared to the constant coupling, thus producing a bigger effect on the modified expansion rate. This is  reflected in the amplitude of the PGWs as well.  
 In Fig.~\ref{fig:sensitivity-curves-pheno1}, we compare the PGW spectra for $D=D_0$ (dashed line) and $D=D_0\varphi$ (dotted line) for the  initial conditions specified in Table~\ref{Table5}. 

As we can see from Fig.~\ref{fig:sensitivity-curves-pheno1}, the phenomenological disformal couplings give rise to very characteristic peaks in the PGW spectrum. In particular,  
the spectrum rises with a slope proportional to $\tilde{f}_0^2$ at the beginning,  subsequently changing slope to $\tilde{f}_0^{25}$ for the case with constant disformal factor, and $\tilde{f}_0^{20}$ for the case with field dependent disformal factor. The spectrum then drops as $\tilde{f}_0^{-3}$ for both cases. Note that this behaviour, namely the difference in the slopes for the two cases, has important implications as it makes these two couplings in principle distinguishable by future GW experiments. Interestingly, the slopes in the D-brane case are very different (see below), thus again being distinguishable from a phenomenological case.  

The shifting effect of the initial temperature condition is illustrated in Fig.~\ref{fig:sensitivity-curves-pheno2}. In this plot, we use the same initial conditions as mentioned in Table~\ref{Table5}, but with an initial temperature given by $\tilde{T}_i=10^{11}\,{\rm GeV}$, instead of $\tilde{T}_i=10^7\,{\rm GeV}$. The slopes of the curves are preserved, but the peak occurs at larger frequencies, which are relevant for the Einstein telescope \cite{Maggiore:2019uih}. Note that, for earlier initial temperatures, the peak can access ultra high frequencies, accessible by future experiments \cite{Aggarwal:2020olq}.

\subsubsection{D-brane disformal enhancement}\label{Sec:DbraneGW}
In the purely disformal D-brane-like case, $C=1$ and thus $D=1/M^4$. As we have  seen, the  enhancement of the Hubble parameter is larger than  in the phenomenological case (see Fig.~\ref{fig:phi-gammafunc-pheno1}). 

\begin{figure}[H]
\begin{center}
\includegraphics[width=10.00cm]{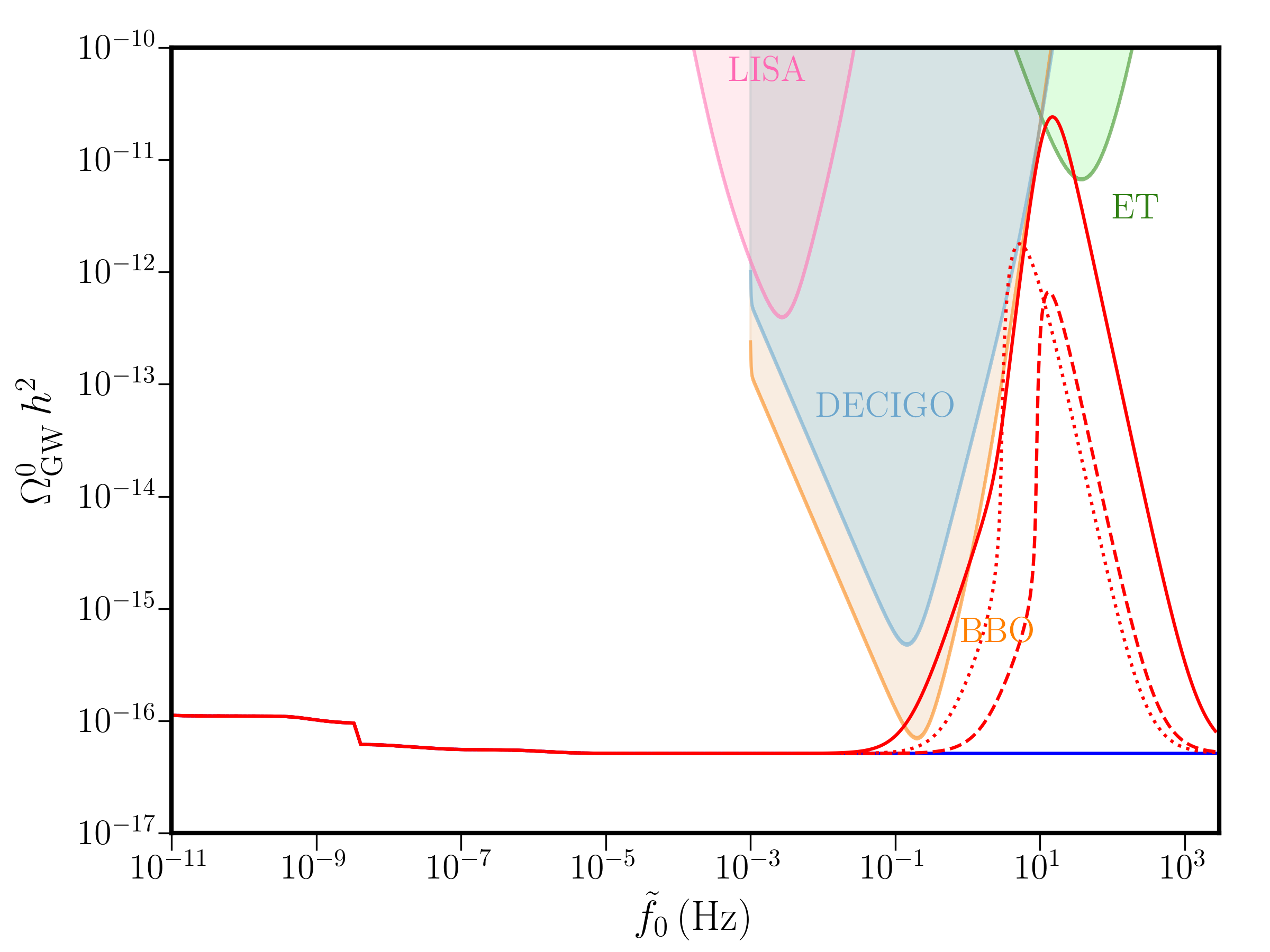}
\end{center}
\vskip -15pt
\caption{The gravitational wave spectra for the  purely disformal scenarios ($C=1$)  in the phenomenological and D-brane cases have been plotted as functions of frequency. 
The dashed and dotted lines correspond to the phenomenological cases with $D=D_0$ and $D=D_0\varphi^2$ respectively, while the solid line corresponds to the D-brane case with $D=1/M^4$.
The initial conditions have been chosen as mentioned in Table~\ref{Table7}.
The experimental sensitivity curves have been plotted using the notion of broken power-law sensitivity curve (see section \ref{sec_BPLS} for details).}\label{fig:sensitivity-curves-pheno2}
\end{figure}

This is thus reflected in the rise  of the PGW spectrum, as can be seen in  Fig.~\ref{fig:sensitivity-curves-pheno1}, where the solid line corresponds to the D-brane disformal case. Indeed in this case, the rise of the PGW spectrum is much larger than that in the constant and field dependent phenomenological cases. It could be possible that a power law disformal coupling, $D=D_0\,\varphi^r$ with $r>2 $ for the phenomenological case  might reach to and above the D-brane case. However, interestingly, the frequency dependence of the PGW enhancement is very different in both cases. The D-brane disformal coupling produces a very characteristic enhancement with a slope  proportional to $\tilde{f}_0^2$ at the beginning, subsequently  changing to $\tilde{f}_0^5$, in contrast to the $\tilde{f}_0^{25}$ or $\tilde{f}_0^{20}$ behaviour in the phenomenological cases. Moreover, the spectrum then drops as $\tilde{f}_0^{-3}$, \ie with the same slope as in the phenomenological case. This behaviour  can be directly understood from the behaviour of the Lorentz factor $\gamma$ for the three cases. We can see from Fig.~\ref{fig:phi-gammafunc-pheno1} that the rise in $\gamma$ differs between the phenomenological cases (dashed and dotted lines), and the D-brane case (solid line), while it drops down with the same slope in all the cases. 
In Fig.~\ref{fig:sensitivity-curves-pheno2}, we show the rise in the PGWs when the initial condition for temperature is set at a higher value. Again, the D-brane case gives the largest enhancement, thus crossing the sensitivity curves of Einstein Telescope. Setting even higher initial temperatures will be relevant for the ultra high frequency experiments.  
Thus, a detection of this characteristic peaked spectrum by either ET, LISA or ultra high frequency experiments will tell us the epoch of  scalar-tensor domination in the early universe.

\begin{table}[H] 
\begin{center}
\centering
\begin{tabular}{| l | c | c | }
\hline
\cellcolor[gray]{0.9}      & \cellcolor[gray]{0.9} $D_0 ({\rm GeV}^{-4})$  &  \cellcolor[gray]{0.9} $H_i$(GeV)  \\
\hline \hline
Pheno case: $D=D_0$  & $5.000\times 10^{-38}$  &   $1.413\times 10^{4}$   \\
\hline
Pheno case: $D=D_0\varphi^2$  & $6.000\times 10^{-37}$   &   $1.408\times10^{4}$  \\
\hline
D-brane case: $D=1/M^4$ & $4.822\times 10^{-37}$   &   $1.516\times10^{4}$  \\
\hline
\end{tabular}
\end{center} 
\caption{Initial conditions for the disformal coupling models in the phenomenological and D-brane scalar-tensor theories. The other initial conditions in all cases are $\varphi^i=0.200$, $\varphi^i_{\tilde N}=2.000\times 10^{-5}, $ $\tilde  T_i = 10^{11}$(GeV).    }
\label{Table7}
\end{table}   

\subsubsection{Einstein frame rise and spinning axions}
\label{subsec:spinaxion}

\begin{figure}[H]
\begin{center}
\includegraphics[width=7.50cm]{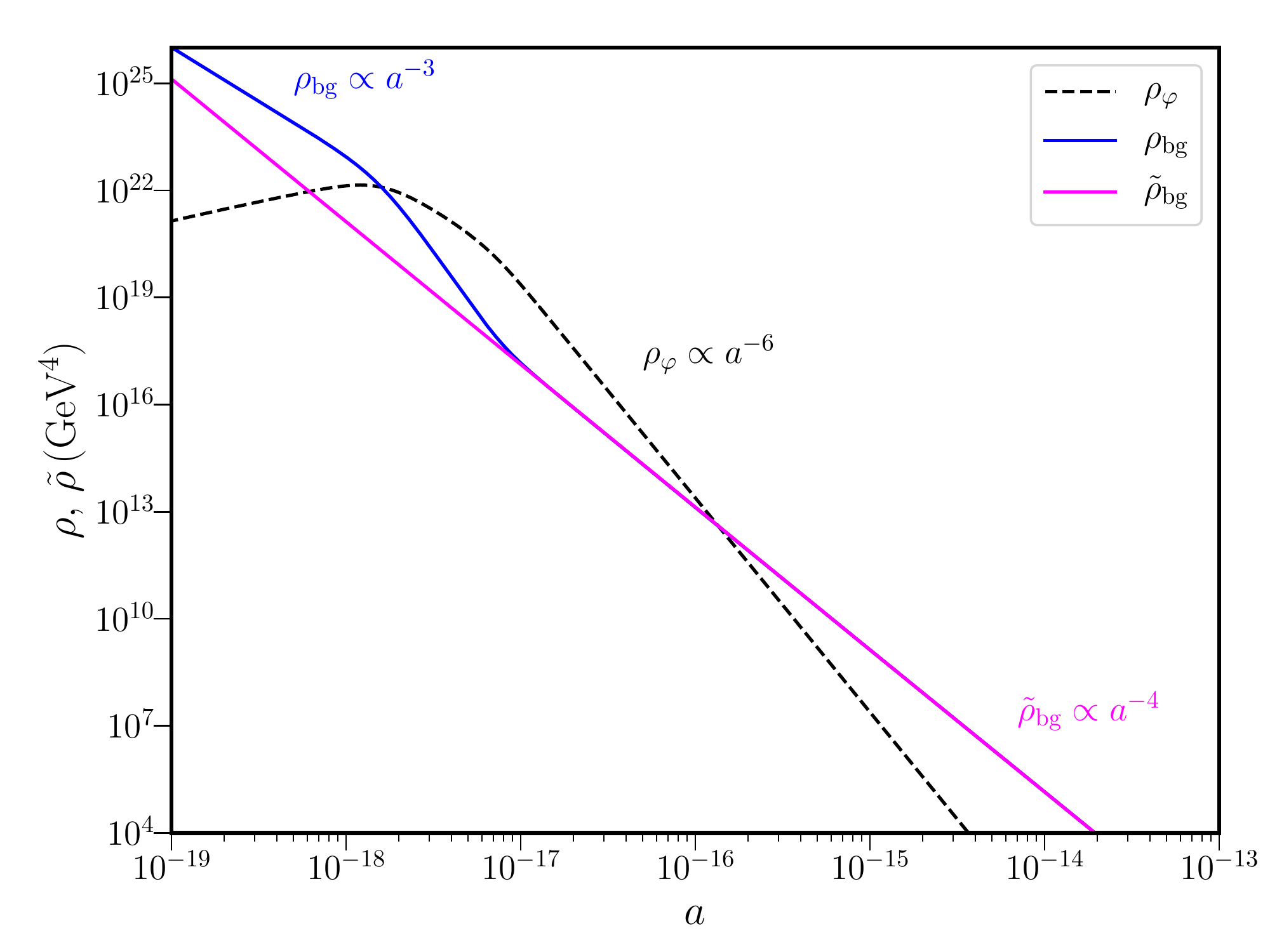}
\includegraphics[width=7.50cm]{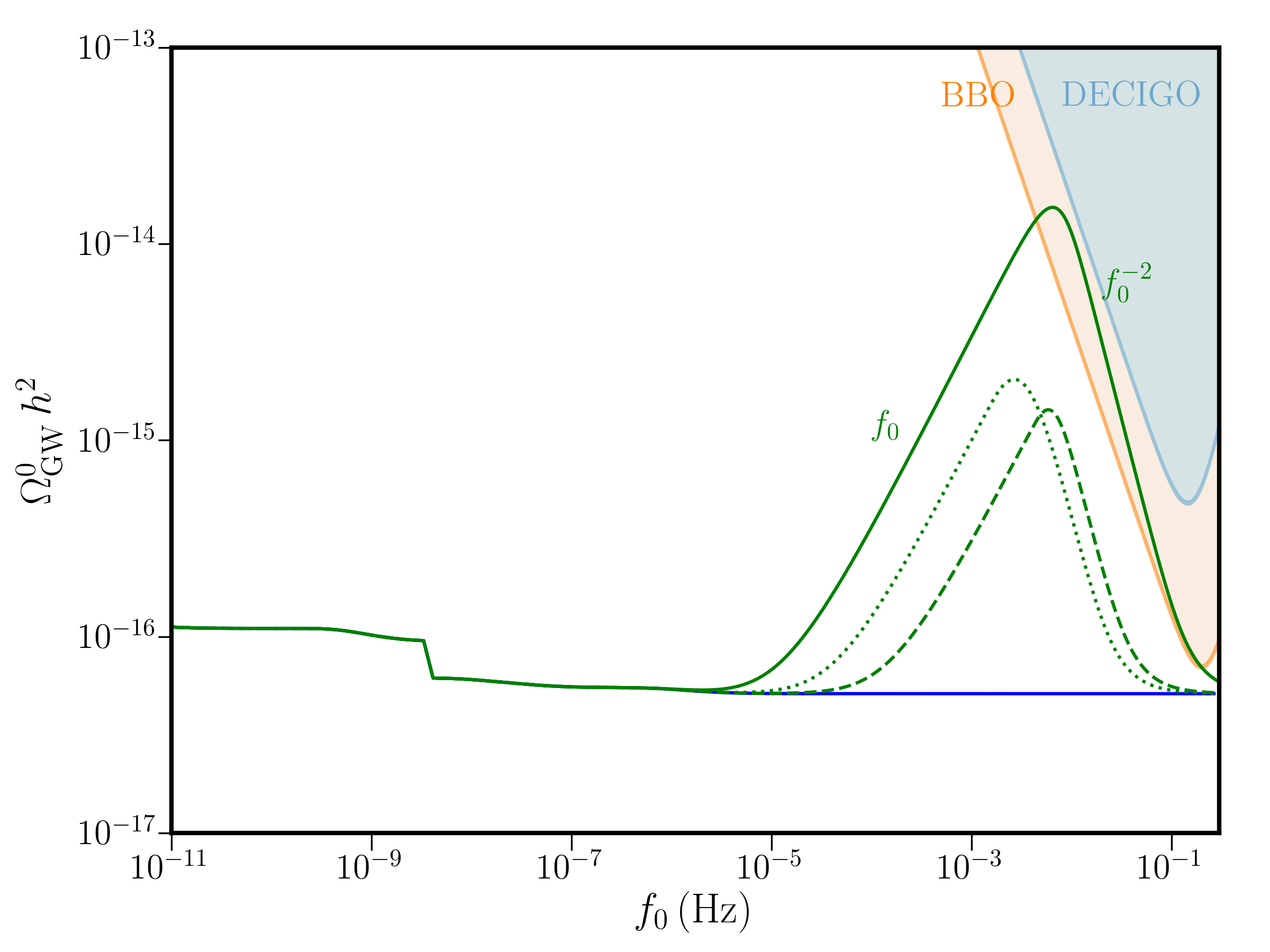}
\end{center}
\vskip -15pt
\caption{Einstein frame energy densities (left panel) \eqref{eq:einstjor}  and $\Omega_{\rm GW}h^2$ for the pure disformal cases. Solid line corresponds to D-brane scalar tensor case, dashed line to constant  and dotted line to the field dependent phenomenological  disformal cases, respectively. Initial conditions are as in Table~\ref{Table5}.}\label{fig:Einstein_rho_Omega}
\end{figure}

We finish this section with a discussion on the  peaked enhancement  of the PGW spectrum, when computed using the Einstein frame Hubble parameter \eqref{eq:H-HGR}. In this case, the spectrum grows and falls with a frequency slope that goes as $\Omega_{GW}^0 h^2\propto  f_0^\beta$, with $\beta=-2\frac{(1-3\omega)}{(1+3\omega)}$ and $\omega$ the Einstein frame equation of state of the  total energy density $\rho_{\rm total}$. The peak arises since the energy density of the universe in the Einstein frame (see \eqref{eq:einstjor} with $C=1$), behaves as matter with $\omega=0$   ($\rho\propto a^{-3}$)  at early times,
subsequently changing to a  period of `kination' domination due to the scalar field energy density with $\omega=1$ ($\rho\propto a^{-6}$), to finally start behaving as radiation when $\gamma$ drops to one. 
 This non-standard evolution of the energy density in the Einstein frame  causes the peaked enhancement in the PGW spectrum. In Fig.~\ref{fig:Einstein_rho_Omega} (left panel), we show  the energy density of the scalar field, $\rho_\varphi$, and the background,  $\rho_{\rm bg}$,
  in the Einstein frame (see \eqref{eq:einstjor} with $C=1$),  while in the right panel we show  the peaked  PGW spectrum. A similar behaviour, although  in a very different set-up, has been recently considered in 
\cite{Co:2021lkc,Gouttenoire:2021wzu}, where a short kination era is implemented due to a spinning axion  in field space. Let us stress however that the systems are very different, and in the present case, it is the Jordan frame where entropy  and energy density are conserved (see \eqref{eq:tilderhocons}), and thus in this  frame $\tilde \rho\propto a^{-4}$ before the onset of BBN, as it should.

\subsection{The concept of  broken power-law sensitivity curve}
\label{sec_BPLS}

We  introduce the concept of  broken power-law sensitivity  curve (BPLS) as a graphical  tool to show the sensitivity of GW experiments 
 to scenarios producing stochastic  GW background (SGWB) spectra 
with   broken power-law profiles, as the ones we met in section \ref{section_disfrise}. The BPLS generalizes the methods used to obtain the nominal and the power-law (PLS) sensitivity curves. 

Nominal sensitivity curves -- see e.g. \cite{Moore:2014lga} for a general discussion -- 
  provide a visual understanding on whether
a  GW event, characterised by a given frequency and amplitude, 
 can or can not be detected with sufficiently high signal-to-noise
ratio  (SNR) by a GW experiment.

 Power-law sensitivity curves (PLS) \cite{Thrane:2013oya} make visually manifest the fact that, by integrating over frequencies, we  can exploit the broadband nature of a  SGWB signal, by  accumulating  SNR  and   making a detection of  the SGWB more feasible even when  its  amplitude does not fit within the nominal noise curves. 
 Assuming that the SGWB is described by a power-law profile in frequency, scaling as $f^\beta$ with $\beta$ being a constant slope, the PLS is made by the envelope  of experimental limits which can be placed for each slope $\beta$, varying $\beta$ over a given interval, and integrating the signal over frequencies.  As 
 clearly explained in \cite{Thrane:2013oya}, this allows one to gain orders of magnitude in sensitivity, at the price of making hypothesis  on the slope of the SGWB frequency profile.
  Moreover, additional sensitivity to a SGWB is accumulated integrating the
  signal over time.

\begin{figure}
\begin{center}
  \includegraphics[width = 0.45 \textwidth]{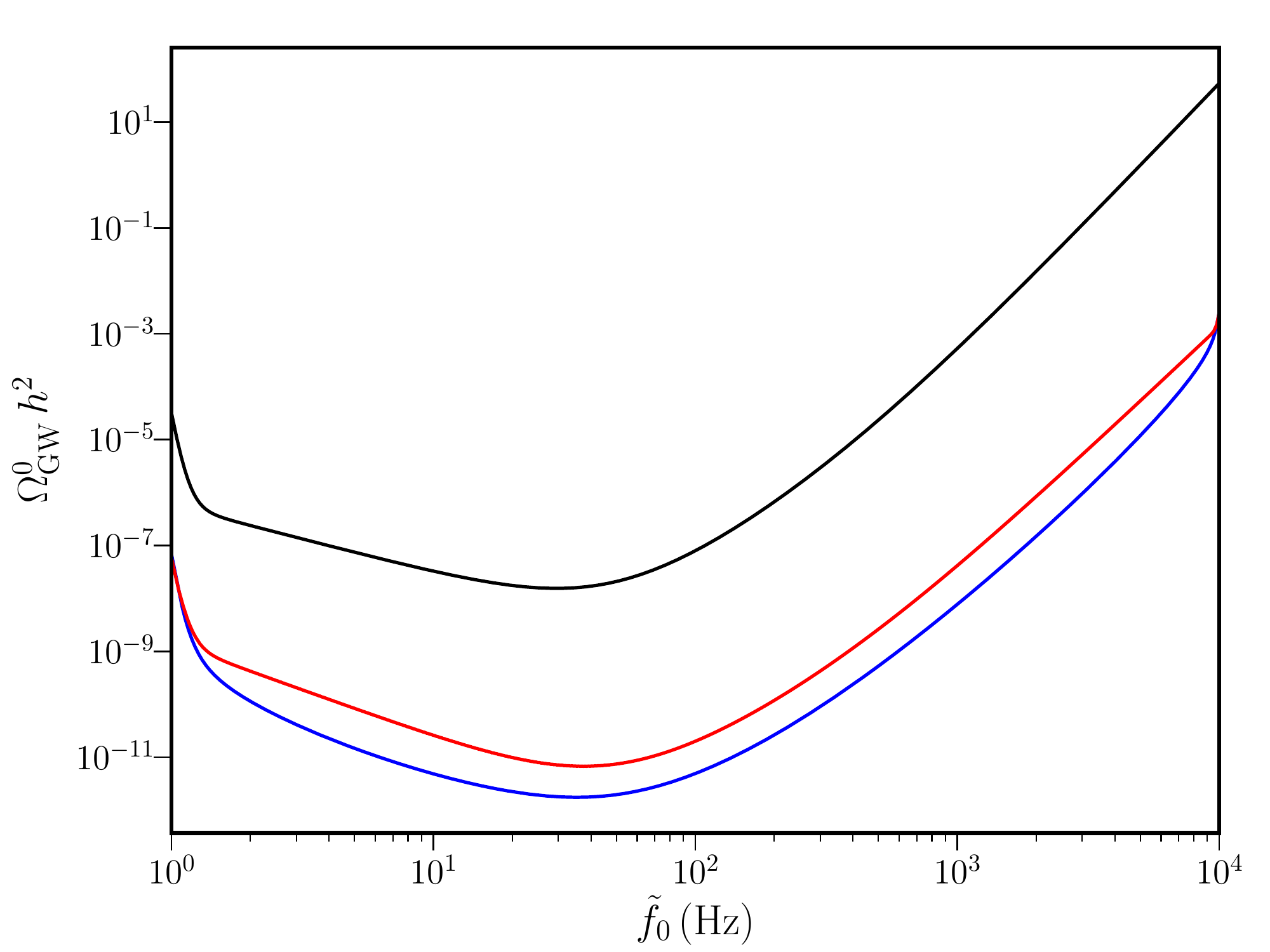}
    \includegraphics[width = 0.45 \textwidth]{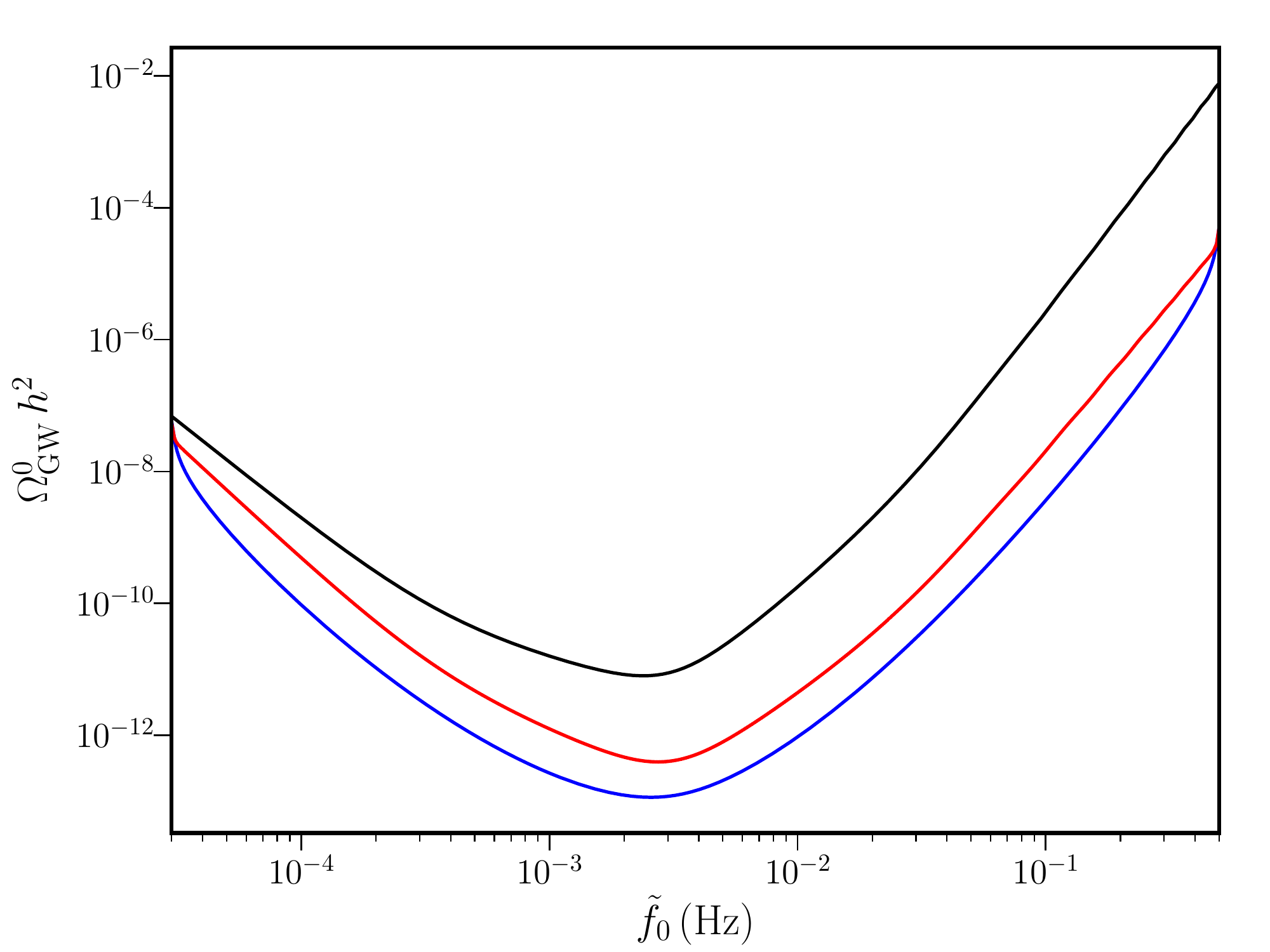}
        \includegraphics[width = 0.45 \textwidth]{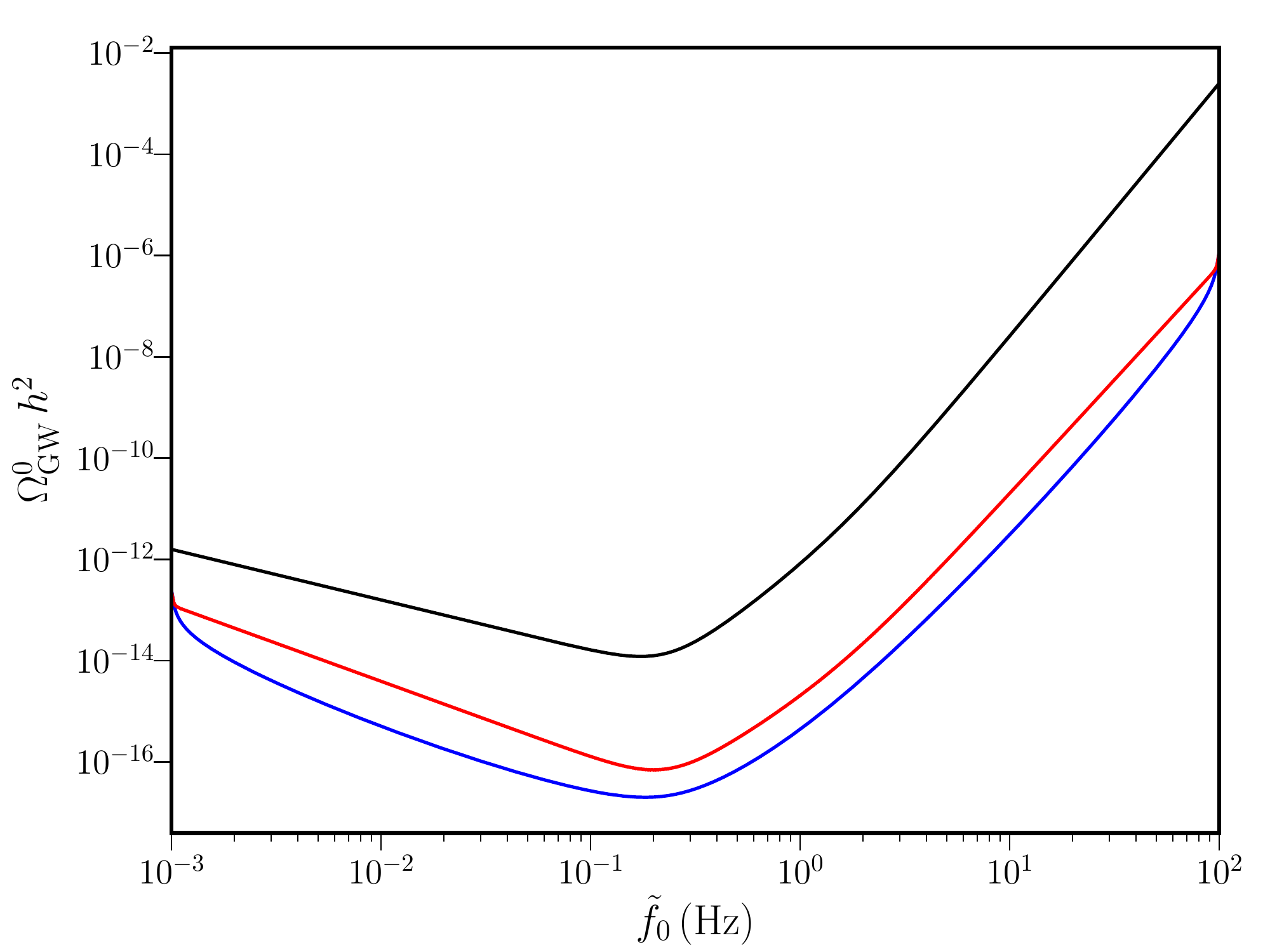}
                \includegraphics[width = 0.45 \textwidth]{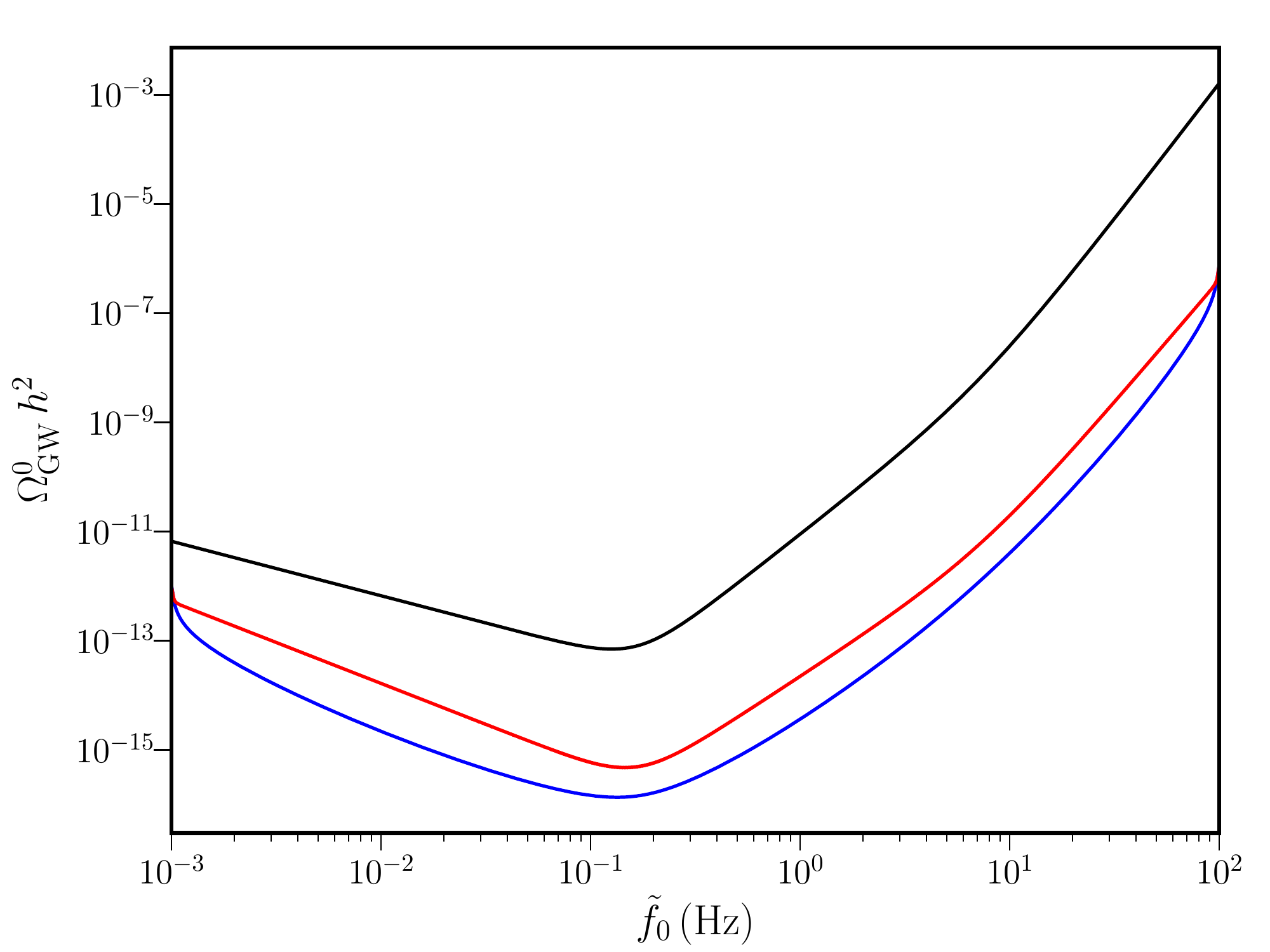}
    \end{center}
 \caption{ \small Nominal (black), BPLS (red),  and PLS (blue) sensitivity curves for four GW experiments:  Einstein Telescope (upper left), LISA (upper right), BBO (lower left), and DECIGO (lower right). For drawing the BPLS we vary the slopes $\beta_1$ and $\beta_2$ of the broken power-laws within the intervals $(- 25, 25)$, while the frequency $f_\star$ where the slope changes  vary over the entire band of each GW experiment.  The duration of the experiment is set to $T=3$ years.}
  \label{fig_BPLS}
\end{figure}

 However, the frequency profiles of  various examples of SGWBs encountered in the literature are often not well described by a single power-law -- they
  can instead be  described  by a broken power-law (see e.g. the analysis in \cite{Kuroyanagi:2018csn,Bartolo:2016ami}). A broken-power law profile changes slope at a  value of GW frequency within the band of a given GW experiment. In this case, by using the standard PLS of \cite{Thrane:2013oya} one  tends to {\it overestimate} the sensitivity of an experiment to the  SGWB signal. In fact,  when assuming that the frequency profile is a single power-law over the entire instrumental band, one misses the possibility that  the frequency profile changes slope and drops  within the experiment frequency range. In this way, the signal   `leaves' the sensitivity region of the experiment sooner than what is expected for a single power-law.
    For this reason, we propose to generalise the  same method of \cite{Thrane:2013oya} to the case of broken power-laws, and obtain more realistic sensitivity curves adapted for this class of models.  We follow the  discussion 
   of \cite{Thrane:2013oya} step by step,  adapting it to our context. We start from the noise spectral densities of specific GW experiments. We assume
   that the signal has a broken power-law functional form, denoted by  $\Omega_{\beta_1,\beta_2}(f/f_\star)$. The  broken power-law  is  characterised by two slopes parameterised by the indices $\beta_1$ and $\beta_2$, as well as
   a reference frequency $f_\star$ at which the slope of the spectrum changes. For any given choice  of the set ${\cal S}=(\beta_1, \beta_2, f_\star)$ we determine the overall amplitude $\bar \Omega_{\cal S}$
   of the spectrum which is necessary for achieving an SNR$=$10 by integrating over the entire instrumental frequency band. Then, the
   broken power-law sensitivity curve 
     is
   the envelope of the profiles   $\Omega_{\beta_1,\beta_2}(f/f_\star)$, with amplitude  $\bar \Omega_{\cal S}$, drawn as a function of the frequency $f$. This method for building the BPLS requires an extremisation process over the three-dimensional space of the set  ${\cal S}=(\beta_1, \beta_2, f_\star)$ (while the PLS requires extremisation only over a single parameter).  We wrote a parallelised Fortran code for performing the procedure
   quickly and efficiently, which is available upon request.  We plot the resulting BPLS curve in Fig.~\ref{fig_BPLS} for the case of four GW experiments:
   Einstein Telescope (we use ET-B version presented in  \cite{Regimbau:2012ir,Sathyaprakash:2009xs} as nominal curve), LISA (we use the noise curves of \cite{Flauger:2020qyi}), 
  BBO, and DECIGO (we use \cite{Yagi:2011wg}).

   Fig.~\ref{fig_BPLS} shows that the BPLS curves lose sensitivity with respect to the PLS ones, by a factor of order one. However, the loss in sensitivity is quite dependent on the interval over which we allow the parameters $\beta_{1,2}$ to vary. The wider  the interval of variation for $\beta_{1,2}$, the larger is the loss  in sensitivity with respect to the PLS curves, because the broken power-laws can  be more peaked, and we are left with  a smaller frequency band wherein to integrate and accumulate SNR. We have chosen  $\beta_{1,2}$ to vary over the interval $(-25, 25)$ because it is the range of slopes we met in the scenarios of  section \ref{section_disfrise}. 
   
   Notice that PLS and BPLS in general tend to converge at the extrema of the frequency band of each  experiment. This is understandable, since  if the change in slope of BPL occurs at those extrema, then
    the signal effectively acts as a single power law over the rest of the frequency band, with the slight change in slope close to the extrema contributing very little to the SNR.
   
   As with the PLS, the BPLS offers a practical visual aid for understanding whether the predictions of a given model
  can be detected by a GW experiment. If the model profile enters within the BPLS, then further sophisticated
  tools are needed for understanding whether the fine details of the model profile can be reconstructed -- see
  e.g. the proposals and analyses in \cite{Caprini:2019pxz,Flauger:2020qyi,Banagiri:2021ovv}. 

\section{Conclusions}\label{sec:end}

Cosmological inflation  predicts the existence of a primordial gravitational wave (PGW) background  spanning a large range of frequencies with an almost scale-invariant profile, whose amplitude is too small to be detected by future gravitational wave experiments. 
The fractional energy density in primordial gravitational waves, as observed today, depends on the  primordial tensor spectrum generated during inflation,  and crucially, on the the expansion rate of the Universe from the end of inflation until today. 

In this work we showed that if a period of well motivated scalar-tensor theories dominated epoch occurs before the onset of BBN, a large enhancement of the PGW signal can occur at frequencies probed by future GW interferometers such as LISA, ET, BBO and DECIGO.

\smallskip

We considered the most general physically consistent relation between two metrics and a single scalar field, which  includes a conformal as well as a disformal (or derivative) coupling~\cite{Bekenstein:1992pj}, $\tilde g_{\mu\nu} = C(\varphi)\,g_{\mu\nu} + D(\varphi)\,\partial_\mu\varphi\,\partial_\nu\varphi$. 
These couplings arise naturally  from (D-)branes in (D-)brane models of cosmology \cite{Koivisto:2013fta}.  In this case, the functions $C$ and $D$ are not independent ($M^4CD=1$) and thus, they cannot simply be turned on 
and off as in a phenomenological treatment of these theories.

We first considered the effect of a  conformal coupling (conformal-disformal in the D-brane case) on the PGW spectrum. 
In the phenomenological case, a conformal  coupling was first considered in \cite{Bernal:2020ywq}. In the present work we explored different conformal functions  \eqref{eq:conformal-factor} motivated by previous work in \cite{Catena:2004ba,Dutta:2016htz,Dutta:2017fcn}, and different initial conditions,  to illustrate  the conformal effect more clearly. Generically, the enhancement of the PGW has a step-like behaviour, similar to that found in \cite{Bernal:2020ywq} for the conformal function, $C=e^{\beta}\phi^2$, and depends on the choice of conformal factor and initial conditions (see Figs.~\ref{fig:phi-Cphi-conformal-zerovel}, \ref{fig:phi-Cphi-conformal},  \ref{fig:phi-Cphi-conf-disf-zerovel}, \ref{fig:phi-Cphi-confdisf}). In the phenomenological case, we saw that a suitable conformal factor  can  enhance the flat spectrum,  reaching the sensitivity curves of LISA and ET, and thus can be tested by performing correlations between these experiments. Interestingly, the conformal-disformal effect in D-brane-like scalar tensor theories is less effective at enhancing the PGW spectrum (see Fig.~\ref{fig:conf-conf-dis}). This may be due to the disformal effect, which although subdominant could affect the enhancing effect. 

We next considered the more interesting purely disformal effect, again both in a   phenomenological set-up and in the non-trivial D-brane-like scalar tensor set-up. In the phenomenological approach, one can simply switch off the conformal factor, $C=0$, while keeping $D\ne 0$. The pure disformal factor gives rise to a  distinct  peaked spectrum. We showed that the rising slope of this peaked spectrum behaves as ($\tilde f_0^{2} \to \tilde f_0^{25}$) for the case with constant disformal factor, while it behaves as ($\tilde f_0^{2} \to \tilde f_0^{20}$) for the case with field dependent disformal factor. However, for both cases, the spectrum falls as $\tilde f_0^{-3}$. The amplitude can be enhanced by a suitable choice of disformal function (see Fig.~\ref{fig:sensitivity-curves-pheno1}).   

On the other hand, the D-brane-like pure disformal coupling, obtained by setting $C=1$ and thus $D=M^{-4}$, gives rise to a substantial enhancement, larger than the phenomenological case, and more remarkably with a characteristic  peaked spectrum with very different rising slopes ($\tilde f_0^{2} \to \tilde f_0^{5}$), but same falling slopes ($\tilde f_0^{-3}$)!  In this case, the  pure disformal rise  reaches well within the LISA and ET sensitivity curves, depending on the initial conditions (see Figs.~\ref{fig:sensitivity-curves-pheno1}, \ref{fig:sensitivity-curves-pheno2}). Indeed, as discussed in \cite{Dutta:2017fcn}, the enhancement of the expansion rate, and thus the PGW, can be shifted to earlier/larger times/frequencies, by changing the initial conditions. 
This peaked-like spectrum with distinct slopes in the frequency is a feature of the disformal effect, which does not arise in  other modified cosmologies and  can thus be a smoking gun signal of a period of D-brane disformal scalar-tensor dominated epoch.  In order to study the sensitivity of GW experiments to spectra with these features more accurately, we used the tool  of  broken power-law sensitivity curves, introduced
 in section 
\ref{sec_BPLS}.

Throughout this paper we have assumed $\lambda\sim0$, namely, the contribution of the scalar potential to  the energy density being negligible during the whole evolution. In particular, we assume that the present day acceleration is due to a pure cosmological constant $\Lambda$. This does not need to be the case and the scalar field could contribute to the dark energy density, as quintessence. We leave the exploration of this possibility for future work. Another interesting aspect for future exploration is the possible implications of the scalar field displacement in the disformal case. As we saw, the disformal coupling drives the field quickly to large values, which may be imply the presence of additional fields, according to recent quantum gravity constraints (see \cite{Palti:2019pca,vanBeest:2021lhn,Grana:2021zvf} for recent reviews). 

\bigskip

\section*{Acknowledgements}
D.~C.~acknowledges the support of the Supercomputing Wales project, which is part-funded by the European Regional Development Fund (ERDF) via Welsh Government.
D.~C.~would also like to thank H.~V.~Ragavendra for discussions.
This work is part-funded by the STFC grant ST/T000813/1. For the purpose of open access, the authors have applied a Creative Commons Attribution (CC BY) licence to any Author Accepted Manuscript version arising.

\newpage
\begin{appendix}

\section{Determining the initial conditions: disformal models}\label{sec:ICsDisform}

\subsection{Phenomenological case, $D=D_0$}
The initial condition for the Hubble parameter can be set by solving the following cubic equation:
\begin{equation}
d_1\,H^6 - H^4 +d_2^2 = 0,
\end{equation}
where
\begin{subequations}
\begin{align}
    d_1 &= \frac{D_0 \varphi_{_{\Nt}}^2}{\kappa^2}, \\
    d_2 &= \frac{\kappa^2}{3 B}\,\tilde{\rho}.
\end{align}
\end{subequations}
One of the solutions for the cubic equation is given by
\begin{equation}
    H^2 = \frac{1}{3 d_1}\left[1+\left(\frac{2}{\Delta}\right)^{1/3}+\left(\frac{\Delta}{2}\right)^{1/3}\right],
\end{equation}
where $\Delta=2-27 d_1^2 d_2^2 + i d_1 d_2 \sqrt{27\left(4-27 d_1^2 d_2^2\right)}$.
The other two solutions can be obtained by replacing
\begin{subequations}
 \begin{align}
     \left(\frac{2}{\Delta}\right)^{1/3} &\to {\rm e}^{2\pi i/3}\left(\frac{2}{\Delta}\right)^{1/3}, \\
     \left(\frac{2}{\Delta}\right)^{1/3} &\to {\rm e}^{4\pi i/3}\left(\frac{2}{\Delta}\right)^{1/3}.
 \end{align}
\end{subequations}
In order to obtain real positive solutions for $H^2$, we need $27 d_1^2 d_2^2 \leq 4$, which leads to the following condition:
\begin{equation}
    D_0 \leq \left(\frac{2}{\varphi_{_{\Nt}}^2}-\frac{1}{3}\right)\frac{30}{\sqrt{3} \pi^2 g_* \tilde{T}^4}.
\end{equation}
Further, the causality condition~(\ref{eq:causality-condition}) implies that we must have 
\begin{equation}
    \left(D_0\,H^2\,\varphi_{_{\Nt}}^2/\kappa^2\right)-1 < 0.\label{eq:causality-pheno1}
\end{equation}

\subsection{Phenomenological case, $D=D_0 \varphi^2$}

 The initial condition for the Hubble parameter can be set by solving the following cubic equation:
\begin{equation}
d_1\,H^6 - H^4 +d_2^2 = 0,
\end{equation}
where
\begin{subequations}
\begin{align}
    d_1 &= \frac{D(\varphi) \varphi_{_{\Nt}}^2}{\kappa^2}, \\
    d_2 &= \frac{\kappa^2}{3 B}\,\tilde{\rho}.
\end{align}
\end{subequations}
One of the solutions for the cubic equation is given by
\begin{equation}
    H^2 = \frac{1}{3 d_1}\left[1+\left(\frac{2}{\Delta}\right)^{1/3}+\left(\frac{\Delta}{2}\right)^{1/3}\right],
\end{equation}
where $\Delta=2-27 d_1^2 d_2^2 + i d_1 d_2 \sqrt{27\left(4-27 d_1^2 d_2^2\right)}$.
The other two solutions can be obtained by replacing
\begin{subequations}
 \begin{align}
     \left(\frac{2}{\Delta}\right)^{1/3} &\to {\rm e}^{2\pi i/3}\left(\frac{2}{\Delta}\right)^{1/3}, \\
     \left(\frac{2}{\Delta}\right)^{1/3} &\to {\rm e}^{4\pi i/3}\left(\frac{2}{\Delta}\right)^{1/3}.
 \end{align}
\end{subequations}
In order to obtain real positive solutions for $H^2$, we need $27 d_1^2 d_2^2 \leq 4$, which leads to the following condition:
\begin{equation}
    D_0 \leq \left(\frac{2}{\varphi_{_{\Nt}}^2}-\frac{1}{3}\right)\frac{30}{\sqrt{3} \pi^2 \varphi^2 g_* \tilde{T}^4}.
\end{equation}
Further, the causality condition~(\ref{eq:causality-condition}) implies that we must have 
\begin{equation}
    \left(D_0\,\varphi^2\,H^2\,\varphi_{_{\Nt}}^2/\kappa^2\right) -1 < 0.\label{eq:causality-pheno2}
\end{equation}

\subsection{D-brane pure disformal case}\label{subsec:ic-choice}

The cubic equation for $H^2$ that needs to be solved is \cite{Dutta:2016htz,Dutta:2017fcn}:
\begin{equation}
    A_1\,H^6 + A_2\,H^4 + A_3\,H^2 + A_4=0,\label{eq:H-cubiceqn-d}
\end{equation}
where
\begin{subequations}
 \begin{align}
     A_1 &= \frac{\varphi_{_{\Nt}}^2}{M^4\,\kappa^2}, \\
     A_2 &= \frac{2\,\varphi_{_{\Nt}}^2}{3}-1, \\
     A_3 &= \frac{M^4\,\kappa^2}{3}\left(\frac{\varphi_{_{\Nt}}^2}{3}-2\right), \\
     A_4 &= \left(\frac{M^4\,\kappa^2}{3}\right)^2\,\frac{\left(1+\lambda\right)\,\tilde{\rho}}{M^4}\left[\frac{\left(1+\lambda\right)\,\tilde{\rho}}{M^4}+2\right].
 \end{align}
\end{subequations}
One of the solutions to Eq.~(\ref{eq:H-cubiceqn-d}) is given by
\begin{equation}
    H^2 = \frac{1}{3\,A_1}\left[-A_2 + \left(A_2^2 - 3\,A_1\,A_3\right)\left(\frac{2}{\Delta}\right)^{1/3} + \left(\frac{\Delta}{2}\right)^{1/3}\right],
\end{equation}
where
\begin{equation}
    \Delta = L + \sqrt{L^2-4\,{\ell}^3},
\end{equation}
with
\begin{subequations}
 \begin{align}
     L &= -27\,A_1^2\,A_4 + 9\,A_1\,A_2\,A_3 - 2\,A_2^3, \\
     \ell &= A_2^2 - 3\,A_1\,A_3.
 \end{align}
\end{subequations}
The two other solutions to Eq.~(\ref{eq:H-cubiceqn-d}) are given by
\begin{subequations}
 \begin{align}
     \left(\frac{2}{\Delta}\right)^{1/3} &\to {\rm e}^{2\pi i/3}\left(\frac{2}{\Delta}\right)^{1/3}, \\
     \left(\frac{2}{\Delta}\right)^{1/3} &\to {\rm e}^{4\pi i/3}\left(\frac{2}{\Delta}\right)^{1/3}.
 \end{align}
\end{subequations}
Since we are looking for real positive solutions of $H^2$, we need to consider values such that $\Delta$ is complex, \ie $4\ell^3 > L^2$. This implies that
\begin{equation}
    \frac{\tilde{\rho}}{M^4}\left(\frac{\tilde{\rho}}{M^4}+2\right) \leq \frac{\left(3+4\varphi_{_{\Nt}}^{i 2}\right)\left(\varphi_{_{\Nt}}^{i 2}-6\right)^2}{81\varphi_{_{\Nt}}^{i 4}}.
\end{equation}
Since we set the initial conditions on the field during the radiation dominated era, we know that $\tilde{\rho}(\tilde{T})=\left(\pi^2/30\right)\,g_*(\tilde{T})\,\tilde{T}^4$. Therefore, for a particular set of initial conditions $\left(\varphi_i, \varphi_{_{\Nt}}^i\right)$, the value of $M$ is bounded by the following interval:
\begin{equation}
    \left[\left(\frac{3\pi^2 g_*(\tilde{T}) \varphi_{_{\Nt}}^{i 2}}{-90 \varphi_{_{\Nt}}^{i 2} + 20\sqrt{(3+\varphi_{_{\Nt}}^{i 2})^3}}\right)^{1/4}\tilde{T}_i, \,+\infty\right].
\end{equation}
Using the above condition, for a particular set of values of $\left(\varphi_i, \varphi_{_{\Nt}}^i, M\right)$, we can obtain the real positive solutions for Eq.~(\ref{eq:H-cubiceqn-d}). This gives us the initial condition $H_i$ for the Hubble parameter. These initial conditions can be used together to obtain the initial value of the Lorentz factor, $\gamma_i$. We find that, in order to satisfy the constraints on the present value of the Hubble parameter, we need to choose suitable initial conditions such that $\gamma_i \sim 1$.

\subsection{D-brane conformal-disformal}\label{subsec:ic-choiceD+C}

The cubic equation for $H^2$ that needs to be solved in this case is give by \cite{Dutta:2016htz,Dutta:2017fcn}:
\begin{equation}
    A_1\,H^6 + A_2\,H^4 + A_3\,H^2 + A_4=0,\label{eq:H-cubiceqn}
\end{equation}
where
\begin{subequations}
 \begin{align}
     A_1 &= \frac{D\,\varphi_{_N}^2}{C\,\kappa^2}, \\
     A_2 &= \frac{2\,M^4\,C\,D\,\varphi_{_N}^2}{3}-1, \\
     A_3 &= \frac{M^4\,C^2\,\kappa^2}{3}\left(\frac{M^4\,C\,D\,\varphi_{_N}^2}{3}-2\right), \\
     A_4 &= \left(\frac{M^4\,C^2\,\kappa^2}{3}\right)^2\,\frac{\left(1+\lambda\right)\,\tilde{\rho}}{M^4}\left[\frac{\left(1+\lambda\right)\,\tilde{\rho}}{M^4}+2\right].
 \end{align}
\end{subequations}
One of the solutions to Eq.~(\ref{eq:H-cubiceqn}) is given by
\begin{equation}
    H^2 = \frac{1}{3\,A_1}\left[-A_2 + \left(A_2^2 - 3\,A_1\,A_3\right)\left(\frac{2}{\Delta}\right)^{1/3} + \left(\frac{\Delta}{2}\right)^{1/3}\right],
\end{equation}
where
\begin{equation}
    \Delta = L + \sqrt{L^2-4\,{\ell}^3},
\end{equation}
with
\begin{subequations}
 \begin{align}
     L &= -27\,A_1^2\,A_4 + 9\,A_1\,A_2\,A_3 - 2\,A_2^3, \\
     \ell &= A_2^2 - 3\,A_1\,A_3.
 \end{align}
\end{subequations}
The two other solutions to Eq.~(\ref{eq:H-cubiceqn}) are given by
\begin{subequations}
 \begin{align}
     \left(\frac{2}{\Delta}\right)^{1/3} &\to {\rm e}^{2\pi i/3}\left(\frac{2}{\Delta}\right)^{1/3}, \\
     \left(\frac{2}{\Delta}\right)^{1/3} &\to {\rm e}^{4\pi i/3}\left(\frac{2}{\Delta}\right)^{1/3}.
 \end{align}
\end{subequations}
Since we are looking for real positive solutions of $H^2$, we need to consider values such that $\Delta$ is complex, \ie $4\ell^3 > L^2$. This implies that
\begin{equation}
    \frac{\tilde{\rho}}{M^4}\left(\frac{\tilde{\rho}}{M^4}+2\right) \leq \frac{\left(3+4\varphi_{_N}^{i 2}\right)\left(\varphi_{_N}^{i 2}-6\right)^2}{81\varphi_{_N}^{i 4}}.
\end{equation}
Since we set the initial conditions on the field during the radiation dominated era, we know that $\tilde{\rho}(\tilde{T})=\left(\pi^2/30\right)\,g_*(\tilde{T})\,\tilde{T}^4$. Therefore, for a particular set of initial conditions $\left(\varphi_i, \varphi_{_N}^i\right)$, the value of $M$ is bounded by the following interval:
\begin{equation}
    \left[\left(\frac{3\pi^2 g_*(\tilde{T}) \varphi_{_N}^{i 2}}{-90 \varphi_{_N}^{i 2} + 20\sqrt{(3+\varphi_{_N}^{i 2})^3}}\right)^{1/4}\tilde{T}_i, \,+\infty\right].
\end{equation}
Using the above condition, for a particular set of values of $\left(\varphi_i, \varphi_{_N}^i, M\right)$, and by using the normalization condition $M^4 C D=1$, we can obtain the real positive solutions for Eq.~(\ref{eq:H-cubiceqn}). Note that, in order to solve the equations~(\ref{eq:conformal-disformal-diffeqs}) in terms of $\Nt$, we have to convert the initial conditions obtained by the above procedure into initial conditions on the set of values of $\left(\varphi_i, \varphi_{_{\Nt}}^i, M\right)$.

For the scenario wherein the field evolution is started with zero initial velocity, we find that $A_1=0$. In this case, the cubic equation for $H^2$ reduces to the following quadratic equation:
\begin{equation}
 A_2\,H^4+A_3\,H^2+A_4=0,
\end{equation}
where
\begin{subequations}
 \begin{align}
     A_2 &= -1, \\
     A_3 &= -\frac{2\,M^4\,C^2\,\kappa^2}{3}, \\
     A_4 &= \left(\frac{M^4\,C^2\,\kappa^2}{3}\right)^2\,\frac{\left(1+\lambda\right)\,\tilde{\rho}}{M^4}\left[\frac{\left(1+\lambda\right)\,\tilde{\rho}}{M^4}+2\right].
 \end{align}
\end{subequations}
The above equation can be solved to obtain
\begin{equation}
 H^2=-\frac{1}{3}\,M^4\,C^2\,\kappa^2 \pm \frac{1}{2}\,\sqrt{A_3^2+4\,A_4}.\label{eq:Hsolve}
\end{equation}
To ensure we get only real and positive values for $H^2$, we must have $A_3^2+4\,A_4 > 0$, which implies that
\begin{equation}
 \left(\frac{2\,M^4\,C^2\,\kappa^2}{3}\right)^2\,\left(1+\frac{\tilde{\rho}\left(1+\lambda\right)}{M^4}\right)^2 > 0.
\end{equation}
This condition is satisfied for all values of $M$. Hence, we see that the choice of initial condition for $H$ does not depend on the value of $M$. When $\varphi_{_N}^i=0$, we have $B_i=1$ and $\gamma_i=1$, so that
\begin{equation}
    H_i^2 = \frac{\kappa^2}{3}\,C_i^2\,\tilde{\rho}_i.\label{eq:Hi-zerovel}
\end{equation}
This equation also illustrates that the choice of initial condition for $H$ is independent of $M$ and depends only on the initial temperature through $\tilde{\rho}_i$ and the initial field value through $C_i$.


\end{appendix}
\bibliographystyle{JHEP}
\bibliography{gwmc}

\providecommand{\href}[2]{#2}\begingroup\raggedright\begin{thebibliography}{10}

\bibitem{Will:2014kxa}
C.~M. Will, {\it {The Confrontation between General Relativity and
  Experiment}},  {\em Living Rev. Rel.} {\bf 17} (2014) 4,
  [\href{http://arxiv.org/abs/1403.7377}{{\tt arXiv:1403.7377}}].

\bibitem{Sakstein:2014isa}
J.~Sakstein, {\it {Disformal Theories of Gravity: From the Solar System to
  Cosmology}},  {\em JCAP} {\bf 12} (2014) 012,
  [\href{http://arxiv.org/abs/1409.1734}{{\tt arXiv:1409.1734}}].

\bibitem{Ip:2015qsa}
H.~Y. Ip, J.~Sakstein, and F.~Schmidt, {\it {Solar System Constraints on
  Disformal Gravity Theories}},  {\em JCAP} {\bf 10} (2015) 051,
  [\href{http://arxiv.org/abs/1507.00568}{{\tt arXiv:1507.00568}}].

\bibitem{Wang:2016lxa}
B.~Wang, E.~Abdalla, F.~Atrio-Barandela, and D.~Pavon, {\it {Dark Matter and
  Dark Energy Interactions: Theoretical Challenges, Cosmological Implications
  and Observational Signatures}},  {\em Rept. Prog. Phys.} {\bf 79} (2016),
  no.~9 096901, [\href{http://arxiv.org/abs/1603.08299}{{\tt
  arXiv:1603.08299}}].

\bibitem{LIGOScientific:2017zic}
{\bf LIGO Scientific, Virgo, Fermi-GBM, INTEGRAL} Collaboration, B.~P. Abbott
  et~al., {\it {Gravitational Waves and Gamma-rays from a Binary Neutron Star
  Merger: GW170817 and GRB 170817A}},  {\em Astrophys. J. Lett.} {\bf 848}
  (2017), no.~2 L13, [\href{http://arxiv.org/abs/1710.05834}{{\tt
  arXiv:1710.05834}}].

\bibitem{Catena:2004ba}
R.~Catena, N.~Fornengo, A.~Masiero, M.~Pietroni, and F.~Rosati, {\it {Dark
  matter relic abundance and scalar - tensor dark energy}},  {\em Phys. Rev. D}
  {\bf 70} (2004) 063519, [\href{http://arxiv.org/abs/astro-ph/0403614}{{\tt
  astro-ph/0403614}}].

\bibitem{Catena:2009tm}
R.~Catena, N.~Fornengo, M.~Pato, L.~Pieri, and A.~Masiero, {\it {Thermal Relics
  in Modified Cosmologies: Bounds on Evolution Histories of the Early Universe
  and Cosmological Boosts for PAMELA}},  {\em Phys. Rev. D} {\bf 81} (2010)
  123522, [\href{http://arxiv.org/abs/0912.4421}{{\tt arXiv:0912.4421}}].

\bibitem{Gelmini:2013awa}
G.~B. Gelmini, J.-H. Huh, and T.~Rehagen, {\it {Asymmetric dark matter
  annihilation as a test of non-standard cosmologies}},  {\em JCAP} {\bf 08}
  (2013) 003, [\href{http://arxiv.org/abs/1304.3679}{{\tt arXiv:1304.3679}}].

\bibitem{Rehagen:2014vna}
T.~Rehagen and G.~B. Gelmini, {\it {Effects of kination and scalar-tensor
  cosmologies on sterile neutrinos}},  {\em JCAP} {\bf 06} (2014) 044,
  [\href{http://arxiv.org/abs/1402.0607}{{\tt arXiv:1402.0607}}].

\bibitem{Wang:2015gua}
S.-z. Wang, H.~Iminniyaz, and M.~Mamat, {\it {Asymmetric dark matter and the
  scalar-tensor model}},  {\em Int. J. Mod. Phys. A} {\bf 31} (2016), no.~07
  1650021, [\href{http://arxiv.org/abs/1503.06519}{{\tt arXiv:1503.06519}}].

\bibitem{Lahanas:2006hf}
A.~B. Lahanas, N.~E. Mavromatos, and D.~V. Nanopoulos, {\it {Dilaton and
  off-shell (non-critical string) effects in Boltzmann equation for species
  abundances}},  {\em PMC Phys. A} {\bf 1} (2007) 2,
  [\href{http://arxiv.org/abs/hep-ph/0608153}{{\tt hep-ph/0608153}}].

\bibitem{Pallis:2009ed}
C.~Pallis, {\it {Cold Dark Matter in non-Standard Cosmologies, PAMELA, ATIC and
  Fermi LAT}},  {\em Nucl. Phys. B} {\bf 831} (2010) 217--247,
  [\href{http://arxiv.org/abs/0909.3026}{{\tt arXiv:0909.3026}}].

\bibitem{Salati:2002md}
P.~Salati, {\it {Quintessence and the relic density of neutralinos}},  {\em
  Phys. Lett. B} {\bf 571} (2003) 121--131,
  [\href{http://arxiv.org/abs/astro-ph/0207396}{{\tt astro-ph/0207396}}].

\bibitem{Arbey:2008kv}
A.~Arbey and F.~Mahmoudi, {\it {SUSY constraints from relic density: High
  sensitivity to pre-BBN expansion rate}},  {\em Phys. Lett. B} {\bf 669}
  (2008) 46--51, [\href{http://arxiv.org/abs/0803.0741}{{\tt
  arXiv:0803.0741}}].

\bibitem{Iminniyaz:2013cla}
H.~Iminniyaz and X.~Chen, {\it {Relic Abundance of Asymmetric Dark Matter in
  Quintessence}},  {\em Astropart. Phys.} {\bf 54} (2014) 125--131,
  [\href{http://arxiv.org/abs/1308.0353}{{\tt arXiv:1308.0353}}].

\bibitem{Meehan:2014zsa}
M.~T. Meehan and I.~B. Whittingham, {\it {Asymmetric dark matter in braneworld
  cosmology}},  {\em JCAP} {\bf 06} (2014) 018,
  [\href{http://arxiv.org/abs/1403.6934}{{\tt arXiv:1403.6934}}].

\bibitem{Meehan:2014bya}
M.~T. Meehan and I.~B. Whittingham, {\it {Dark matter relic density in
  Gauss-Bonnet braneworld cosmology}},  {\em JCAP} {\bf 12} (2014) 034,
  [\href{http://arxiv.org/abs/1404.4424}{{\tt arXiv:1404.4424}}].

\bibitem{Meehan:2015cna}
M.~T. Meehan and I.~B. Whittingham, {\it {Dark matter relic density in
  scalar-tensor gravity revisited}},  {\em JCAP} {\bf 12} (2015) 011,
  [\href{http://arxiv.org/abs/1508.05174}{{\tt arXiv:1508.05174}}].

\bibitem{Dutta:2016htz}
B.~Dutta, E.~Jimenez, and I.~Zavala, {\it {Dark Matter Relics and the Expansion
  Rate in Scalar-Tensor Theories}},  {\em JCAP} {\bf 06} (2017) 032,
  [\href{http://arxiv.org/abs/1612.05553}{{\tt arXiv:1612.05553}}].

\bibitem{Dutta:2017fcn}
B.~Dutta, E.~Jimenez, and I.~Zavala, {\it {D-brane Disformal Coupling and
  Thermal Dark Matter}},  {\em Phys. Rev. D} {\bf 96} (2017), no.~10 103506,
  [\href{http://arxiv.org/abs/1708.07153}{{\tt arXiv:1708.07153}}].

\bibitem{Bernal:2020ywq}
N.~Bernal, A.~Ghoshal, F.~Hajkarim, and G.~Lambiase, {\it {Primordial
  Gravitational Wave Signals in Modified Cosmologies}},  {\em JCAP} {\bf 11}
  (2020) 051, [\href{http://arxiv.org/abs/2008.04959}{{\tt arXiv:2008.04959}}].

\bibitem{Watanabe:2006qe}
Y.~Watanabe and E.~Komatsu, {\it {Improved Calculation of the Primordial
  Gravitational Wave Spectrum in the Standard Model}},  {\em Phys. Rev. D} {\bf
  73} (2006) 123515, [\href{http://arxiv.org/abs/astro-ph/0604176}{{\tt
  astro-ph/0604176}}].

\bibitem{Saikawa:2018rcs}
K.~Saikawa and S.~Shirai, {\it {Primordial gravitational waves, precisely: The
  role of thermodynamics in the Standard Model}},  {\em JCAP} {\bf 05} (2018)
  035, [\href{http://arxiv.org/abs/1803.01038}{{\tt arXiv:1803.01038}}].

\bibitem{Bernal:2019lpc}
N.~Bernal and F.~Hajkarim, {\it {Primordial Gravitational Waves in Nonstandard
  Cosmologies}},  {\em Phys. Rev. D} {\bf 100} (2019), no.~6 063502,
  [\href{http://arxiv.org/abs/1905.10410}{{\tt arXiv:1905.10410}}].

\bibitem{Planck:2018jri}
{\bf Planck} Collaboration, Y.~Akrami et~al., {\it {Planck 2018 results. X.
  Constraints on inflation}},  {\em Astron. Astrophys.} {\bf 641} (2020) A10,
  [\href{http://arxiv.org/abs/1807.06211}{{\tt arXiv:1807.06211}}].

\bibitem{BICEP:2021xfz}
{\bf BICEP, Keck} Collaboration, P.~A.~R. Ade et~al., {\it {Improved
  Constraints on Primordial Gravitational Waves using Planck, WMAP, and
  BICEP/Keck Observations through the 2018 Observing Season}},  {\em Phys. Rev.
  Lett.} {\bf 127} (2021), no.~15 151301,
  [\href{http://arxiv.org/abs/2110.00483}{{\tt arXiv:2110.00483}}].

\bibitem{Kamionkowski:1993fg}
M.~Kamionkowski, A.~Kosowsky, and M.~S. Turner, {\it {Gravitational radiation
  from first order phase transitions}},  {\em Phys. Rev. D} {\bf 49} (1994)
  2837--2851, [\href{http://arxiv.org/abs/astro-ph/9310044}{{\tt
  astro-ph/9310044}}].

\bibitem{Maggiore:1999vm}
M.~Maggiore, {\it {Gravitational wave experiments and early universe
  cosmology}},  {\em Phys. Rept.} {\bf 331} (2000) 283--367,
  [\href{http://arxiv.org/abs/gr-qc/9909001}{{\tt gr-qc/9909001}}].

\bibitem{Bekenstein:1992pj}
J.~D. Bekenstein, {\it {The Relation between physical and gravitational
  geometry}},  {\em Phys. Rev. D} {\bf 48} (1993) 3641--3647,
  [\href{http://arxiv.org/abs/gr-qc/9211017}{{\tt gr-qc/9211017}}].

\bibitem{Erickcek:2013dea}
A.~L. Erickcek, N.~Barnaby, C.~Burrage, and Z.~Huang, {\it {Chameleons in the
  Early Universe: Kicks, Rebounds, and Particle Production}},  {\em Phys. Rev.
  D} {\bf 89} (2014), no.~8 084074, [\href{http://arxiv.org/abs/1310.5149}{{\tt
  arXiv:1310.5149}}].

\bibitem{Coc:2006rt}
A.~Coc, K.~A. Olive, J.-P. Uzan, and E.~Vangioni, {\it {Big bang
  nucleosynthesis constraints on scalar-tensor theories of gravity}},  {\em
  Phys. Rev. D} {\bf 73} (2006) 083525,
  [\href{http://arxiv.org/abs/astro-ph/0601299}{{\tt astro-ph/0601299}}].

\bibitem{Damour:1993id}
T.~Damour and K.~Nordtvedt, {\it {Tensor - scalar cosmological models and their
  relaxation toward general relativity}},  {\em Phys. Rev. D} {\bf 48} (1993)
  3436--3450.

\bibitem{Allahverdi:2013noa}
R.~Allahverdi, M.~Cicoli, B.~Dutta, and K.~Sinha, {\it {Nonthermal dark matter
  in string compactifications}},  {\em Phys. Rev. D} {\bf 88} (2013), no.~9
  095015, [\href{http://arxiv.org/abs/1307.5086}{{\tt arXiv:1307.5086}}].

\bibitem{Thrane:2013oya}
E.~Thrane and J.~D. Romano, {\it {Sensitivity curves for searches for
  gravitational-wave backgrounds}},  {\em Phys. Rev. D} {\bf 88} (2013), no.~12
  124032, [\href{http://arxiv.org/abs/1310.5300}{{\tt arXiv:1310.5300}}].

\bibitem{Maggiore:2019uih}
M.~Maggiore et~al., {\it {Science Case for the Einstein Telescope}},  {\em
  JCAP} {\bf 03} (2020) 050, [\href{http://arxiv.org/abs/1912.02622}{{\tt
  arXiv:1912.02622}}].

\bibitem{Aggarwal:2020olq}
N.~Aggarwal et~al., {\it {Challenges and Opportunities of Gravitational Wave
  Searches at MHz to GHz Frequencies}},
  \href{http://arxiv.org/abs/2011.12414}{{\tt arXiv:2011.12414}}.

\bibitem{Co:2021lkc}
R.~T. Co, D.~Dunsky, N.~Fernandez, A.~Ghalsasi, L.~J. Hall, K.~Harigaya, and
  J.~Shelton, {\it {Gravitational Wave and CMB Probes of Axion Kination}},
  \href{http://arxiv.org/abs/2108.09299}{{\tt arXiv:2108.09299}}.

\bibitem{Gouttenoire:2021wzu}
Y.~Gouttenoire, G.~Servant, and P.~Simakachorn, {\it {Revealing the Primordial
  Irreducible Inflationary Gravitational-Wave Background with a Spinning
  Peccei-Quinn Axion}},  \href{http://arxiv.org/abs/2108.10328}{{\tt
  arXiv:2108.10328}}.

\bibitem{Moore:2014lga}
C.~J. Moore, R.~H. Cole, and C.~P.~L. Berry, {\it {Gravitational-wave
  sensitivity curves}},  {\em Class. Quant. Grav.} {\bf 32} (2015), no.~1
  015014, [\href{http://arxiv.org/abs/1408.0740}{{\tt arXiv:1408.0740}}].

\bibitem{Kuroyanagi:2018csn}
S.~Kuroyanagi, T.~Chiba, and T.~Takahashi, {\it {Probing the Universe through
  the Stochastic Gravitational Wave Background}},  {\em JCAP} {\bf 11} (2018)
  038, [\href{http://arxiv.org/abs/1807.00786}{{\tt arXiv:1807.00786}}].

\bibitem{Bartolo:2016ami}
N.~Bartolo et~al., {\it {Science with the space-based interferometer LISA. IV:
  Probing inflation with gravitational waves}},  {\em JCAP} {\bf 12} (2016)
  026, [\href{http://arxiv.org/abs/1610.06481}{{\tt arXiv:1610.06481}}].

\bibitem{Regimbau:2012ir}
T.~Regimbau et~al., {\it {A Mock Data Challenge for the Einstein
  Gravitational-Wave Telescope}},  {\em Phys. Rev. D} {\bf 86} (2012) 122001,
  [\href{http://arxiv.org/abs/1201.3563}{{\tt arXiv:1201.3563}}].

\bibitem{Sathyaprakash:2009xs}
B.~S. Sathyaprakash and B.~F. Schutz, {\it {Physics, Astrophysics and Cosmology
  with Gravitational Waves}},  {\em Living Rev. Rel.} {\bf 12} (2009) 2,
  [\href{http://arxiv.org/abs/0903.0338}{{\tt arXiv:0903.0338}}].

\bibitem{Flauger:2020qyi}
R.~Flauger, N.~Karnesis, G.~Nardini, M.~Pieroni, A.~Ricciardone, and
  J.~Torrado, {\it {Improved reconstruction of a stochastic gravitational wave
  background with LISA}},  {\em JCAP} {\bf 01} (2021) 059,
  [\href{http://arxiv.org/abs/2009.11845}{{\tt arXiv:2009.11845}}].

\bibitem{Yagi:2011wg}
K.~Yagi and N.~Seto, {\it {Detector configuration of DECIGO/BBO and
  identification of cosmological neutron-star binaries}},  {\em Phys. Rev. D}
  {\bf 83} (2011) 044011, [\href{http://arxiv.org/abs/1101.3940}{{\tt
  arXiv:1101.3940}}]. [Erratum: Phys.Rev.D 95, 109901 (2017)].

\bibitem{Caprini:2019pxz}
C.~Caprini, D.~G. Figueroa, R.~Flauger, G.~Nardini, M.~Peloso, M.~Pieroni,
  A.~Ricciardone, and G.~Tasinato, {\it {Reconstructing the spectral shape of a
  stochastic gravitational wave background with LISA}},  {\em JCAP} {\bf 11}
  (2019) 017, [\href{http://arxiv.org/abs/1906.09244}{{\tt arXiv:1906.09244}}].

\bibitem{Banagiri:2021ovv}
S.~Banagiri, A.~Criswell, T.~Kuan, V.~Mandic, J.~D. Romano, and S.~R. Taylor,
  {\it {Mapping the gravitational-wave sky with LISA: a Bayesian spherical
  harmonic approach}},  {\em Mon. Not. Roy. Astron. Soc.} {\bf 507} (2021),
  no.~4 5451--5462, [\href{http://arxiv.org/abs/2103.00826}{{\tt
  arXiv:2103.00826}}].

\bibitem{Koivisto:2013fta}
T.~Koivisto, D.~Wills, and I.~Zavala, {\it {Dark D-brane Cosmology}},  {\em
  JCAP} {\bf 06} (2014) 036, [\href{http://arxiv.org/abs/1312.2597}{{\tt
  arXiv:1312.2597}}].

\bibitem{Palti:2019pca}
E.~Palti, {\it {The Swampland: Introduction and Review}},  {\em Fortsch. Phys.}
  {\bf 67} (2019), no.~6 1900037, [\href{http://arxiv.org/abs/1903.06239}{{\tt
  arXiv:1903.06239}}].

\bibitem{vanBeest:2021lhn}
M.~van Beest, J.~Calder\'on-Infante, D.~Mirfendereski, and I.~Valenzuela, {\it
  {Lectures on the Swampland Program in String Compactifications}},
  \href{http://arxiv.org/abs/2102.01111}{{\tt arXiv:2102.01111}}.

\bibitem{Grana:2021zvf}
M.~Gra\~na and A.~Herr\'aez, {\it {The Swampland Conjectures: A Bridge from
  Quantum Gravity to Particle Physics}},  {\em Universe} {\bf 7} (2021), no.~8
  273, [\href{http://arxiv.org/abs/2107.00087}{{\tt arXiv:2107.00087}}].

\end{thebibliography}\endgroup
\end{document}